\documentclass[12pt]{article}
\usepackage{epsfig,here}
\def\gappeq{\mathrel{\rlap {\raise.5ex\hbox{$>$}}
{\lower.5ex\hbox{$\sim$}}}}

\def\lappeq{\mathrel{\rlap{\raise.5ex\hbox{$<$}}
{\lower.5ex\hbox{$\sim$}}}}

\begin{document}
\setcounter{footnote}{0}
\newcommand{\mycomm}[1]{\hfill\break
 $\phantom{a}$\kern-3.5em{\tt===$>$ \bf #1}\hfill\break}
\newcommand{\mycommA}[1]{\hfill\break
$\phantom{a}$\kern-3.5em{\tt***$>$ \bf #1}\hfill\break}
\renewcommand{\thefootnote}{\fnsymbol{footnote}}

\catcode`\@=11 % This allows us to modify PLAIN macros.
\def\lsim{\mathrel{\mathpalette\@versim<}}
\def\gsim{\mathrel{\mathpalette\@versim>}}
\def\@versim#1#2{\vcenter{\offinterlineskip
        \ialign{$\m@th#1\hfil##\hfil$\crcr#2\crcr\sim\crcr } }}
\catcode`\@=12 % at signs are no longer letters
\def\beq{\begin{equation}}
\def\eeq{\end{equation}}
\def\MSbar {\hbox{$\overline{\hbox{\tiny MS}}\,$}}
\def\eff{\hbox{\scriptsize eff}}
\def\res{\hbox{\scriptsize res}}
\def\FP{\hbox{\tiny FP}}
\def\PV{\hbox{\tiny PV}}
\def\IR{\hbox{\tiny IR}}
\def\UV{\hbox{\tiny UV}}
\def\ECH{\hbox{\tiny ECH}}
\def\NP{\hbox{\tiny NP}}
\def\APT{\hbox{\tiny APT}}
\def\QCD{\hbox{\tiny QCD}}
\def\CMW{\hbox{\tiny CMW}}
\def\SDG{\hbox{\tiny SDG}}
\def\pinch{\hbox{\tiny pinch}}
\def\brem{\hbox{\tiny brem}}
\def\V{\hbox{\tiny V}}
\def\BLM{\hbox{\tiny BLM}}
\def\NLO{\hbox{\tiny NLO}}
\def\PT{\hbox{\tiny PT}}
\def\PA{\hbox{\tiny PA}}
\def\1loop{\hbox{\tiny 1-loop}}
\def\2loop{\hbox{\tiny 2-loop}}
\def\mysim{\kern -.1667em\lower0.8ex\hbox{$\tilde{\phantom{a}}$}}
\def\a{\bar{a}}

\begin{titlepage}
%\begin{flushleft}
%\end{flushleft}
\begin{flushright}
{\small CERN-TH/2001-083}\\
{\small March, 2001}

\end{flushright}
\vspace{.13in}

\begin{center}
{\Large {\bf Renormalon resummation and exponentiation }}\\
{\Large{\bf of soft and collinear gluon radiation}}\\
{\Large {\bf
in the thrust distribution\footnote{Research
supported in part by the EC
program ``Training and Mobility of Researchers'', Network
``QCD and Particle Structure'', contract ERBFMRXCT980194.}}}

\vspace{.4in}

{\bf Einan Gardi}\,\,\,\,\,and\,\,\,\,\,{\bf Johan Rathsman}

\vspace{0.25in}

TH Division, CERN, CH-1211 Geneva 23, Switzerland\\

\vspace{.4in}

\end{center}
\noindent {\small {\bf Abstract:} The thrust distribution in
$e^+e^-$ annihilation is calculated exploiting its
 exponentiation property in the two-jet region $t=1-T\ll 1$.
We present a general method~(DGE) to calculate a large class of
logarithmically enhanced terms, using the dispersive approach in
renormalon calculus. Dressed Gluon Exponentiation is based on the
fact that the exponentiation kernel is associated primarily with a
single gluon emission, and therefore the exponent is naturally
represented as an integral over the running coupling. Fixing the
definition of $\Lambda$  is enough to guarantee consistency with
the exact exponent to next-to-leading logarithmic accuracy.
Renormalization scale dependence is avoided by keeping all the
logs. Sub-leading logs, that are usually neglected, are
factorially enhanced and are therefore important.
Renormalization-group invariance as well as infrared renormalon
divergence are recovered in the sum of all the logs. The
logarithmically enhanced cross-section is evaluated by Borel
summation. Renormalon ambiguity is then used to study power
corrections in the peak region $Qt\gsim \Lambda$, where the
hierarchy between the renormalon closest to the origin
($\sim1/Qt$) and others ($\sim1/(Qt)^n$) starts to break down. The
exponentiated power-corrections can be described by a
shape-function, as advocated by Korchemsky and Sterman. Our
calculation suggests that the even central moments of the
shape-function are suppressed. Good fits are obtained yielding
$\alpha_s^{\MSbar}({\rm M_Z})=0.110\pm 0.001$, with a theoretical uncertainty of 
$\sim 5\%$. }

\vspace{.25in}

\end{titlepage}

\section {Introduction}

Event-shape distributions, as well as other observables that are
not completely inclusive, tend to have significant contributions
from soft and collinear gluon emission. Perturbatively, this
appears in the form of logarithmically enhanced terms, the well-known 
Sudakov logs. Another, more general source of enhancement of
perturbative coefficients, which is associated with the running
coupling, is infrared renormalons. These appear due to integration
over particularly small momenta and therefore reflect the
sensitive of the observable to large distance physics. The
perturbative expansion of an event-shape distribution has quite a
rich structure incorporating both sources of enhancement.
Resummation techniques exist for both Sudakov logs and
renormalons, separately. Yet, in many cases, both are important.
The Dressed Gluon Exponentiation (DGE) method advocated here takes into account
both these aspects.

It is understood how infrared renormalons are related to
power-corrections through sensitivity to the phase-space
boundaries~\cite{beneke}. It is also known that Sudakov logs
appear due to the same type of sensitivity
\cite{Shape_function1,DW_dist}. However, the precise relation
between the two phenomena was not addressed before, except in the
specific case of Drell-Yan production \cite{Shape_function1,BB_DY}, where the
emphasis was on the identification of the leading
power-correction. DGE allows one to address this relation directly,
and thus analyse the distribution in the peak region where the hierarchy
between the renormalons starts to break down.

Consider a generic event-shape variable $y$ which is infrared and
collinear safe~\cite{Sterman_Weinberg}, where the limit $y
\longrightarrow 0$ corresponds to the extreme two-jet
configuration. The logarithmically  enhanced part of the
differential cross-section has the form \cite{CTTW}:
 \beq \left.\frac{1}{\sigma} \frac{d\sigma}{d
y}\right\vert_{\log}\sim \frac1y \sum_{n=1}^{\infty}
\sum_{m=0}^{2n-1} c_{n,m} \left(\ln \frac1y\right)^m\, a^n,
\label{log_enhanced} \eeq where $a\equiv \alpha_s(Q^2)/\pi$. When
$y\ll 1$ a naive fixed-order calculation is not adequate and the
large logs must be resummed. However, this does not imply that
the full perturbative calculation has to be carried out to all
orders: the factorisation of soft and collinear radiation allows
one to calculate the log-enhanced part of the cross-section,
roughly speaking, ``exponentiating the one-loop result''.

It is useful to define the integrated cross-section
\[
R(y)\equiv \int_0^y\, \frac{1}{\sigma}\frac{d\sigma}{d y}\, dy
\]
where the singularity in the two-jet limit, $y\longrightarrow 0$,
is cancelled by virtual corrections, so that $R(0)=0$. The logarithmically enhanced
part of $R(y)$ has the form $\left.R(y)\right\vert_{\log} \sim
1+\sum_{n=1} R_{n}(y) a^n$ with
\[
 R_{n_{\,}}(y) = \sum_{m=1}^{2n} r_{n,m} L^m
\]
where $L\equiv \ln \frac1y$ and $r_{n,m}=-c_{n,m-1}/m$. The full
$R(y)$ can\footnote{There are various different ways to
make this separation. The one we use here is motivated by the
so-called log-R matching scheme. See~\cite{CTTW}.} be written
\cite{CTTW} as
\beq
\ln R(y)=
\left.\ln R(y)\right\vert_{\log}+\left.\ln R(y)\right\vert_{\rm {non-log}},
\label{log_match_separation}
\eeq
where the first term contains only logarithmic terms that diverge
in the two-jet limit whereas the second is finite in this limit.

Exponentiation means that $\left.\ln R(y)\right\vert_{\log}$ has
the following perturbative expansion,
\begin{eqnarray} \label{Exponentiation}
\left.\ln
R(y)\right\vert_{\log} &=&\sum_{n=1}^{\infty} \sum_{m=1}^{n+1}
G_{n,m} L^m\, a^n \\
&=& Lg_1(aL)+g_2(aL)+ag_3(aL)+a^2g_4(aL)+\cdots \nonumber
\end{eqnarray}
where the function $g_1(aL)$ resums the leading-logs (LL)
$G_{n,n+1}L^{n+1}a^n$ to all orders, $g_2(aL)$ resums the
next-to-leading logs (NLL) $G_{n,n}L^{n}a^n$, etc. Exponentiation
is a universal feature which follows from the factorisation
property of QCD matrix elements (see e.g.
\cite{Contopanagos:1997nh}), provided that the phase-space
integration also factorizes. The latter condition makes the
calculation of the exact coefficients $G_{n,m}$ rather difficult.
However, thanks to an intensive research effort in the last two
decades, there is today a detailed understanding of Sudakov logs
in a large class of QCD observables. In particular, in the case of
event-shape observables in $e^+e^-$ annihilation, like the thrust,
the jet mass parameters, the C-parameter, and jet broadening
parameters, the state of the art is resummation of $\left.\ln
R(y)\right\vert_{\log}$ up to NLL accuracy~\cite{CTTW,C-par,Broad}.

Current phenomenology of event shape 
distributions (see~\cite{Biebel:2001dm} for a recent review) 
is based on NLL resummation, neglecting $g_k(aL)$ for $k\geq 3$ in
(\ref{Exponentiation}). The resummed cross-section is matched with
next-to-leading order calculations that are available numerically
\cite{thrust-nlo,EVENT,EVENT2}. This combination allows a good qualitative
description of the observables, with the main caveat being the
fact that ``hadronization corrections'', which are not under
control theoretically, are required to bridge the gap between the
calculation and the measurements.

Before alluding to non-perturbative physics it is worthwhile to
re-examine the perturbative calculation. It was already shown in
the example of the average thrust, that a major part of the
discrepancy between the next-to-leading order calculation and the
data can be explained by resummation of higher-order perturbative
terms associated with infrared renormalons~\cite{average_thrust}.
Physically the Single Dressed Gluon (SDG) resummation
of~\cite{average_thrust} allows a better description of the
branching of a {\em single} gluon emitted from the primary
quark--anti-quark pair. Following the spirit of
BLM~\cite{BLM,conf}, this resummation takes into account the fact
that the physical scale of the emitted gluon is its virtuality,
rather than the centre of mass energy $Q$, which is usually used
as the scale of the coupling. As emphasized in~\cite{conf}, the
essence of this approach is the idea that the perturbative
expansion should be reorganized in a ``skeleton expansion'' in
analogy with the Abelian theory. Then, the separation between
running coupling effects and conformal effects is unique and
physical. We shall see here, that in the specific context of the
Sudakov region, one has access to sub-leading terms in the
``skeleton expansion'', corresponding to multiple gluon emission,
through exponentiation.

In order to examine the approximations in the standard NLL perturbative
treatment of the thrust distribution, let us first recall two main
features of the resummation~(\ref{Exponentiation}). The first is
the fact that the calculation of the functions $g_k(aL)$ is
based on an integral over the running
coupling. The second is that each $g_k(aL)$ resums a {\em
convergent} series in $aL$ (for $aL\ll 1$). At first sight these facts
seem contradictory: usually integrals over the running coupling
translate into renormalon factorial divergence. The resolution is
that the expansion (\ref{Exponentiation}) itself is divergent. The
{\em functions} $g_k(aL)$ increase factorially with $k$, thus endangering
the validity of the truncation at a certain logarithmic accuracy,
e.g. NLL.
Consequently, when examining the logs $G_{n,m}L^{m}$
at a fixed order $n$ in eq.~(\ref{Exponentiation}), one finds that
the sub-leading logs are enhanced by a relative factor of $m!$
compared to the leading ones. In the region where the
perturbative treatment holds ($aL\ll 1$) this numerical factor
can over-compensate the power of the log and thus invalidate
the expansion.

Another (related) argument for keeping track of sub-leading logs
is the fact that integrals over the running coupling yield at once
all powers of $L$. Therefore, truncation at a certain logarithmic
accuracy implies renormalization scheme and scale dependence. A
renormalization-group invariant calculation of $\left.\ln
R(y)\right\vert_{\log}$ requires keeping {\em all the logs}. Of
course, the exact calculation of all the logs is far beyond reach.
However, a specific class of logs is fully accessible: these are
the logs that are leading in $\beta_0$, $G_{n,m}\sim
\beta_0^{n-1}$, namely the ones associated with a single dressed
gluon (SDG) emission. Within this class all the sub-leading logs,
i.e. the functions $g_k(aL)$, are calculable. Moreover, since the
exponentiation kernel is primarily associated with a single gluon,
the entire logarithmically enhanced cross-section, which includes
any number of gluons, can be generated in some approximation by
exponentiating the SDG cross-section. The result is fully
consistent with NLL accuracy, provided an adequate choice of the
coupling (the ``gluon bremsstrahlung'' scheme~\cite{CMW}) is made.
Our exponentiation procedure is based on the standard assumption that
gluons emitted from the primary quark and anti-quark are
completely independent and contribute additively to the thrust
(this assumption holds for soft gluons). Given this assumption
the SDG cross-section exponentiates under the Laplace transform at any
logarithmic accuracy. Thus, through integrals over the running
coupling we are able to calculate $\left.\ln
R(y)\right\vert_{\log}$ with the same formal logarithmic accuracy
(NLL), yet avoiding any truncation of sub-leading logs, which
would inevitably violate renormalization group invariance. In this
way we resum at once Sudakov logs and renormalons.

Comparison of the perturbatively calculated event-shape
distribution with data requires some phenomenological model to
deal with hadronization. Traditionally, hadronization effects are
included using detailed Monte-Carlo hadronization models such as HERWIG~\cite{HERWIG} or
PYTHIA~\cite{PYTHIA}. An
attractive alternative that became popular recently is to
parametrise the effects in terms of a small number of
non-perturbative parameters which control power-suppressed
contributions. The basic assumption, which we adopt, is that the
form of the most important non-perturbative corrections can be
deduced from the ambiguities of the perturbative
result~\cite{wise,W,DW,AZ,DMW}.
 It was shown~\cite{Shape_function1,Shape_function2,DW_dist} that the primary
non-perturbative effect is a shift of the perturbatively calculated
distribution to larger values of $y$:
\beq
\left.\frac{1}{\sigma}
\frac{d\sigma}{d y}\right\vert_{\PT}(y)\longrightarrow
\left.\frac{1}{\sigma} \frac{d\sigma}{d
y}\right\vert_{\PT}(y-\lambda_1\Lambda/Q)
\label{shift1}
\eeq where $\lambda_1$
is a non-perturbative parameter. According to the model of
\cite{DW_dist}, this parameter is expressed as an observable
dependent factor which is perturbatively calculable times the
``effective infrared coupling'' (see \cite{DMW}). The shift~(\ref{shift1})
allows a fair description of the data at
all measured energies and in a wide range of $y$ values for
several event-shapes in terms of a single parameter -- the
``effective infrared coupling''. The success of this
perturbatively motivated approach suggests that the perturbative
calculation itself, if pushed further, may be much closer to the
data. It should be emphasised though that non-perturbative effects
are essential in the two-jet limit when $aL \sim {\cal O}(1)$, and
in particular in determining the precise location of the
distribution peak and its shape. In this region the above-mentioned 
approach, which uses a single non-perturbative
parameter, is insufficient.
In~\cite{Shape_function1,Shape_function2,Shape_function3,Shape_function4,Korchemsky_Tafat}
a more general approach was suggested, namely to replace the infinite sum of
power-corrections of the form $\lambda_n\,\Lambda^n/(yQ)^n$ which exponentiate
by a single shape-function whose central moments are $\lambda_n$.

As a consequence of infrared renormalons
the resummation of $\left.\ln R(y)\right\vert_{\log}$ contains
power-suppressed ambiguities which signal the necessity of
non-perturbative corrections. The fact that
renormalons occur in the exponent strongly suggest that
non-perturbative power corrections also exponentiate, in
accordance with the shift and the shape-function approaches.
We show that the renormalon ambiguity
contains valuable information on the non-perturbative corrections,
which is not restricted to the level of the leading power
correction. In particular, it is possible to deduce the $y$ dependence\footnote{The idea that the
dependence of power-corrections on external variables can be deduced from the functional form
of the renormalon ambiguity is not new. It was used e.g. in~\cite{DMW} for the case of higher-twist in
deep-inelastic structure functions and in~\cite{Beneke:1997sr} for the case of power corrections to
fragmentation functions in $e^+e^-$ annihilation. In the latter reference (see also~\cite{beneke}) 
it was justified by the `ultraviolet dominance' assumption. Here we use this idea at the level of 
the exponent.} of each power correction $1/Q^n$ from the corresponding residue of the Borel transform 
of the exponent.
This allows us to make definite
predictions concerning the form of the non-perturbative corrections. In the specific case of the
thrust distribution the main conclusion from this analysis is that the even central moments
of the shape-function are suppressed. Moreover, in the case of the second central moment, an additional
cancellation occurs between large-angle contributions, which are included in the shape-function
of refs.~\cite{Shape_function4,Korchemsky_Tafat}, and a collinear contribution which was not
addressed before.
 
Another important aspect which our approach incorporates
is the complementarity between perturbative and non-perturbative contributions, namely renormalon
resummation and explicit parametrisation of power corrections~\cite{average_thrust}.
It was clearly shown in~\cite{average_thrust} that a consistent treatment of
non-perturbative power-corrections as well as a reliable determination of
$\alpha_s$ requires renormalon resummation. Such resummation is
performed here for the first time at the level of the differential cross-section.

This paper deals with a specific observable, the thrust
distribution. Nevertheless, DGE is
a general method that can be applied in any case where Sudakov
logs appear. It makes a direct link between Sudakov resummation and
renormalons. This paper is organised as follows: we begin in
section~2 by recalling the calculation of the thrust distribution
in the single dressed gluon approximation~\cite{higher_moments}.
In principle, this calculation does not require any specific
kinematic approximations, however, it is technically simpler if
one neglects~\cite{CMW,average_thrust,higher_moments}
non-inclusive decay~\cite{Nason_Seymour,Beneke:1997sr} of the gluon into
opposite hemispheres. Here we adopt this approximation and provide
further evidence for its validity at small values of $1\,-$
thrust. We then observe that log-enhanced terms dominate the
single dressed gluon result in a large range of thrust values.
This justifies concentrating on the log-enhanced cross-section of
the form (\ref{log_enhanced}). In section 3 we calculate the
resummed cross-section by exponentiating the logarithmically
enhanced part of the SDG result. We also analyse there the
renormalon structure of the exponent and the relation with
power-correction models mentioned above. In section 4 we compare our approach
with known results. First we refer to the NLL calculation which is based on 
solving an evolution equation for the jet mass distribution~\cite{CTTW}.
We show that the two calculations coincide up to NLL accuracy, provided an appropriate choice of the
coupling is made~\cite{CMW}. Next we briefly recall the matching procedure of the resummed result with 
the full NLO calculation.  In section 5 we study the
phenomenological implication of the suggested resummation by fitting to data
in a large range of energies and in section 6 we summarise our conclusions.

\section {Thrust distribution from Single Dressed Gluon}

The purpose of this section is to calculate the thrust distribution in case of a single gluon emission.
The result will be used in section 3 to calculate the physical multi-gluon cross-section, through exponentiation.
We begin by recalling the calculation of the Single Dressed Gluon characteristic function (sec.~2.1). Then, in
sec.~2.2, we isolate the logarithmically enhanced terms and study their structure and finally, in sec~2.3,
we construct the Borel representation of the result.

\subsection{Single Dressed Gluon (SDG) characteristic function}

In renormalon calculations~\cite{Beneke:1995qe,Ball:1995ni,DMW,beneke} 
the all-order perturbative result, which resums all
the terms that are leading in $\beta_0$, can be expressed as a 
time-like momentum integral of an observable dependent
Single Dressed Gluon (SDG) characteristic function, times an 
observable independent effective-charge:
\beq
\int_0^{\infty}\frac{d\epsilon}{\epsilon}
 \dot{\cal F}(\epsilon)\,\bar{A}_{\eff}(\epsilon Q^2)\,=\,
\int_0^{\infty}\frac{d\epsilon}{\epsilon} 
\left[{\cal F}(\epsilon)-{\cal F}(0)\right]\,\bar{\rho}
(\epsilon Q^2)
\label{dispersive}
\eeq
where $\bar{A}(k^2)\equiv  \beta_0{\a}(k^2)=\beta_0\bar{\alpha}_s(k^2)/\pi$, 
and the bar stands for a specific renormalization-scheme. 
In the large $\beta_0$ limit (large $N_f$ limit)
$\bar{A}(k^2)$ is related to the $\rm {\overline{MS}}$ coupling
${A}(k^2)\equiv \beta_0{\alpha}^{\MSbar}_s(k^2)/\pi$ by
\beq
\bar{A}(k^2)=\frac{{A}(k^2)}{1-\frac53 {A}(k^2)}.
\label{A_bar}
\eeq
Going beyond this limit is discussed in sec.~4,
where we identify $\bar{A}$ with the ``gluon bremsstrahlung''
coupling~\cite{CMW}, thus making the DGE result exact to NLL accuracy.

The two integrals in (\ref{dispersive}) are related by integration by parts:
\beq
\dot{\cal F}(\epsilon)\equiv-\epsilon\frac{d}{d\epsilon}\,
{\cal F}(\epsilon),
\eeq
and the function $\bar{\rho}(\mu^2)$ is
identified as the time-like discontinuity of the
coupling,
\begin{equation}
\bar{\rho}(\mu^2) = \frac{1} {2\pi i}
 {\rm Disc}\left\{\bar{A}(-\mu^2)\right\}
\equiv
\frac{1}{2\pi i}\left[\bar{A}\left(-\mu^2+i\epsilon\right)
-\bar{A}\left(-\mu^2-i\epsilon\right)\right].
\label{discpt}
\end{equation}
The ``time-like coupling'' $\bar{A}_{\eff}(\mu^2)$ in (\ref{dispersive}) obeys
\beq
\mu^2\,\frac{d\bar{A}_{\eff}(\mu^2)}{d\mu^2}=\bar{\rho}(\mu^2).
\eeq
For example, in the one-loop case $\bar{A}(k^2)=1/(\ln k^2/\bar{\Lambda}^2)$ 
and~\cite{Beneke:1995qe,Ball:1995ni}
\beq
\bar{A}_{\eff}(\mu^2)
=\frac12-\frac{1}{\pi}\arctan\left(\frac{1}{\pi}\ln\frac{\mu^2}{\bar{\Lambda}^2}\right).
\label{1_loop_time_like}
\eeq

Eq.~(\ref{dispersive}) which represents the all-order perturbative sum is not yet defined to power
accuracy. The precise definition will be given in sec.~3.3 using Borel summation.  
A delicate issue, which was addressed 
in~\cite{Beneke:1995qe,Ball:1995ni,Grunberg:1998ix,average_thrust} is the fact that the dispersive
integral~(\ref{dispersive}) over the time-like perturbative coupling (e.g.~(\ref{1_loop_time_like}))
is formally well-defined but it differs from the Borel sum by power corrections, which are not entirely 
related to large distances. In~\cite{DMW} the dispersive integral is given non-perturbative meaning
by replacing the perturbative coupling by a non-perturbative coupling which is regular in the space-like
region (i.e. it does not have a Landau singularity) and obeys a dispersion relation. 
Here we do not make this replacement. We refrain from attaching 
any non-perturbative meaning to~(\ref{dispersive}). Large distance power-corrections will be deduced 
directly from the ambiguity of the Borel sum (sec.~3.5). Eventually, the normalisation of 
these power corrections can be expressed as an integral over an infrared-finite coupling, which is
conjectured~\cite{DMW} to be universal. 

The characteristic function ${\cal F}(\epsilon)$ itself can be
calculated either using the Abelian large $N_f$ limit followed by
naive non-Abelianization~\cite{beneke}, or in the dispersive
approach~\cite{Beneke:1995qe,Ball:1995ni,DMW}, referring to $\mu^2=\epsilon Q^2$ as the
squared gluon mass. In case of inclusive observables, such as the
total cross-section in $e^+e^-$ annihilation, the two calculations
coincide. The two
differ~\cite{Nason_Seymour,DMW,Beneke:1997sr,average_thrust,higher_moments,Korchemsky_Tafat},
however, in the case of observables such as event-shape variables,
that are not completely inclusive with respect to the decay
products of the gluon. Here we shall use the massive-gluon
characteristic function for the thrust distribution, which was
calculated in \cite{higher_moments}. It corresponds to an
inclusive integration over the decay products of the gluon, so
that the possibility that a gluon decays into partons that end up
in opposite hemispheres is not taken into account. It is important
to emphasise, though, that the steps that follow can be repeated
in very much the same way starting with the large $N_f$ result,
once available. In the case of the thrust, the inclusive
approximation is expected to be good (some evidence is given
below). Note that for some event-shapes, e.g. the heavy jet mass,
this is probably not so.

Let us now briefly recall the calculation of the thrust distribution characteristic function of
ref.~\cite{higher_moments}.
At the SDG level, the distribution of $t=1\,-$ thrust is
\beq
\left.\frac{1}{\sigma}\frac{d\sigma}{dt}(Q^2,t)\right\vert_{\SDG}\!\!\!\!
=\frac{C_F}{\beta_0}\int_0^{1}\frac{d\epsilon}{\epsilon}\,\bar{A}_{\eff}
(\epsilon Q^2) \dot{\cal F}(\epsilon,t)=
\frac{C_F}{\beta_0}\int_0^{1}\frac{d\epsilon}{\epsilon}\,\bar{\rho}
(\epsilon Q^2) \left[{\cal F}(\epsilon,t)-{\cal F}(0,t)\right].
\label{res}
\eeq
The characteristic function ${\cal F}(\epsilon,t)$ is obtained
from the following integral over phase-space,
\beq
{\cal F}(\epsilon,t)=\int_{\rm {phase\,\, space}} dx_1 dx_2 \, {\cal
M}(x_1,x_2,\epsilon) \, \delta\left(1-T(x_1,x_2,\epsilon)-t\right)
\label{F_def}
\eeq
where
$C_F\,a \,{\cal M}$ is the squared tree level matrix
element for the production of a quark--anti-quark
pair and a gluon of virtuality
$\mu^2\equiv \epsilon Q^2$, and
\beq
{\cal M}(x_1,x_2,\epsilon)=\frac12\left[
\frac{(x_1+\epsilon)^2+(x_2+\epsilon)^2}{(1-x_1)(1-x_2)}
-\frac{\epsilon}{(1-x_1)^2}-\frac{\epsilon}{(1-x_2)^2}\right].
\label{M}
\eeq
The integration variables $x_{1,2}$ represent the energy fraction
of each of the quarks in the centre-of-mass frame. The energy fraction
of the gluon is $x_3=2-x_1-x_2$.

We use the following~\cite{DMW,average_thrust,higher_moments} definition of the thrust
\beq
T=\frac{\sum_i \left\vert \vec{p}_i \cdot \vec{n}_T\right\vert}
{\sum _i E_i }=\frac{\sum_i \left\vert \vec{p}_i \cdot \vec{n}_T\right\vert}
{Q}.
\label{T_def2}
\eeq
In case of three partons (a quark, an anti-quark  and a ``massive''
gluon) it yields~\cite{average_thrust},
\beq
1-T(x_1,x_2,\epsilon)={\rm min}\left\{ 1-x_1\,, \,1-x_2\,
,\,1-\sqrt{(2-x_1-x_2)^2-4\epsilon}
\right\}.
\label{thrust}
\eeq

Evaluating (\ref{F_def}) one obtains \cite{higher_moments} the characteristic function for the
thrust distribution,
\begin{equation}
{\cal F}(\epsilon,t)= \left\{\begin{array}{lll}
{\cal F}_Q^l(\epsilon,t)+{\cal F}_G(\epsilon,t)&\,\,\,\,\,\,\,\,\,\,
& \epsilon<t<\sqrt{\epsilon}\\
{\cal F}_Q^h(\epsilon,t)+{\cal F}_G(\epsilon,t)&
&\sqrt{\epsilon}<t<\frac23\sqrt{1+3\epsilon}-\frac13
\end{array}\right.
\label{F_dist}
\end{equation}
where the dominant contribution  ${\cal F}_Q(\epsilon,t)$ corresponds to
the phase-space regions where one of the primary quarks carries the
largest momentum ($T=x_{1,2}$) and ${\cal F}_G(\epsilon,t)$ corresponds
to the region where the gluon momentum is the largest. The superscripts $l$ and $h$ on ${\cal
F}_Q(\epsilon,t)$ denote low and high $t$ values, respectively.
These functions are given by
\begin{eqnarray}
\label{F}
{\cal F}^h_{Q}(\epsilon,t) &=&  - \frac{1}{t}\left[(1 -
t + \epsilon )^{2} + (1 + \epsilon )^{2}\right]\,\ln{ \frac {t}{q - t}}
+\left(3 - 2\,{ \frac {q}{t}}  + {
\frac {1}{2}} \,{ \frac {1}{t}}  + {
\frac {1}{2}} \,t - q\right) \nonumber \\
&+& \left(4 - 2\,{ \frac {q}{t}}  + 3\,
{ \frac {1}{t}}  - { \frac {q}{t^{2}}}
 + { \frac {1}{q - t}} \right)\,\epsilon \nonumber \\
{{\cal F}^l_{Q}}(\epsilon,t) &=&  - \frac{1}{t}\left[(1
 - t + \epsilon )^{2} + (1 + \epsilon )^{2}\right]\,\ln
{ \frac {\epsilon }{t\,(q - t)}}
+\left(1 - 2\,{ \frac {q}{t}}  + { \frac {1}{
2}} \,{ \frac {1}{t}}  - q\right) \nonumber \\
&+& \left(3\,{ \frac {1}{t}}  + { \frac {1}{q - t}}  -
{ \frac {q}{t^{2}}}  + 2\,{ \frac {1}{t
^{2}}}  + 2 - 2\,{ \frac {q}{t}} \right)\,\epsilon   +
\left(2\,+\, \frac {1}{2t}  \right)\,\frac{\epsilon ^{2}}{t^2}\\ \nonumber
{\cal F}_{G}(\epsilon,t) &=&
\frac{1 - t}{q^{2}} \left[
\left((1 + \epsilon )^{2} + (1 + \epsilon  - q)^{2}\right)
\ln\frac{q - t}{t} +(2t -q)\,\left(q + \frac {\epsilon}{t}
 +\frac {\epsilon}{q-t} \right) \right]
\end{eqnarray}
where $q \equiv \sqrt{(1 - t)^{2} + 4\,\epsilon}$.

\subsection{Logarithmically Enhanced Terms}

Next we examine the SDG perturbative expression (\ref{res}) with the characteristic function of
eqs. (\ref{F_dist}) and (\ref{F}). For this purpose it is convenient to
expand the time-like coupling $\bar{A}_{\eff} (\epsilon Q^2)$
in terms of a fixed-scale space-like coupling $\bar{A}(Q^2)$,
\beq
\bar{A}_{\eff} (\epsilon Q^2)=\sum_{m=0}^{\infty} \left(\bar{A}(Q^2)\right)^{m+1} \sum_{j=0}^{m}K_{m,j}
\left(\ln\frac{1}{\epsilon}\right)^j,
\label{A_eff_exp}
\eeq
where, in general, $K_{m,j}$ depend on the coefficients of the $\beta$ function which controls the evolution of
$\bar{A}(Q^2)$.
Performing the
integral (\ref{res}) over $\dot{\cal F}(\epsilon,t)\equiv-\epsilon\frac{d}{d\epsilon}\, {\cal F}(\epsilon,t)$ order by order
one gets the sum
\beq
\label{ds_log_mom}
\left.\frac{1}{\sigma}\frac{d\sigma}{dt}(Q^2,t)\right\vert_{\SDG}\!\!\!\!
=\frac{C_F}{\beta_0}\sum_{m=0}^{\infty}\left(\bar{A}(Q^2)\right)^{m+1} \sum_{j=0}^{m}K_{m,j} \, h_j(t)
\eeq
where the functions $h_j(t)$ are the log-moments of the characteristic function,
\beq
h_j(t) = \int_0^1\frac{d\epsilon}{\epsilon} \left(\ln\frac{1}{\epsilon}\right)^j \dot{\cal F}(\epsilon,t).
\label{log_moments}
\eeq
The analytic calculation of $h_j(t)$ with the full characteristic function of eqs.~(\ref{F_dist}) and~(\ref{F})
is rather involved. Alternatively, concentrating in the small $t$ region, we distinguish
in $h_j(t)$ between logarithmically enhanced terms which are not suppressed by any power of $t$ and the rest,
which we eventually neglect.

As we shall see, the logarithmically enhanced terms, which have the form of eq.~(\ref{log_enhanced}),
can be easily computed analytically to all orders.
Using (\ref{F_dist}) for $t<1/3$ we obtain
\beq
h_j(t) = \int_0^t\frac{d\epsilon}{\epsilon} \left(\ln\frac{1}{\epsilon}\right)^j \dot{\cal F}_G(\epsilon,t)\,+\,
\int_0^{t^2}\frac{d\epsilon}{\epsilon} \left(\ln\frac{1}{\epsilon}\right)^j \dot{\cal F}_Q^h(\epsilon,t)
\,+\,\int_{t^2}^t\frac{d\epsilon}{\epsilon} \left(\ln\frac{1}{\epsilon}\right)^j \dot{\cal
F}_Q^l(\epsilon,t).
\label{log_moments_w_limits}
\eeq
We notice that a term of the form $\epsilon^n/t^k$ in $\dot{\cal F}(\epsilon,t)$ contributes to $h_j(t)$ a term proportional to
$t^{n-k}$ (times a polynomial in $\ln \frac{1}{t}$) from the collinear gluon emission limit of
phase-space\footnote{A detailed analysis of the phase-space can be found in \cite{higher_moments}. See figure~1 there.}
where $\epsilon=t$, and a term proportional to $t^{2n-k}$ (times a polynomial in $\ln \frac{1}{t}$) from the large-angle
soft gluon emission limit of phase-space where $\epsilon=t^2$.
Taking the double expansion of (\ref{F}) at small $t$ and $\epsilon$ we have
\begin{eqnarray}
\label{F_dot_expanded}
\dot{\cal F}^h_{Q}(\epsilon,t) &\simeq& \frac{\epsilon}{t^2}+ \frac{{\cal O}\left(\epsilon\right)}{t} \\
\dot{\cal F}^l_{Q}(\epsilon,t) &\simeq&  -\frac{\epsilon^2}{t^3}-\frac{\epsilon}{t^2}+\frac{2}{t}+{\cal O}\left( 1 \right)
\nonumber \\
\dot{\cal F}_{G}(\epsilon,t) &\simeq& \frac{\epsilon}{t}+ {\cal O}\left( 1 \right)\nonumber
\end{eqnarray}
where we ignored higher powers of $\epsilon$ multiplying a given power of $1/t$, which are not relevant.
From (\ref{log_moments_w_limits}) and (\ref{F_dot_expanded}) we conclude that only
$\dot{\cal F}^l_{Q}(\epsilon,t)$ contributes to the logarithmically enhanced cross-section
and thus
\beq
\left. h_j(t)\right\vert_{\log} =\int_{t^2}^t\frac{d\epsilon}{\epsilon} \left(\ln\frac{1}{\epsilon}\right)^j
\left[-\frac{\epsilon^2}{t^3}-\frac{\epsilon}{t^2}+\frac{2}{t}\right].
\label{log_moments_relevant}
\eeq
Performing the integral over $\epsilon$ yields
\begin{eqnarray}
\left. h_j(t)\right\vert_{\log}& =&\frac1t\left[2 \, \frac{2^{j+1}-1}{j+1} L^{j+1}
\,-\, e^L\Gamma\left(j+1,L\right)\,-\, 2^{-j-1}\, e^{2L}\Gamma\left(j+1,2L\right)
\right]
\nonumber \\
& =&\frac1t \left[2 \,\frac{2^{j+1}-1}{j+1} L^{j+1}\,-\,\sum_{k=0}^j\frac{j!}{k!}\,
\left(1+2^{k-j-1}\right)\,
L^{k}\right]
\label{log_moments_explicit}
\end{eqnarray}
where $L\equiv \ln\frac{1}{t}$.

Having evaluated the log-moments we can substitute them into (\ref{ds_log_mom}) to obtain the
logarithmically enhanced SDG differential cross-section to arbitrarily high order.
The result (\ref{log_moments_explicit}) has a few features that deserve attention.
First we note that the leading-logs in $h_j(t)$ depend on both collinear and large-angle soft gluon
limits of phase-space, while the sub-leading logs are associated with the collinear limit alone.
Next, we observe that the leading-logs in $h_j(t)$ originate in the $\epsilon$ independent term $2/t$ in
$\dot{\cal F}^l_{Q}(\epsilon,t)$, while the sub-leading logs originate in the two $\epsilon$ dependent terms,
$-\epsilon^2/t^3-\epsilon/t^2$. This $\epsilon$ dependence is reflected in factorial
increase of the coefficients (see below) which is associated with the sensitivity of the thrust to
collinear gluon emission.

It is important to notice that the factorial growth is associated with sub-leading logs.
Considering the different logs $L^k$ in a given log-moment $h_j(t)$ we find that with
decreasing power~$k$ the numerical coefficient increases as ${j!}/{k!}$.
Note, in particular, that the next-to-leading logs $L^j$ (which are the
only sub-leading logs taken into account in the standard analysis) have a small coefficient ($\frac32$),
while further sub-leading logs (which are usually ignored) are multiplied by large numbers.
It is useful to analyse the square brackets in (\ref{log_moments_explicit}) by considering
two extreme limits for the function $F(L)\equiv e^{L}\Gamma(j+1,L)$. In the limit $L \longrightarrow 0$,
one finds factorial growth $F(L)\sim j!$. In the limit $L \longrightarrow \infty$ one
can keep only the leading logarithmic term in $F(L)$, i.e. $F(L)\sim L^j$, recovering the standard NLL result.
The fact that the latter formally holds may be of little relevance in practice
given that the entire perturbative treatment is adequate only when $L\ll 1/\bar{A}(Q^2)$.
One can now examine the validity of these two extreme approximations for phenomenologically interesting values
of $L$. Consider for example the case $t=0.05$,  $L=\ln(1/0.05)\simeq 3$ at order $j=3$: the value
is $F(L)=77.8$ while the NLL gives $L^j=26.9$ and the naive factorial gives $j!=6$. Clearly, 
none of these extreme approximations is valid. One must retain all the logs.

\begin{figure}[t]
\begin{center}
\mbox{\kern-0.5cm\epsfig{file=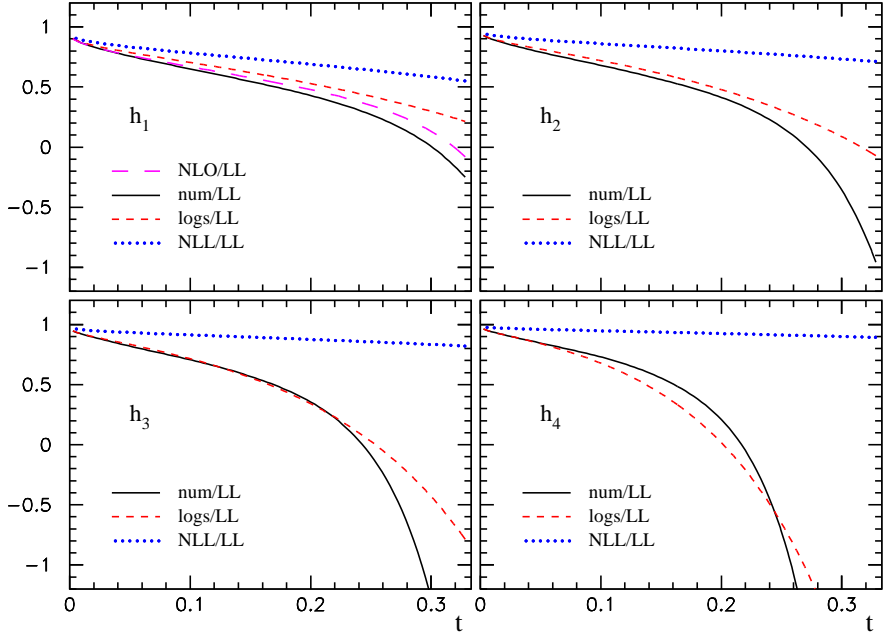,width=12.0truecm,angle=0}
}
\end{center}
\caption{The functions $h_j(t)$, $j=1$ through $4$, in different
approximations normalised to the leading logarithm (LL). The
complete result (calculated numerically) is the full line and the
sum of the logarithmically enhanced terms
$\left.h_j(t)\right\vert_{\rm log}$ (calculated analytically) is
dashed, the NLL approximation is dotted. For the first moment, the
$\beta_0$ dependent part of the NLO result is also shown for
comparison.
 }
\label{log_mom_fig}
\end{figure}

One may wonder to what extent it is justified to consider only the logarithmically enhanced part of
the log moments. Factorial increase is certainly not restricted to the logarithmically enhanced
part. The immediate answer is, of course, that the neglected part is parametrically smaller, by a factor of $t$
compared to the one kept. This answer is not complete: as we know, parametric suppression can be compensated
by large numerical coefficients. To be convinced that this approximation is valid at physically interesting
values of $t$ let us now compare the logarithmic part of the log-moments $h_j(t)$ with the full function
which we evaluate numerically. Writing $h_j(t)=\left.h_j(t)\right\vert_{\rm log}+\Delta h_j(t)$,
where the logarithmically enhanced part $\left.h_j(t)\right\vert_{\rm log}$ is given
by~(\ref{log_moments_explicit})
and $\Delta h_j(t)$ are suppressed by a relative factor of~$t$,
the first few log-moments are
\begin{eqnarray*}
h_1(t) & = & \frac{1}{t}\left[3L^2- \frac{3}{2}L-\frac{5}{4} + \Delta h_1(t) \right] \\
h_2(t) & = & \frac{1}{t}\left[\frac{14}{3}L^3-\frac{3}{2}L^2- \frac{5}{2}L-\frac{9}{4} + \Delta h_2(t) \right] \\
h_3(t) & = & \frac{1}{t}\left[\frac{15}{2}L^4-\frac{3}{2}L^3-\frac{15}{4}L^2-\frac{27}{4}L-\frac{51}{8} + \Delta h_3(t) \right] \\
h_4(t) & = & \frac{1}{t}\left[\frac{62}{5}L^5-\frac{3}{2}L^4-5L^3-\frac{27}{2}L^2-\frac{51}{2}L-\frac{99}{4}
+ \Delta h_4(t) \right]
\end{eqnarray*}
The $h_j(t)$ and thus $\Delta h_j(t)$ are calculated by
integrating~(\ref{log_moments}) numerically where $\dot {\cal F}(\epsilon,t)$ is the derivative
of the {\em full} characteristic function~(\ref{F}).
The comparison for the first four log-moments is presented in fig.~\ref{log_mom_fig}.
We see that while the NLL approximation is quite far from the
actual values of $h_j(t)$ even at rather small~$t$, there is a
fairly wide range in which the logarithmic approximation
$\left.h_j(t)\right\vert_{\rm log}$ is good. At large $t$ values
the gap between the two increases. However, since our interest
here is in improving the calculation in the small $t$ region, we
consider this agreement satisfactory. It should be kept in
mind, though, that our resummation procedure is not the
appropriate way to improve the calculation of the cross-section at
$t\gsim 0.2$. We comment that for higher log-moments $j\geq 5$ the
approximation deteriorates even at smaller $t$ values; however, the resulting
discrepancy at the level of the SDG cross-section~(\ref{ds_log_mom})
is not large. We shall revisit the significance of terms
that are suppressed by $t$ in the phenomenological analysis in sec.~5.

Finally, using~(\ref{A_bar}), the first few orders of the SDG calculation~(\ref{ds_log_mom}) of the
log-enhanced cross-section are given in the $\overline{\rm MS}$ scheme by
\begin{eqnarray}
\label{our_res}
&&\left.\frac{1}{\sigma}\frac{d\sigma}{dt}(Q^2,t)\right\vert_{\SDG}=\frac{C_F^{\,}}{t\,\beta_0}\,\left\{
(2\,L - 1.5)\, {A(Q^2)}+ (3\,L^{2}+1.8333\,L - 3.75 )\,
{A(Q^2)}^{2} \right. \\ \nonumber
 & &+ \,\,   (4.6667\,L^{3} + 8.5000\,L^{2} - 8.5242\,L -
5.6481)\,{A(Q^2)}^{3} \\ \nonumber
 & &+\,\,    (7.5000\,L^{4} + 21.833\,L^{3} - 15.859\,L^{2} -
40.585\,L + 2.025)\,{A(Q^2)}^{4} \\ \nonumber
 & &\left. +\,\,    (12.400\,L^{5} + 48.500\,L^{4} - 29.339\,L^{3} -
175.72\,L^{2} - 47.164\,L + 40.22)\,{A(Q^2)}^{5}+\cdots \right\}
\end{eqnarray}

The leading and next-to-leading logs are consistent with previous
calculations~\cite{CTTW}. In addition, at NLO we can compare the
next-to-next-to-leading log ($-3.75$  in~(\ref{our_res})) with the
exact numerical computation~\cite{EVENT2}. To make an explicit
comparison, one should separate the contributions of
different colour factors in the NLO coefficient (this was done
before using the EVENT program~\cite{EVENT} in 
\cite{Magnoli:1990mt,Kluth_Biebel}) and then fit the numerical results
to a parametric form which includes a log-enhanced part, and terms
which are suppressed by powers of $t$. The result of such a fit,
where the first two logs are fixed to their exact values, gives
the following log-enhanced NLO expression in the large $\beta_0$ limit,
\begin{eqnarray}
\label{exact}
\left.\frac{1}{\sigma}\frac{d\sigma}{dt}(Q^2,t)\right\vert_{\NLO}\!\!\!\!=\frac{C_F^{\,}}{t\,\beta_0}\,\left\{
(2\,L - 1.5)\, {{A(Q^2)}}+ (3\,L^{2}+1.8333\,L -3.77\pm0.07 )\,{{A(Q^2)}}^{2} \right\} .
\end{eqnarray}
Details concerning the fit and the error estimate can be found in the Appendix.
The closeness of the NNLL coefficients is encouraging. It strongly
suggests that the impact of the non-inclusive splitting of the
gluon into partons that end up in opposite
hemispheres~\cite{Nason_Seymour,Beneke:1997sr}, which is not taken into account
in our calculation, is small, at least at the logarithmic level.
A similar comparison, which goes beyond the logarithmic level, has
been performed for the average thrust
in~\cite{average_thrust,higher_moments} and the higher
moments~$\left<t^n \right>$ in~\cite{higher_moments}. It was found
that the non-inclusive effect is rather small for the average
thrust~($4.4\%$) while it increases for the higher moments. This
implies that the effect is mainly important in the large $t$
region. Being interested here in the small $t$ region and in
particular in the logarithmically enhanced part of the cross
section, we shall neglect this effect altogether. The reader
should keep in mind, however, that the evidence provided above
concerns {\rm directly} only the Abelian part of the coefficients.
Only through the assumption that running coupling effects dominate
(making the large~$\beta_0$ approximation legitimate) this
evidence becomes relevant for the non-Abelian non-inclusive
effect. Moreover, in the context of power-corrections within the
infrared-finite coupling approach, the impact of the non-inclusive
effect in the non-Abelian case was shown to be more significant
than in the Abelian one~\cite{Mil,DMS}.

\subsection{Borel representation of the SDG result}

A convenient way to deal with perturbation theory to all orders is Borel summation.
In this section we construct the scheme-invariant Borel representation of the logarithmically enhanced part of the
SDG result. This will be used in the next section to calculate the multiple emission cross-section.
We start with the scheme-invariant Borel representation of the coupling~\cite{BY,B2,G3},
\beq
\bar{A}(k^2)=\int_0^{\infty}d{z} \,\exp\left({-{z} \ln k^2/\Lambda^2}\right)
\bar{A}_B({z}).
\label{A_Borel}
\eeq
In the case of one-loop running coupling
$\bar{A}_B({z})=\exp\left(-z\ln\Lambda^2/{\bar{\Lambda}}^2\right)$.
Taking the time-like discontinuity \cite{BY} we get
\beq
\bar{A}_{\eff} ( Q^2)=\int_0^{\infty}d{z}\,\exp\left(-{{z} \ln Q^2/\Lambda^2}\right)\,
 \frac{\sin\pi{z}}{\pi{z}}\, \bar{A}_B({z}).
\label{Borel_A_eff}
\eeq
In eq.~(\ref{res}) we use the part of $\dot{\cal F}(\epsilon,t)$ that 
contributes to log-enhanced terms (as in (\ref{log_moments_relevant})),
\beq
\left.\frac{1}{\sigma}\frac{d\sigma}{dt}(Q^2,t)\right\vert_{\SDG}=
\frac{C_F}{\beta_0}\int_{t^2}^{t}\frac{d\epsilon}{\epsilon}\,\bar{A}_{\eff}
(\epsilon Q^2) \left[-\frac{\epsilon^2}{t^3}-\frac{\epsilon}{t^2}+\frac{2}{t}\right].
\label{ours}
\eeq
Substituting $\bar{A}_{\eff} (\epsilon Q^2)$ from (\ref{Borel_A_eff}) 
and integrating over $\epsilon$ we obtain the Borel representation of 
the logarithmically enhanced SDG cross-section,
\beq
\left.\frac{1}{\sigma}\frac{d\sigma}{dt}(Q^2,t)\right\vert_{\SDG}
=\frac{C_F}{\beta_0}\,\frac1t \,\int_0^{\infty} d{z} B_{\SDG}({z},t) \,\exp\left(-{{z} \ln Q^2/\Lambda^2}\right)\,
 \frac{\sin\pi{z}}{\pi{z}}\, \bar{A}_B({z})
\label{Borel_rep}
\eeq
with
\beq
B_{\SDG}({z},t)= \frac{2}{{z}} \exp\left(2{z} L\right)-\left(\frac{2}{{z}}+\frac{1}{1-{z}}+\frac{1}{2-{z}}\right)
 \exp\left({z} L\right)
\label{Borel}
\eeq
where $L\equiv \ln\frac{1}{t}$. Note that the first exponent
in the Borel transform is associated with the large-angle soft gluon limit of phase-space, while the second is associated
with the collinear limit. The Borel integral is well-defined at the perturbative level thanks to the
cancellation of the
pole at ${z}=0$ between these two contributions. Moreover, provided that the coupling does not
enhance the singularity (e.g. the case of one-loop coupling), the poles at ${z}=1$ and ${z}=2$
are regulated by the $\frac{\sin\pi{z}}{\pi{z}}$ factor,
which originates in the analytic continuation to the time-like
region.

Note that the Borel integral converges at infinity only provided that
$2 L<\ln Q^2/{\bar{\Lambda}}^2$, or $Qt>{\bar{\Lambda}}$, reflecting
large-angle gluon emission sensitivity. A weaker condition appears
from the collinear limit: $L<\ln Q^2/{\bar{\Lambda}}^2$, or $Q^2t>{\bar{\Lambda}}^2$.
These convergence conditions signal the fact that the small $t$ region
is not under control in perturbation theory, and that
power corrections of the form $\Lambda^n/(Qt)^n$ (and eventually also $\Lambda^{2n}/(Q^2t)^n$)
should be included~\cite{DW,Shape_function4,Korchemsky_Tafat}.
We will return to this point in more detail in the next section.

\section{Exponentiation}

The calculation of the previous section concerns a {\em single}
gluon emission from the primary quark--anti-quark pair. Close to
the two-jet limit, resummation of multiple soft and collinear
gluons is essential. This resummation is achieved by
exponentiating the SDG cross-section. As mentioned in the
introduction, our basic assumption is that emission of several
soft or collinear gluons can be regarded as independent. The total
contribution to the thrust is then computed as the sum of the
individual contributions of these emissions: $t=\sum_{k=1}^n t_k$.
For large-angle soft gluons, which give the dominant contribution
at small $t$, additivity holds since the contribution of each
gluon to $t$ is proportional to its transverse momentum. This
assumption fails, of course, for sufficiently hard gluons. In
general, the contribution of several gluons to the thrust, as any
other event-shape, are {\em correlated}. Such a correlation
emerges from the modification of the phase-space limits in the
integration over the gluon momenta (the quarks do recoil) as well
as from additional interactions between the emitted gluons. These
correlations are expected to be important only at large values of
$1\,-$ thrust, away from the two-jet configuration. Since they may
have an effect at the logarithmic level, our calculation of
sub-leading logs (beyond NLL accuracy) that are sub-leading in
$\beta_0$\footnote{Such correlations do not affect the logs that
are leading in $\beta_0$. These terms are not modified by the
exponentiation and their calculation in the previous section is
exact barring the inclusive approximation made in the definition
of the thrust.} is only partial. It remains for future work to
quantify the effect of these correlations. Here we simply assume
that it is small.

\subsection{Exponentiation Formula}

The result of the previous section,
$\left.\frac{1}{\sigma}\frac{d\sigma}{dt}(Q^2,t)\right\vert_{\SDG}$,
has a probabilistic interpretation as the probability
distribution, of the random variable $t\in (0,1)$, given that only
one gluon is emitted. Since, in reality any number of gluons $n$
can be emitted, we sum over these possibilities 
\beq
\left.\frac{1}{\sigma}\frac{d\sigma}{dt}(Q^2,t)\right\vert_{\res} =
\sum_{n=0}^{\infty}\frac{1}{n!} \prod_{k=1}^n \int_0^1 dt_k
\left.\frac{1}{\sigma}\frac{d\sigma}{dt}(t_k)\right\vert_{\SDG}
 \left[\, \delta \left(t-\sum_{k=1}^n
t_k\right)-\delta(t)\right] \label{comb} 
\eeq 
where the singularity due to zero gluon emission probability $n=0$ is
cancelled by virtual corrections, $\delta(t)$. The combinatorial
$n!$ is introduced to avoid multiple counting -- the expression is
symmetric for any permutation of the $n$ gluons. Using the Laplace
conjugate variable~$\nu$, the $\delta$ function can be written in
a factorized form as (see e.g.~\cite{DW}), \beq
\delta(t-\sum_{k=1}^n t_k)=\int_{\cal C} \frac{d\nu}{2\pi
i}\exp\left\{-\nu\left(\sum_{k=1}^n t_k-t\right)\right\}. \eeq
where the contour $\cal C$ runs parallel to the imaginary axis (to
the right of all the singularities of the integrand, if any).
Changing the order of the $\nu$ integration and the sum over $n$
in (\ref{comb}) we get the exponentiated form,
\begin{eqnarray}
\label{ds_res}
\left.\frac{1}{\sigma}\frac{d\sigma}{dt}(Q^2,t)\right\vert_{\res}
&=&\int_{\cal C} \frac{d\nu}{2\pi i}
e^{\nu t}\sum_{n=0}^{\infty}\frac{1}{n!}
\left[\int_0^1 \left.\frac{1}{\sigma}\frac{d\sigma}{dt}(t_k)\right\vert_{\SDG} 
\left(e^{-\nu t_k}-1\right) dt_k\right]^n \nonumber \\
&=&\int_{\cal C} \frac{d\nu}{2\pi i} e^{\nu t}\exp{
\left\{\int_0^1 \left.
\frac{1}{\sigma}\frac{d\sigma}{dt}(Q^2,\tilde{t})\right\vert_{\SDG} 
\left(e^{-\nu \tilde{t}}-1\right)d\tilde{t}\right\}}
\end{eqnarray}
Defining
\beq
{\cal S}_{\PT}(Q^2,\ln\nu)\equiv \left< e^{-\nu t}\right>_{\SDG}\equiv
\int_0^1 
\left.\frac{1}{\sigma}\frac{d\sigma}{dt}(Q^2,\tilde{t})\right\vert_{\SDG}
 \left(e^{-\nu \tilde{t}}-1\right) d\tilde{t},
\label{exp_nut_def}
\eeq
we have
\beq
\left.\frac{1}{\sigma}\frac{d\sigma}{dt}(Q^2,t)\right\vert_{\res}=\int_{\cal C}
 \frac{d\nu}{2\pi i}  e^{\nu t} \exp\left\{ \left< e^{-\nu t}\right>_{\SDG}\right\},
\label{ds_res_1}
\eeq
and finally, integrating over $t$,
\beq
\left.R(Q^2,t)\right\vert_{\res}=\int_{\cal C} \frac{d\nu}{2\pi i\, \nu}
e^{\nu t} \exp\left\{ \left< e^{-\nu t}\right>_{\SDG}\right\}.
\label{R_res}
\eeq
Note that the terms which are leading in $\beta_0$ in the coefficients of
$\left.\frac{1}{\sigma}\frac{d\sigma}{dt}(Q^2,t)\right\vert_{\res}$, or in the log-enhanced part of the exact result,
are the {\em same} as the ones in $\left.\frac{1}{\sigma}\frac{d\sigma}{dt}(Q^2,t)\right\vert_{\SDG}$. These coefficients correspond to
a single emission.

\subsection{The Exponent in the Conjugate Variable}

Starting with the Borel representation~(\ref{Borel_rep}) with the Borel 
function~(\ref{Borel}) we now calculate the exponent
$\left< e^{-\nu t}\right>_{\SDG}$ as defined in (\ref{exp_nut_def}),
\begin{eqnarray}
\label{Borel_res_nu}
\left< e^{-\nu t}\right>_{\SDG}&=&\frac{C_F}{\beta_0}
\int_0^{\infty} d{z}\,B_{\nu}({z}) \exp\left(-{{z} \ln Q^2/\Lambda^2}\right)\,
 \frac{\sin\pi{z}}{\pi{z}}\, \bar{A}_B({z})
\end{eqnarray}
with
\beq
B_{\nu}({z})\equiv \int_0^1 \frac{dt}{t} \left[\frac{2}{{z}}
e^{2{z} \ln\frac1t}-\left(\frac{2}{{z}}+\frac{1}{1-{z}}+\frac{1}{2-{z}}\right)
 e^{{z} \ln\frac1t} \right]\,\left(e^{-\nu t}-1\right).
\eeq
To evaluate $B_{\nu}({z})$ we use the integral
\beq
\int_0^1\frac{dt}{t}e^{{z}\ln \frac 1t}\,\left(e^{-\nu t}-1\right) 
=\nu^{z}\gamma(-{z},\nu)+\frac 1{z}
\label{int_gamma}
\eeq
where
\beq
\gamma(-{z},\nu)\equiv \Gamma(-{z})-\Gamma(-{z},\nu).
\eeq
Note that (\ref{int_gamma}) is regular at ${z}=0$ due to
cancellation between the $\gamma$ function
and the simple pole -- the first
originates in the $e^{-\nu t}$ part, i.e. in real contributions,
while the latter in the unity, namely in virtual corrections.
We thus find
\beq
B_{\nu}({z})=\frac{2}{{z}}\left[e^{2{z}\ln\nu}\gamma(-2{z},\nu)
+\frac{1}{2{z}}\right]-\left(\frac{2}{{z}}+\frac{1}{1-{z}}
+\frac{1}{2-{z}}\right)\left[e^{{z}\ln \nu}\gamma(-{z},\nu)
+\frac{1}{{z}}\right].
\label{Borel_nu}
\eeq
Similarly to (\ref{Borel}), this Borel function is well defined at the perturbative 
level thanks to cancellation of the pole at
${z}=0$ between the two terms in (\ref{Borel_nu}); the first corresponds 
to the large-angle soft gluon emission and the
second to collinear gluon emission. 
In contrast\footnote{At first sight it seems surprising 
that the Laplace-transformed $B_{\nu}({z})$
has a completely different renormalon structure
from $B_{\SDG}({z},t)$. The reason is, of course, 
that $B_{\nu}({z})$, thanks to the integration over $t$,
is sensitive to the small $t$ region,
where the exponents $\exp\left(2{z} L\right)$ and $\exp\left({z} L\right)$ in (\ref{Borel})
are large and impose convergence constraints on (\ref{Borel_rep}).}
with $B_{\SDG}({z},t)$ in~(\ref{Borel}), $B_{\nu}({z})$ has a rich renormalon structure,
 making the integral (\ref{Borel_res_nu}) ambiguous at the level of power accuracy.  Note that
$\gamma(-{z},\nu)$ has the same singularity structure as $\Gamma(-{z})$: a simple pole at any non-negative integer.
It follows that $B_{\nu}({z})$ in (\ref{Borel_nu}) has {\em two sets} of infrared renormalons: the first
at all positive integers and half-integers due to large-angle emission sensitivity and the second at all positive integers
-- due to collinear sensitivity. In the latter, the renormalons at ${z}=1$ and ${z}=2$ become double poles.
It is important to note that the
${\sin(\pi{z})}$ factor in (\ref{Borel_res_nu}) regulates simple poles
and transforms double poles into simple poles at all integer ${z}$ values, while it does not affect the singularities at half
integer ${z}$ values. We will return to this point in the context of power-corrections in sec.~3.5.

An alternative representation of $\left< e^{-\nu t}\right>_{\SDG}$ in terms of 
a characteristic function can be obtained using eq.~(\ref{ours}), namely
\begin{eqnarray}
\label{char_function_rep_nu}
\left< e^{-\nu t}\right>_{\SDG}&=&\frac{C_F}{\beta_0}\,\int_0^1 dt\,
(e^{-\nu t}-1)\int_{t^2}^{t}\frac{d\epsilon}{\epsilon}\,\bar{A}_{\eff}
(\epsilon Q^2) \left[-\frac{\epsilon^2}{t^3}-\frac{\epsilon}{t^2}+\frac{2}{t}\right] \nonumber \\
&=&\frac{C_F}{\beta_0}\int_0^{1}\frac{d\epsilon}{\epsilon}\,\bar{A}_{\eff}
(\epsilon Q^2) \dot{\cal F}_{\nu}(\epsilon),
\end{eqnarray}
where the order of integrations over $t$ and $\epsilon$ has been changed. The characteristic function is
given by,
\begin{eqnarray}
\label{char_function_nu}
\dot{\cal { F}}_{\nu}(\epsilon) &=& \left(2 + \nu \,\epsilon  - {\
\frac {1}{2}} \,\nu ^{2}\,\epsilon ^{2}\right)\,\rm{Ei}(1, \,
\sqrt{\epsilon }\,\nu ) + \left( - \nu \,\epsilon  - 2 +
{\  \frac {1}{2}} \,\nu ^{2}\,\epsilon ^{2}\right)\,
\rm{Ei}(1, \,\nu \,\epsilon ) \nonumber \\ &+&  \left( - \sqrt{
\epsilon } - {\  \frac {1}{2}} \,\epsilon  +
{\  \frac {1}{2}} \,\nu \,\epsilon ^{3/2}\right)\,e^{
 - \sqrt{\epsilon }\,\nu } +\left({\  \frac {3}{2}}
 - {\  \frac {1}{2}} \,\nu \,\epsilon \right)\,e^{ - \nu
 \,\epsilon } \\ \nonumber &+& 2\,\rm{ln}(\sqrt{
\epsilon }\,\nu )
 - 2\,\rm{ln}(\nu \,\epsilon )+{\  \frac {1}{2}} \,\epsilon
 + \sqrt{\epsilon } - {\  \frac {3}{2}}.
\end{eqnarray}

\setcounter{footnote}{0}

Given a model\footnote{An example is provided by~\cite{Webber_coupling}.} for
the coupling which is regular in the infrared~\cite{DMW,DW_dist}, the integral
(\ref{char_function_rep_nu}) could be readily evaluated. We prefer, however, to
complete the perturbative treatment without any additional assumptions on
non-perturbative physics. For this purpose we find the Borel representation 
(\ref{Borel_nu}) most suitable. In the next section
we calculate this integral explicitly, but before doing so it is  worthwhile to
examine more closely the structure of the sum, term by term.

Since we are interested in large $\nu$ we can replace~$\gamma(-{z},\nu)$
in~(\ref{Borel_nu}) by~$\Gamma(-{z})$. The logarithmically enhanced terms are
not modified by such a replacement. Expanding the integrand in
(\ref{Borel_res_nu}) in powers of ${z}$,  we obtain
\begin{eqnarray}
\label{Borel_nu_expanded}
B_\nu({z})=\sum_{n=0}^{\infty}
\frac{(\ln \nu)^n}{n!}\!\!\!\!\!\!\!\!&&\left[\,\,\,2\!\!\!\!\!\!
%\sum_{{ \, \begin{array}{ll}{m\geq 1,}\\{m\geq n-1}\end{array}}}
\sum_{m\geq \max\{1,n-1\}}
\!\!\!\!\!\!\left(1-2^m\right)\, c_{m-n}\,z^{m-1} \right. \\
\nonumber
&&\left. \,+\,\,\sum_{l\geq 0}\left(1+2^{-l-1}\right)\!\!\!\!\!\!
%\sum_{{\begin{array}{ll}{m\geq n+l}\\{m\geqn+1}\end{array}}}
\sum_{m\geq \max\{n+l,n+1\}}
\!\!\!\!\!\!\!\,c_{m-n-l-1}
\,z^{m-1}\right],
\end{eqnarray}
where the first term in the square brackets
originates in the $2/t$ part in $\dot{\cal  F}(\epsilon,t)$ from both the soft
and collinear limits of phase-space, while the second originates
in the $\epsilon$ dependent terms in $\dot{\cal  F}(\epsilon,t)$ from the collinear limit alone.
The numbers $c_k$ are defined by $\Gamma(-{z})=-\sum_{k=-1}^{\infty}\,c_k\,{z}^k$. This means that
$c_{-1}=1$, $c_0=\gamma_E$, $c_1={\pi^2}/{12}\,
+\,{\gamma_E^2}/{2}$, etc. Higher $c_k$ contain higher $\zeta_i$ numbers ($i\leq k+1$), yet numerically $c_k$ ($k\geq 1$)
are all close to $1$.

Next, performing the Borel integral term by term
we get a series in the coupling~${\bar{A}(Q^2)}$.  It is then
straightforward to rewrite the sum as
\beq
\left< e^{-\nu t}\right>_{\SDG}=\frac{C_F}{\beta_0}
 \sum_{k=1}^{\infty}{{\bar{A}(Q^2)}^{k-2}}\,\, f_{k}\!\left({\bar{A}}(Q^2)\ln\nu\right)
\label{log_expansion}
\eeq
and perform the summation over all powers of $\xi\equiv {\bar{A}}(Q^2)\ln\nu$. 
The functions $f_{k}$ are given~by
\beq
f_k(\xi)=\sum_{u=0}^{(k-1)/2}\,\frac{(-\pi^2)^u}{(2u+1)!}
\,\left[g_k^{(u)}(\xi)\,M_{\rm col}(k)\,+\,g_k^{(u)}(2\xi)\,M_{\rm la}(k)
\right]
\label{f_k}
\eeq
where
\begin{eqnarray}
\label{M_col_la}
M_{\rm col}(k)&=&2\,c_{k-2}+\sum_{m=0}^{k-2}(1+2^{-m-1})\,c_{k-3-m} \\ \nonumber
M_{\rm la}(k)&=&-2^{k-1}\,c_{k-2}
\end{eqnarray}
and
\begin{eqnarray*}
g_k^{(u)}(\xi)=\xi^{n_0}\,\frac{\Gamma(k-2+n_0)}{\Gamma(1+n_0)}\,{\rm Hypergeom}([1,k-2+n_0],[1+n_0],\xi)
\end{eqnarray*}
with $n_0=\max(0,2u+3-k)$.  In (\ref{M_col_la}), $M_{\rm col}(k)$ and $M_{\rm la}(k)$ correspond to 
the collinear and large-angle contributions to the Borel function, respectively. 
Explicitly, this yields
\begin{eqnarray}
\label{f_k_expml}
f_{1}(\xi)&=& 2 (1- \,\xi)\,\ln(1 - \xi) - (1 - 2\,\xi)\,
\ln(1 - 2\,\xi)   \nonumber \\ 
f_{2}(\xi)&=& - 2\,\gamma_E \,(\ln(1 - \xi) - \ln(1 - 2\,\xi)) -
{\  \frac {3}{2}} \,\ln(1 - \xi) 
\end{eqnarray}
and
\begin{eqnarray}
\label{f_k_expml1} 
\begin{array}{ll}
f_{3}(\xi)=   { 0.804}/{(1-\xi ) }  - {  {2.32}/({1 -2\xi   })}  &
f_{4}(\xi)= {0.779}/{({1-\xi  } )^{2}} - {  {2.68}/{(1 -2\xi )^{2}}}  \\ 
f_{5}(\xi)=   { 2.32}/{({1-\xi  })^{3}}  - {  {5.00}/{(1 -2\xi )^{3}}} &
f_{6}(\xi)= {6.12}/{({1-\xi  } )^{4}}  - { {15.32}/{(1 -2\xi )^{4}}} \\ 
f_{7}(\xi)=  { 23.8}/{({1-\xi  } )^{5}}  - {  {61.12}/{(1 -2\xi )^{5}}} &
f_{8}(\xi)=  {120.9}/{({1-\xi  } )^{6}} - {  {305.52}/{(1 -2\xi )^{6}}}, \\ 
\end{array}
\end{eqnarray}
etc. Here the function $f_{1}$ resums the leading-logs (LL), $f_2$ resums the next-to-leading
logs (NLL) and so on. 
The effect of the analytic continuation to the
time-like region is taken into account through the summation over $u$ in
(\ref{f_k}). Note that we omitted a constant term in eq.~(\ref{f_k_expml1}). Such a constant does not
necessarily exponentiate and it can be shifted into the ``remainder function''
$\left.\ln R(y)\right\vert_{\rm {non-log}}$ in~(\ref{log_match_separation}).

The first two functions $f_{1}(\xi)$ and $f_{2}(\xi)$ are already known from~\cite{CTTW}. When comparing
$f_{2}(\xi)$ in~(\ref{f_k_expml}) to~\cite{CTTW} one should be careful concerning the definition of the coupling.
Strictly speaking, our calculation is done in the large $\beta_0$ limit, so $\beta_1$ terms as
well as other terms\footnote{Such terms originate~\cite{CTTW} in the singular part of the
NLO splitting function.} which are sub-leading in $\beta_0$ are missing.
However, these terms can be generated upon replacing the large $\beta_0$ coupling, which is
one-loop, by a two-loop coupling in the ``gluon bremsstrahlung''
scheme~\cite{CMW} (see sec.~4) inside the integral~(\ref{Borel_res_nu}).
Such a replacement will be made in sec.~3.3 and then in the phenomenological analysis in sec.~5.

The functions $f_{k}(\xi)$, for $k\geq 2$ are singular at
$\xi=1/2$ and at $\xi=1$; the first is associated with large-angle
emission $\nu{\bar{\Lambda}}/Q\sim 1$ and the second with collinear
emission $\nu{\bar{\Lambda}}^2/Q^2\sim 1$. The singularity (at both
limits) becomes stronger as the order $k$ increases, in particular
$f_k\sim 1/(1-2\xi)^{k-2}$ close to $\xi=1/2$ and $f_k\sim
1/(1-\xi)^{k-2}$ close to $\xi=1$, for any $k\geq 3$. Since soft
gluon resummation has a universal structure, we expect that this
singularity structure in $\xi$ will be common to different
physical processes. An example is provided by a recent calculation
of NNLL for deep-inelastic scattering and the Drell-Yan
process~\cite{vogt}. There, the singularity structure of
$f_k(\xi)$ for $k=1$ through $3$ is indeed similar to~(\ref{f_k_expml}) 
and~(\ref{f_k_expml1}). In particular $f_3(\xi)$ has simple poles.

As usual in perturbation theory the expansion~(\ref{log_expansion}) is divergent.
In principle it can be given sense e.g.
by truncation at the minimal term. Note that this truncation is cumbersome since the minimal term will
correspond to different orders depending
on the value of $\nu$. The standard procedure~\cite{CTTW} amounts to truncating this expansion
at NLL, so $f_k(\xi)$ for $k\geq 3$ are just neglected. To see whether this is legitimate let us now examine
the numerical values of $f_k(\xi)$.

\begin{figure}[t]
\begin{center}
\mbox{\kern-0.5cm\epsfig{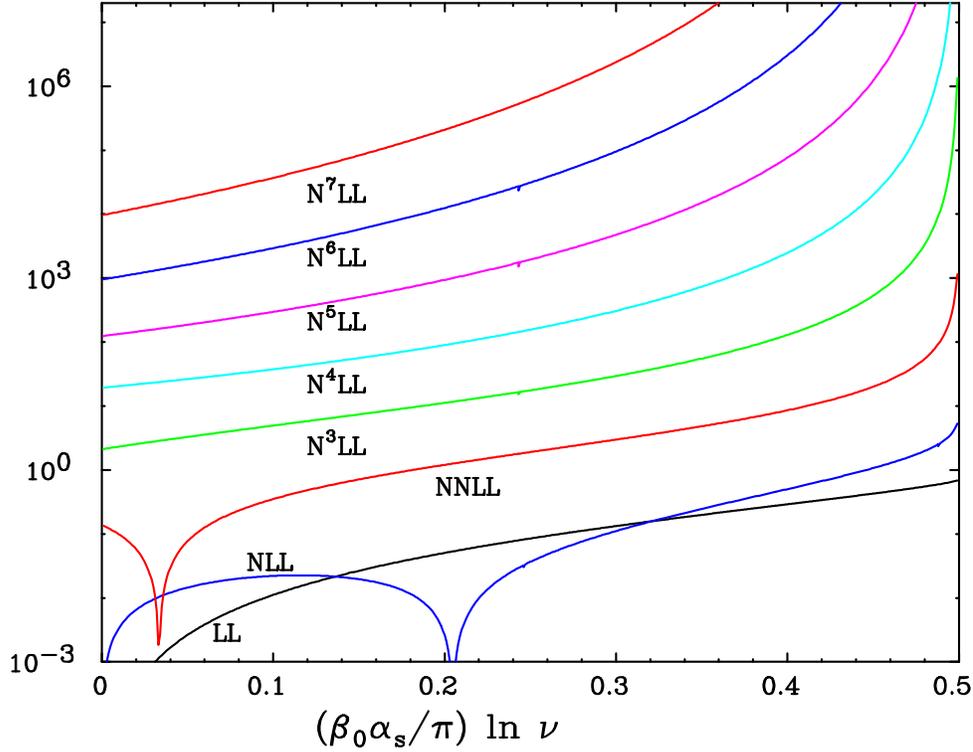}
}
\end{center}
\caption{$f_k(\xi)$ for $k=1$ (LL) through $k=8$ (${\rm N}^7{\rm LL}$) as a function of
$\xi\equiv {\bar{A}}(Q^2)\ln\nu$. The absolute value $\left\vert f_k(\xi)\right\vert$ is plotted on a
logarithmic scale.
 }
\label{log_func}
\end{figure}

\begin{figure}[t]
\begin{center}
\mbox{\kern-0.5cm\epsfig{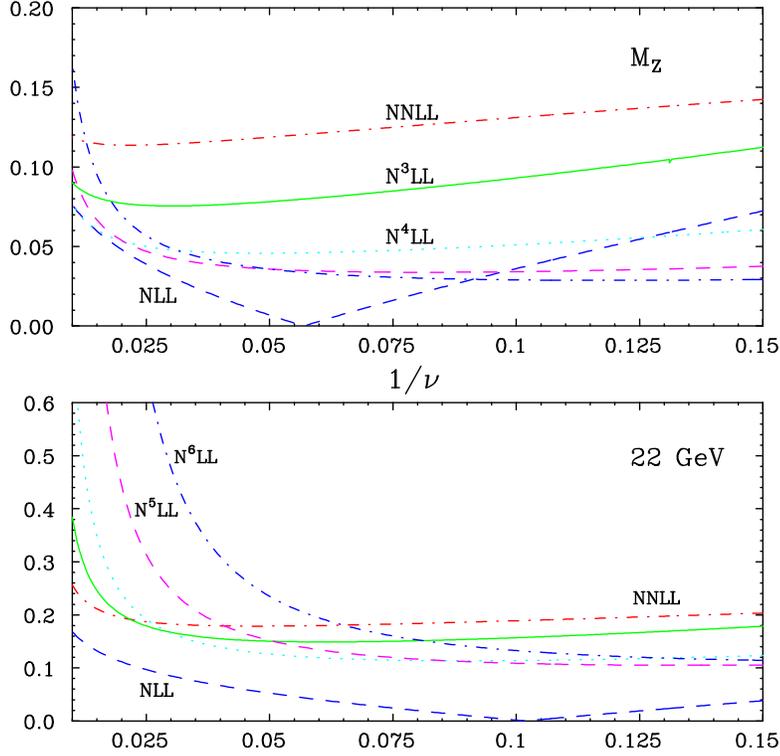}
}
\end{center}
\caption{The relative significance (in absolute value) of terms
${{\bar{A}(Q^2)}^{k-2}}\,\,
f_{k}\!\left({\bar{A}}(Q^2)\ln\nu\right)$
in~(\ref{log_expansion}) for $k=2$ (NLL) through $k=7$ (${\rm
N}^6{\rm LL}$) as a function of $1/\nu$ ($1/\nu\sim t$ in the
small $t$ region). In both plots, the terms are normalized by the
LL function ($1$ on the vertical~axis).} \label{Sub_logs}
\end{figure}

The functions $f_k(\xi)$ are plotted in fig.~\ref{log_func}. It is clear from the plot that the
expansion~(\ref{log_expansion}) is divergent. In particular, the
ratios~$f_k(\xi)/f_{k+1}(\xi)$
increase with~$k$, reflecting factorial growth.
The actual effect of these increasing coefficients depends on the value of the
expansion parameter, here $\bar{A}(Q^2)=\beta_0 a(Q^2)$.
$\bar{A}(Q^2)$ is defined in the ``gluon bremsstrahlung'' scheme~\cite{CMW}
(the reason for this choice is explained in section~4).
In fig.~\ref{Sub_logs} we show the individual terms in~(\ref{log_expansion}) at two
representative centre-of-mass energy scales: $Q=91$ and $22\, {\rm GeV}$.
First note that the NNLL ($f_3$) gives a significant
contribution in all cases. It should be emphasised that the
leading part of its contribution is taken into account in the
standard NLL calculation in the process of matching with the NLO
result. Next note that $\rm N^3LL$ ($f_4$) also gives
quite an important correction: $\sim 8\%$ at $\rm M_Z$ and $\sim
18\%$ at $22\, {\rm GeV}$. A closer examination shows that the
hierarchy between the contributions of sub-leading logs depends
strongly on both the energy (the value of the coupling) and
$1/\nu$. At sufficiently large $1/\nu$ values, i.e. far enough
from the two-jet limit, the expansion~(\ref{log_expansion})
appears to be convergent at the first few orders. Consider for
example the case $1/\nu=0.075$ at $22\, {\rm GeV}$: the contribution of
$f_4$ (full line) is smaller than that of $f_3$ and 
$f_5$ (dotted) is yet smaller. But then the
trend changes: $f_6$ (dashed) contributes slightly more than $f_5$ 
and the series starts diverging. 
In this particular example the contribution at the
minimal term is about $7.8\%$. Thus a resummed perturbative
calculation should not be expected to be better than that. For
smaller $1/\nu$ values, and likewise at lower energies, the
expansion diverges faster: the order at which the minimal term is
reached becomes smaller while its contribution becomes larger. At
some point, close to the distribution peak, the expansion becomes
meaningless. Technically, this behaviour is a direct consequence
of the enhanced singularity of the functions $f_k$ and their
factorially increasing numerical coefficients. This is the way
infrared renormalons show up in this framework. Physically it is
well expected that the perturbative expansion would gradually
become less predictive and eventually irrelevant as a quantitative
evaluation of the cross-section, since it misses non-perturbative
soft gluon contributions or, in other words, hadronization
effects. We shall return to this issue in sec.~3.5. Until then,
our goal is to complete the perturbative evaluation of the cross
section. To achieve this we leave the
expansion~(\ref{log_expansion}) and revert to
Borel summation~(\ref{Borel_res_nu}) which allows us, in rough
terms, to resum the renormalons first, and then complete the
summation over the logs.

\subsection{Evaluation of the Borel sum}

Our next task is to evaluate explicitly the Borel sum~(\ref{Borel_res_nu}).
An exact analytic integration is difficult. However, it is possible to
make a systematic approximation of the Borel function and then perform
the integration analytically.

We begin by writing the Borel function $B_\nu({z})$ as
\beq
B_{\nu}({z})=\frac{2}{{z}}V(2{z})-\left(\frac{2}{{z}}+\frac{1}{1-{z}}
+\frac{1}{2-{z}}\right)V({z})
\label{Borel_nu_app}
\eeq
with
\beq
V({z})\equiv e^{{z}\ln\nu}\Gamma(-{z})+\frac{1}{{z}}.
\label{V_def}
\eeq
Since $V({z})$ has only simple poles at positive integer ${z}$ values,
it is natural to approximate it by a rational function.
Clearly, we would like the approximation to coincide with the function at least at the first few orders in the expansion around
${z}=0$. One possibility is then to use Pad\'e approximants based on some truncated sum. However, since we know the
singularity structure, it is better to use an approximation of the
form\footnote{A similar approximation for the
Borel function was used in~\cite{PAP} in a different application.
Comparison with Pad\'e approximants showed consistent results.}
\beq
V({z})\simeq \sum_{p=1}^{p_{\max}} \frac{r_p}{p-{z}},
\label{V_sum}
\eeq
where $r_p$ are set such that the expansion of this approximant coincides with that of (\ref{V_def})
up to order ${z}^{p_{\max}-1}$. This means that $r_p$, for every $1\leq p\leq {p_{\max}}$, is a polynomial of
order $p_{\max}$ in $\ln \nu$,
\beq
r_p=\sum_{q=0}^{p_{\max}}\, \rho_{p,q}\,\left(\ln \nu\right)^q
\label{r_p_def}
\eeq
where $\rho_{p,q}$ are (irrational) numerical coefficients.

Next, $B_{\nu}({z})$  can be written as a sum of simple and double poles,
\beq
B_{\nu}({z})\simeq \sum_{n=1}^{p_{\max}} \frac{\bar{r}^{\rm la}_{n/2}}{n/2-{z}}+
\sum_{p=1}^{p_{\max}} \frac{\bar{r}^{\rm col}_{p}}
{p-{z}}-\frac{r_1}{\left(1-{z}\right)^2}-\frac{r_2}{\left(2-{z}\right)^2}
\label{B_nu_poles}
\eeq
where $\bar{r}^{\rm la}_p$ and $\bar{r}^{\rm col}_p$ are linear combinations of $\left\{r_p\right\}_{p=1}^{p_{\max}}$,
corresponding to the
large-angle (first term) and collinear (second term) parts of~(\ref{Borel_nu_app}), respectively.
Simple poles appear at
all positive integers and half integers $1/2 \leq p\leq p_{\max}/2$ due to the
large-angle part, and at all integers $3 \leq p \leq p_{\max}$ due to the collinear part. The double poles
at ${z}=1$ and ${z}=2$ appear due to the collinear part. As an example, consider the case of ${p_{\max}}=3$ 
where one has
\[
B_{\nu}({z})\simeq
 {\ \frac {{2r_{1}}}{{\ \frac {1}{2}} - {z} }}  + {\ \frac { - 3\,{r
_{1}} - {\ \frac {1}{2}} \,{r_{3}}}{1 - {z} }}  +
{\ \frac {{\ \frac {2}{3}} \,{r_{3}}}{
{\ \frac {3}{2}}  - {z}}}  + {\ \frac {{r_{
1}} - {r_{3}}}{2 - {z}}}  + {\ \frac {{\
\frac {5}{6}} \,{r_{3}}}{3 - {z}}}  - {\ \frac {{r_{1}}
}{(1 - {z})^{2}}}  - {\ \frac {{r_{2}}}{(2 - {z})^{2}}
}.
\]

Finally, given a specific coupling and the function (\ref{B_nu_poles})
the Borel integral (\ref{Borel_res_nu}) can be evaluated analytically.
The simple pole integrals are of the form
\begin{eqnarray}
I_p(s)&=&\int_0^{\infty}dz \,\exp\left({-{z}\ln Q^2/\Lambda^2}\right)
\frac{1}{p-{z}}\,\frac{\sin \pi {z}}{\pi{z}}\, \bar{A}_B({z}) \\
\nonumber
&=&\int_0^{\infty}dz \,\exp\left({-{z}s}\right)\frac{1}{p-{z}}\,\frac{\sin \pi {z}}{\pi{z}}\,A_B({z})
\label{sp}
\end{eqnarray}
where $A_B({z})$ is the Borel transform of the ${\rm \overline{MS}}$ coupling and
$s\equiv \ln Q^2/{\bar{\Lambda}}^2$. 
Regarding $p$ as a complex number, $I_p(s)$ can be written in terms of $\tilde{I}_p(s)$,
\beq
\tilde{I}_p(s)=\int_0^{\infty}dz \,\exp\left({-{z}s}\right)\frac{1}{p-{z}}\, A_B({z}),
\label{tilde_I}
\eeq
by analytically continuing in $s$, $\bar{I}_p(s)=\frac{1}{2\pi i}\left(\tilde{I}_p(s-i\pi)-\tilde{I}_p(s+i\pi)\right)$,
and
\beq
I_p(s)=\frac{1}{p}\left[ \bar{I}_p(s) - \lim_{p=0^-}\bar{I}_p(s)\right].
\eeq
The double pole integrals
\beq
I^{(1)}_p(s)=\int_0^{\infty}dz\,\exp\left({-{z}s}\right)\frac{1}{\left(p-{z}\right)^2}
\,\frac{\sin \pi {z}}{\pi{z}}\, A_B({z})
\eeq
can be simply obtained using $I^{(1)}_p(s)=-\frac{d}{dp}I_p(s)$.

In case of infrared renormalons in space-like observables~$\tilde{I}_p(s)$ 
should be evaluated for real $s$ and $p$ where it is ill-defined. Then Cauchy 
principal value can be defined taking the real part of (\ref{tilde_I}).  Our
definition for the time-like renormalon integral  is motivated by this
regularisation. Here there is no need to take the real part: thanks to
the analytic continuation procedure, (\ref{tilde_I}) is evaluated 
above and below the cut in a symmetric way (the relevant complex parameter is~$ps$), 
yielding a real value for $I_p(s)$ and
$I^{(1)}_p(s)$. The ambiguity of these quantities is  directly related to the
ambiguity in~$\tilde{I}_p(s)$. The definition of the Borel-sum through the
space-like renormalon integral also  allows to introduce a cut-off
regularisation~\cite{average_thrust}.

$\tilde{I}_p(s)$ can be written in terms of known analytic functions.
The simplest case is the one-loop running coupling, where
$A_B({z})\equiv 1$. Here we have 
\beq
\left.\tilde{I}_p(s)\right\vert_{\rm one-loop}=-{\rm
Ei}(1,-ps)\,e^{-ps}. \label{Ip_1loop} 
\eeq 
The more difficult two-loop case was dealt with in \cite{average_thrust} (see
sec.~3.4 there) using results from~\cite{G4andDU}. It is useful to
express the coupling using the Lambert W function~\cite{Lambert},
\beq 
\left.\bar{A}(Q^2)\right\vert_{\rm two-loop}=-\frac{1}{\delta}\frac{1}{1+w(s)} 
\eeq 
where $\delta\equiv \beta_1/\beta_0^2$ and \footnote{The analytic
continuation in the calculation of $\bar{I}_p(s)$ should be done
carefully: one should use the correct branch $n$ of the Lambert W
function, $W_n$. In practice this is simple since for $N_c=3$,
$N_f\leq 6$ the time-like axis is always contained in the $n=\pm
1$ branches (see \cite{Lambert} for more details). 
Also note that with the current definition of the two-loop coupling
the Landau branch point is at $Q^2=\bar{\Lambda}^2$.
}
$w(s)\equiv W_{-1}\left(-e^{-{s}/{\delta}-1}\right)$. 
In these variables the two-loop renormalon integral (\ref{tilde_I}) is 
simply
\beq
\left.\tilde{I}_p(s)\right\vert_{\rm two-loop} =
-\left[w(s)p\delta\right]^{p\delta}\,e^{w(s)p\delta}
\,\Gamma\left(-p\delta,w(s)p\delta\right). \label{Ip_2loop} \eeq
Altogether the Borel
integral (\ref{Borel_res_nu}) is given by \beq \left< e^{-\nu
t}\right>_{\SDG}\simeq
\frac{C_F}{\beta_0}\,\left[\sum_{n=1}^{p_{\max}} {\bar{r}^{\rm
la}_{n/2}}I_{n/2}(s)+ \sum_{p=1}^{p_{\max}} {\bar{r}^{\rm
col}_{p}}I_p(s)-{r_1}I^{(1)}_1(s)-{r_2}I^{(1)}_2(s)\right],
\label{R_res_sum_I} \eeq where the (polynomial) dependence on $\ln
\nu$ is fully contained in the coefficients $\bar{r}_p$ and $r_p$,
given by eqs.~(\ref{V_def}) through~(\ref{r_p_def}), while the
dependence on the external energy scale $s\equiv \ln Q^2/{\bar{\Lambda}}^2$ 
is fully contained in
the renormalon integrals $I_p(s)$. Eventually, the log-resummation
order $p_{\max}$ should be set large enough, so that higher orders
can be safely neglected.  In sec.~5 we show that in practice
$p_{\max}\simeq 6$ is enough to make the result stable.

\subsection{Evaluation of the inverse Laplace transform}

The last stage of the calculation is to evaluate the inverse Laplace transform (\ref{R_res}),
\beq
\left.R(Q^2,t)\right\vert_{\res}=\int_{\cal C} \frac{d\nu}{2\pi i\, \nu}
e^{\nu t} \exp\left\{{\cal S_{\PT}}(Q^2, \ln \nu)\right\}\
\label{R_res_PT}
\eeq
with ${\cal S}_{\PT}(Q^2, \ln \nu)\equiv\left< e^{-\nu t}\right>_{\SDG}$.
We apply the same technique used in~\cite{CTTW} (see section 5 there).
However, at a difference with the latter, we must keep track of sub-leading logs so we refrain from making any approximation.
Defining $u=\nu t$ and expanding ${\cal S}(Q^2, \ln \nu)={\cal S}(Q^2, L+\ln u)$ (where $L=\ln\frac 1t$)
in powers of $\ln u$, one has
\begin{eqnarray}
\left.R(Q^2,t)\right\vert_{\res}&=&\int_{\cal C} \frac{d u}{2\pi i\, u}\exp\left\{u+{\cal S}(Q^2, L+\ln u)\right\}\\ \nonumber
&=&e^{{\cal S}(Q^2, L)}\int_{\cal C} \frac{d u}{2\pi i\, u}\exp\left\{u+d_1(Q^2,L)\ln u+\sum_{k=2}^{\infty}\frac{1}{k!}
d_k(Q^2,L)(\ln u)^k)\right\}
\end{eqnarray}
where $d_k(Q^2,L)\equiv\frac{d^k}{dL^k}{ {\cal S}(Q^2,L)}$.
Finally, we calculate  $\left.R(Q^2,t)\right\vert_{\res}$ by expanding \[\exp\left\{\sum_{k=2}^{\infty}\frac{1}{k!}
d_k(Q^2,L)(\ln u)^k)\right\}\] in powers of $\ln u$ and using the integral
\beq
\int_{\cal C}\frac{d u}{2\pi i\, u}(\ln u)^n \exp\left\{u+g\ln u\right\}=\frac{d^n}{dg^n}\frac{1}{\Gamma(1-g)}.
\label{der_GAMMA}
\eeq

Our final expression for $\left.R(Q^2,t)\right\vert_{\res}$ includes several special functions:
derivatives of the $\Gamma$ function from (\ref{der_GAMMA}) and  the expressions for
$I_p(\ln Q^2/{\bar{\Lambda}}^2)$, written in terms of the exponential integral function
 in the one-loop case (\ref{Ip_1loop}),
or the incomplete $\Gamma$ function and the
Lambert W function in the two-loop case~(\ref{Ip_2loop}). We therefore chose to evaluate it using the computer algebra program Maple.

\subsection{Power-Corrections}

Since hadronization is not described by perturbation theory it is a priori clear that there are non-perturbative
corrections to the resummed cross-section. In this section we study the structure of these corrections.
We already saw that the perturbative sum is, in general, ambiguous due to infrared renormalons.
Having used a specific regularization of the renormalon singularities we are bound to address the question
of the related power-suppressed ambiguity. The appearance of infrared renormalons does not only
signal the existence of non-perturbative power-corrections, but also suggests their functional dependence
on the energy and the event-shape variable.
One should not expect, of course, that the entire non-perturbative contribution would be deducible
from a perturbative result. Appropriate non-perturbative parameters will be introduced and eventually fixed by the data.
Upon including these non-perturbative corrections the result becomes well-defined to power accuracy.

If the Borel sum (\ref{Borel_res_nu}) is supplemented by a non-perturbative contribution, eq.~(\ref{R_res_PT}) becomes
\beq
R(Q^2,t)=\int_{\cal C} \frac{d\nu}{2\pi i\, \nu}
e^{\nu t} \exp\left\{{\cal S_{\PT}}(Q^2, \ln \nu)\right\}\,\exp\left\{{\cal S_{\NP}}(Q^2, \ln \nu)\right\}.
\label{R_res_NP}
\eeq
In sec.~3.4 we evaluated the inverse Laplace integral ignoring the non-perturbative part. Thanks to the factorized form of
(\ref{R_res_NP}) non-perturbative corrections can be included by
convoluting~\cite{Shape_function1,Shape_function2,Shape_function3,Shape_function4}
the perturbative result (\ref{R_res_PT}) with a non-perturbative function:
\beq
R(Q^2, t)=\int_0^t d\tilde{t} \left.R(Q^2,t-\tilde{t})\right\vert_{\res} \, f_{\NP}(Q^2,\tilde{t}) \equiv
\left.R(Q^2,t)\right\vert_{\res}\star  f_{\NP}(Q^2,t)
\label{R_res_total}
\eeq
where
\beq
f_{\NP}(Q^2,t)=\int_{\cal C} \frac{d\nu}{2\pi i\, }
e^{\nu t} \exp\left\{{\cal S_{\NP}}(Q^2, \ln \nu)\right\}\ .
\label{R_res_f_NP}
\eeq

In order to learn about the functional form of ${\cal
S_{\NP}}(Q^2, \ln \nu)$, we shall now examine the renormalon
structure of ${\cal S_{\PT}}(Q^2, \ln \nu)$.  Considering the
integrand of~(\ref{Borel_res_nu}) with $B_{\nu}({z})$ of
eq.~(\ref{Borel_nu}) (see description following the equation), we
identify potential singularities at all positive integers and half
integers. On the other hand, the $\sin \pi {z}$ factor which
originates in the analytic continuation to the time-like region
introduces an explicit zero at all integer ${z}$ values. It is
important to realize that the {\em complete} cancellation of some
of the singularities may be an artifact of the large $\beta_0$
approximation. A similar situation occurs in any time-like
observable, e.g. in the total cross-section~\cite{beneke}. There,
having an operator product expansion, one can relate the
singularity strength of the Borel\footnote{Note that the
$p\beta_1/\beta_0^2$ term appears in the standard Borel transform
but not in the scheme-invariant one~\cite{BY,B2,G3}.} pole
$1/\left(p-{z}\right)^{1+\kappa}$ to the anomalous dimension of
the corresponding operator(s) ${\cal O}_p$ of dimension $2p$:
$\kappa=p\beta_1/\beta_0^2+\gamma_0/\beta_0$, where
$\left[d/d\ln\mu^2+\gamma\right]{\cal O}_p=0$ with
$\gamma=\gamma_0\,(\alpha_s/\pi)+\ldots$. Thus, in general, a
large-$\beta_0$ calculation does not yield the correct singularity
strength: not only the residue, but also $\kappa$ is expected to
be modified when sub-leading terms in $\beta_0$ are included.
Being aware of this fact (and having no operator product expansion
to supply additional information) it is reasonable to allow for
modifications of order~$1$ in the singularities of $B_\nu({z})$,
\beq b({z})=\frac{r_p}{\left(p-{z}\right)^m}
\,\,\,\,\,\,\longrightarrow \,\,\,\,\,\,
\tilde{b}({z})=\frac{\tilde{r}_p}{\left(p-{z}\right)^{m+\kappa_p}}
\label{modified_B} \eeq where we assume $\left\vert
\kappa_p\right\vert < 1$. Thus, instead of simple and double poles
($m=1,2$), the Borel integrand has cuts which survive the
suppression by the analytic continuation and introduce some
ambiguity into ${\cal S_{\PT}}(Q^2, \ln \nu)$. Note that, we will
not consider the possibility of a modified $\nu$ dependence
of the exact Borel function compared to the calculated one. We
allow only a modification of the overall coefficient and the
singularity strength.

Consider first a scheme-invariant Borel integral in the space-like region,
\beq
\tilde{I}(s)=\int_0^{\infty}e^{-{z}s}\tilde{b}({z})d{z},
\eeq
where the Borel function behaves as (\ref{modified_B}) with $m=1$ near ${z}=p$.
The Borel integral can be written as a sum of two terms,
\begin{eqnarray*}
\tilde{I}(s)&=&\int_0^{p}d{z}\,e^{-{z}s}\frac{1}{\left(p-{z}\right)^{1+\kappa}}+\int_p^{\infty}d{z}
\,e^{-{z}s}\frac{1}{\left(p-{z}\right)^{1+\kappa}}\\ 
&=&(-s)^{\kappa}\,\gamma(-\kappa,-sp)\,e^{-sp}
-(-s)^{\kappa}\,\Gamma(-\kappa)\,e^{-sp}=-(-s)^{\kappa}\,\Gamma(-\kappa,-sp)\,e^{-sp}.
\end{eqnarray*}
Next we use the fact that the ambiguity $\Delta \tilde{I}(s)$ is purely imaginary and originates uniquely
in the second integral where ${z}\geq p$,
i.e. in the complete $\Gamma$ function part\footnote{A similar mathematical analysis appears in~\cite{average_thrust}
in the context of a renormalon integral with a two-loop running coupling.}. 
Taking $1/\pi$ times the imaginary part,
we obtain
\beq
\Delta \tilde{I}(s) = -\frac{\sin\left( \pi \kappa\right)}{\pi}\, s^{\kappa}\,
\Gamma(-\kappa)\,e^{-sp}=\frac{1}{\Gamma(1+\kappa)}\,s^{\kappa}\,e^{-sp}.
\label{Delta_I_tilde}
\eeq
The ambiguity for the $m=2$ case can be calculated using
$\tilde{I}^{(1)}(s)=-\frac{d}{dp}\tilde{I}(s)$: one finds
that $\Delta \tilde{I}^{(1)}(s)=\Delta \tilde{I}(s)\,s$.
Substituting $s=\ln Q^2/{\bar{\Lambda}}^2$ we see that the ambiguity $\Delta \tilde{I}(s)$ scales as a
power $(Q^2/{\bar{\Lambda}}^2)^p$, up to
a logarithmic factor,  $\left(\ln Q^2/{\bar{\Lambda}}^2\right)^\kappa$, which remains
undetermined as long as $\kappa$ is unknown.
Next, the ambiguity of the corresponding Borel integral in the time-like region
\beq
I(s)=\int_0^{\infty}e^{-{z}s}\tilde{b}({z})\,\frac{\sin{\pi{z}}}{\pi{z}}\,d{z},
\eeq
can be deduced by analytically continuing (\ref{Delta_I_tilde}) in $Q^2\longrightarrow -Q^2$, i.e. $s \longrightarrow s \pm i\pi$,
and taking ($\frac{1}{\pi p}$ times) the imaginary part,
\begin{eqnarray}
\label{Delta_I}
\Delta I(s) &=& \frac{1}{\pi p}\frac{\left(s^2+\pi^2\right)^{\kappa/2}}{\Gamma(1+\kappa)}\,\,e^{-sp}\,
\left[\cos\left({\kappa}\arctan\frac{\pi}{s}\right)\sin(\pi p)-\sin\left({\kappa}\arctan\frac{\pi}{s}\right)\cos(\pi p)\right]
\nonumber \\ &\simeq&\frac{1}{\pi p}
 \frac{1}{\Gamma(1+\kappa)}\,s^{\kappa}\,e^{-sp}\left[\sin(\pi p)-\frac{\kappa\pi}{s}\cos(\pi p)\right],
\end{eqnarray}
where the expression in the second line is obtained expanding in $1/s$ and neglecting $1/s^2$ terms and higher. Similarly,
we obtain
\begin{eqnarray}
\label{Delta_I_1}
\Delta I^{(1)}(s) \simeq \frac{1}{\pi p}\frac{1}{\Gamma(1+\kappa)}\,s^{\kappa}\,e^{-sp}\left[s\sin(\pi p)-
(1+\kappa)\pi\cos(\pi p)\right].
\end{eqnarray}

Given the general singularity structure of the SDG Borel function
$B_{\nu}({z})$ around ${z}=p$, \beq
b({z})=\frac{r^{(1)}_p}{\left(p-{z}\right)^2}\,+\,\frac{r_p}{p-{z}},
\eeq our result for the ambiguity ((\ref{Delta_I}) and
(\ref{Delta_I_1})) suggests the following assignments of
non-perturbative contributions, \beq
I_{\NP}(s)=\frac{\rho_{2p}}{\pi
p}\,\frac{e^{-sp}\,s^{\kappa_p}}{\Gamma(1+\kappa_p)}\,
\left[\left(r^{(1)}_ps+r_p\right) \sin\left( \pi
p\right)-\left(\left({1+\kappa_p}\right)r^{(1)}_p+\kappa_p\,r_p/s
\right)\pi \cos\left( \pi p\right)\right], \label{I_NP} \eeq where
$s=\ln Q^2/{\bar{\Lambda}}^2$ and $\rho_{2p}$ are dimensionless
phenomenological parameters to be fixed, eventually, by the data.
On general grounds we expect that $\rho_{2p}\sim{\cal O}(1)$. It
is important to keep in mind that the actual values of $\rho_{2p}$
depend on the regularization prescription used for the Borel sum.
For example, when the perturbative sum is defined using a cut-off
(on the Euclidean momentum), $\rho_{2p}$ can be interpreted as
moments of an ``infrared finite effective
coupling''~$\bar{A}_{\NP}(k^2)$~\cite{DMW,average_thrust}, which
is assumed to coincide with the perturbative
coupling~$\bar{A}_{\PT}(k^2)$ above some scale $\mu_I$
\hbox{(${\bar{\Lambda}}\ll \mu_I\ll Q$),} \beq
 \rho_{2p}^{\mu_I}= \left(\frac{\mu_I^2}{{\bar{\Lambda}}^2}\right)^p\, m^{\NP}_p(\mu_I^2)
\eeq
where
\beq
m^{\NP}_p(\mu_I^2)\equiv \int_0^{\mu_I^2}\,p\,\frac{dk^2}{k^2}\left(\frac{k^2}{\mu_I^2}\right)^p\,\bar{A}_{\NP}(k^2).
\eeq
Here we define the perturbative sum $\rm \grave{a}$ la Borel. In this case $\rho_{2p}$ is
\beq
 \rho_{2p}=  \left(\frac{\mu_I^2}{{\bar{\Lambda}}^2}\right)^p\, \left[m^{\NP}_p(\mu_I^2)-m^{\PT}_p(\mu_I^2)\right]
\label{rho_PV}
\eeq
where
\beq
m_p^{\PT}(\mu_I^2)\equiv {\rm
P.V.}\int_0^{\mu_I^2}\,p\,\frac{dk^2}{k^2}\left(\frac{k^2}{\mu_I^2}\right)^p\,
\bar{A}_{\PT}(k^2),
\eeq
and P.V. indicates the principal value prescription.
Note that in this case $\rho_{2p}$ are independent of $\mu_I$.
A detailed comparison of the two regularization prescriptions can be found in~\cite{average_thrust}.
The explicit expressions for $m_p^{\PT}(\mu_I^2)$ in the one-loop and two-loop cases can be obtained by substituting
$s\longrightarrow \ln \mu_I^2/{\bar{\Lambda}}^2$ in eqs.~(\ref{Ip_1loop}) and
(\ref{Ip_2loop}) (taking the real part and multiplying by $p$), respectively.

Let us now examine in more detail the singularities in (\ref{Borel_nu}). First
note that double poles at half integer values do not occur.
Thus the term $r^{(1)}_p\, s\, \sin\left( \pi p\right)$ in (\ref{I_NP}) is actually not relevant.
Next, to determine the residues $r_p$ of $B_{\nu}({z})$ we use the fact 
that\footnote{Since our interest is in large $\nu$ values, 
corresponding to small $t$, we can replace $\gamma(-{z},\nu)$ by
$\Gamma(-{z})$, which is a good approximation at any $\nu>1$.}
${\rm Residue}\left\{\Gamma(-{z}) \right\}=(-1)^{p+1}/p!$, and
therefore ${\rm Residue}\left\{\Gamma(-2{z}) \right\} =(-1)^{2p+1}/\left(2(2p)!\right)$.
Consider the first term in (\ref{Borel_nu}) which is related to large-angle emission.
Here $r^{la}_p= \frac{1}{p(2p)!} \nu^{2p} (-1)^{2p}$ and there are no double poles,
so
\begin{eqnarray}
\label{I_p_la} I_p^{la}=\frac{-\rho_{2p}}{\pi p^2(2p)!\,{\Gamma(1+\kappa_p)}} \!
\left(\frac{\nu {\bar{\Lambda}}}{Q}\right)^{2p}\! \left[\left({\ln
\frac{Q^2}{{\bar{\Lambda}}^2}}\right)^{\kappa_p}\!\!\! \sin\left( \pi
p\right)+{\kappa_p\pi}\,\left({\ln
\frac{Q^2}{{\bar{\Lambda}}^2}}\right)^{\kappa_p-1} \!\!\!\cos\left( \pi p\right)\right].
\end{eqnarray}
Summing over $p$ this gives
\beq
\label{NP_large_angle}
\left< e^{-\nu t}\right>_{\NP}^{\rm large-angle}=\,-\!\!\!
\sum_{p=\frac12,1,\frac32,2,\frac52,\,\cdots}\!\!\!
\lambda_{2p}\frac{1}{(2p)!}\left(\frac{\nu{\bar{\Lambda}}}{Q}\right)^{2p}
=\,-\,\sum_{n=1}^{\infty}
\lambda_{n}\frac{1}{n!}\left(\frac{\nu{\bar{\Lambda}}}{Q}\right)^{n}
\eeq
where we defined
\begin{eqnarray}
\label{lambda_def} 2p\,\,\,\,\, {\rm odd}
\,\,\,\,\,\,\,\,\,\,\,\,\,\,\,\,&&
\lambda_{2p}=\frac{C_F}{\beta_0}\,
\frac{\rho_{2p}}{p^2}\,\frac{1}{\Gamma(1+\kappa_p)}\,\left({\ln
Q^2/{\bar{\Lambda}}^2}\right)^{\kappa_p}\,\frac{\sin\left( \pi
p\right)}{\pi}
\nonumber \\
2p\,\,\,\,\, {\rm even}  \,\,\,\,\,\,\,\,\,\,\,\,\,\,\,\,&&
\lambda_{2p}=\frac{C_F}{\beta_0}\,\frac{\rho_{2p}}{p^2} \,
{\kappa_p}\,\frac{1}{\Gamma(1+\kappa_p)}\,\left({\ln
Q^2/{\bar{\Lambda}}^2}\right)^{\kappa_p-1}\,\cos\left( \pi p\right).
\end{eqnarray}

The second term in (\ref{Borel_nu}), associated with collinear emission, is treated in a similar manner.
Note that here there are also double poles at $p=1$ and $2$ and consequently the residues have a more complicated
dependence on $\nu$. The result is
\begin{eqnarray}
\label{NP_col}
\left< e^{-\nu t}\right>_{\NP}^{\rm col}&=& -\lambda^{(1)}_2\,\,\frac{\nu {\bar{\Lambda}}^2}{Q^2}\,-\,\lambda^{(1)}_4
\,\frac{\nu^2 {\bar{\Lambda}}^4}{Q^4}
\nonumber \\&-&
\left[\left(4-\gamma_E-\ln\nu\right)\nu-1\right]\,\lambda_2\,\frac{{\bar{\Lambda}}^2}{Q^2}
-\left[\left(-\frac 32+\gamma_E+ \ln\nu\right)\nu^2-1\right]\,\lambda_4\,\frac{{\bar{\Lambda}}^4}{Q^4}\nonumber \\
&&-\sum_{p\geq 3}^{\infty} \lambda_{2p}\left(\frac 2p+\frac{1}{1-p}+\frac{1}{2-p}\right)\,\frac{(-1)^{p+1}}{(p-1)!}
\left(\frac{\nu{\bar{\Lambda}}^2}{Q^2}\right)^{p},
\end{eqnarray}
where the first line corresponds to the double poles at $p=1$ and
$2$, with \[
\lambda^{(1)}_{2p}=-\frac{C_F}{\beta_0}\,
\frac{\rho_{2p}}{p}\frac{(1+\kappa_p)}{{\Gamma(1+\kappa_p)}}\left({\ln
Q^2/{\bar{\Lambda}}^2}\right)^{\kappa_p}\,;
\]
the second line corresponds to the simple poles at $p=1$ and $2$, and the last line to the
remaining simple poles at integers $p\geq 3$. $\lambda_{2p}$ are
given in (\ref{lambda_def}).

Finally, the total non-perturbative contribution in
(\ref{R_res_NP}) is given by 
\beq {\cal S}_{\NP}(Q^2, \ln
\nu)\equiv\left< e^{-\nu t}\right>_{\NP}^{\rm large-angle} +\left<
e^{-\nu t}\right>_{\NP}^{\rm col}.
\label{S_NP} 
\eeq 
Note that in
both (\ref{NP_large_angle}) and (\ref{NP_col}) we use the same
non-perturbative parameters $\rho_{2p}$. One could introduce a
more general parametrisation of power-corrections by defining
separate parameters for the large-angle and collinear regions of
phase-space. This, however, seems redundant from the current point
of view: compensating the ambiguity in the perturbative sum does
not require separate parameters. Moreover, the delicate relation
between the large-angle and collinear limits of phase-space at the
perturbative level suggests that the two are correlated in the
same way also at the non-perturbative level.

It is now straightforward to see the relation with previously
suggested models for
power-corrections~\cite{Shape_function1,Shape_function2,DW_dist,Shape_function3,Shape_function4,Korchemsky_Tafat}.
As mentioned in the introduction, it was shown
\cite{Shape_function1,Shape_function2,DW_dist} that the primary
non-perturbative effect in the region $t\gg{\bar{\Lambda}}/Q$ is a shift
of the resummed distribution to larger values of $t$: \beq
\left.\frac{1}{\sigma} \frac{d\sigma}{d
t}\right\vert_{\res}(Q^2,t)\longrightarrow \left.\frac{1}{\sigma}
\frac{d\sigma}{d t}\right\vert_{\res}(t-\lambda_1{\bar{\Lambda}}/Q).
\label{dist_shift} \eeq

This is indeed the effect if one keeps only the first term in the
sum in (\ref{NP_large_angle}): then the only non-perturbative
contribution to ${\cal S}_{\NP}(Q^2,\ln \nu)$ in (\ref{R_res_NP})
is $\lambda_1\,\nu\,{\bar{\Lambda}}/Q$, and
\beq
f_{\rm la}=\delta(t-\lambda_1{\bar{\Lambda}}/Q).
\eeq
Assuming that the average thrust is dominated by the two-jet region, the corresponding
correction is $\left<t\right>\longrightarrow \left<t\right> +\lambda_1{\bar{\Lambda}}/Q$, namely it is the
same parameter that controls the shift of the distribution and the leading power-correction
to~$\left<t\right>$. This $1/Q$ correction to the average thrust was discussed
extensively in the literature~\cite{W,DW,AZ,DMW,Mil,Shape_function4,average_thrust,higher_moments}.

As explained in~\cite{Shape_function4,Korchemsky_Tafat}, the
shift~(\ref{dist_shift}) is no longer adequate when $t\simeq
{\bar{\Lambda}}/Q$. The shape-function
\cite{Shape_function1,Shape_function2,Shape_function3,Shape_function4,Korchemsky_Tafat}
allows parametrisation of all the terms in (\ref{NP_large_angle})
making the following identification: \beq \int_0^{\infty}
d\tilde{t} \exp(-\nu \tilde{t})\,(Q/{\bar{\Lambda}})\,f_{\rm
la}(\tilde{t}Q/{\bar{\Lambda}}) =\exp\left\{\left< e^{-\nu
t}\right>_{\NP}^{\rm large-angle}\right\}. \label{SF} \eeq The
single argument function $(Q/{\bar{\Lambda}})\,f_{\rm
la}(tQ/{\bar{\Lambda}})$ is therefore a particular example of the
two-argument non-perturbative function $f_{\NP}(Q^2,t)$ defined in
eqs.~(\ref{R_res_total}) and~(\ref{R_res_f_NP}). Similarly to
\cite{Shape_function4,Korchemsky_Tafat}, the moments of the
large-angle shape-function $\sigma_m\equiv \int_0^{\infty}f_{\rm
  la}(\epsilon)\epsilon^m d\epsilon$ are related to the parameters
$\lambda_{2p}$. Writing (\ref{SF}) as \beq
\int_0^{\infty}\exp\left\{-\frac{\nu\epsilon{\bar{\Lambda}}}{Q}\right\}
f_{\rm la}(\epsilon)d\epsilon=\exp\left\{\,-\,\sum_{n=1}^{\infty}
\lambda_{n}\frac{1}{n!}\left(\frac{\nu{\bar{\Lambda}}}{Q}\right)^{n}\right\}
\label{SF_sum} \eeq and expanding the exponents on both sides one
has: \beq
\sigma_0=1,\,\,\,\,\,\,\,\,\,\sigma_1=\lambda_1,\,\,\,\,\,\,\,\,\,\sigma_2=\lambda_1^2-\lambda_2,\,\,\,\,\,\,\,\,\,\sigma_3=\lambda_1^3-3\lambda_1\lambda_2+\lambda_3.
\eeq The function $f_{\rm la}(\epsilon)$, like the parameters
$\lambda_{2p}$, depends on the regularization prescription of the
perturbative calculation. In the absence of a non-perturbative
calculation (ref. \cite{Shape_function4} gives a field theoretic
definition) the shape-function and thus $\lambda_n$ can only be
determined by fitting the data.

Finally, to deal with the collinear power-corrections, one should
include~(\ref{NP_col}). Here, the most important corrections arise
from the terms that are enhanced by $\ln \nu$. Note that the {\em
remaining} terms can be simply recast in the form \beq
\int_0^{\infty} d\tilde{t} \exp(-\nu
\tilde{t})\,(Q^2/{\bar{\Lambda}}^2)\,f^{\rm sub-leading}_{\rm
col}(\tilde{t}Q^2/{\bar{\Lambda}}^2) \eeq similarly to the shape-function
of (\ref{SF}). These corrections become important only at
$t\sim{\bar{\Lambda}}^2/Q^2$ and can be safely neglected in the
region $t\simeq {\bar{\Lambda}}/Q$. To treat the terms that are enhanced
by $\ln \nu$ we first expand the exponent,
\begin{eqnarray}
\label{col_exponent}
 \exp\left\{\left< e^{-\nu t}\right>_{\NP}^{\rm col}\right\}
&=& 1+\lambda_2\frac{\nu{\bar{\Lambda}}^2}{Q^2}
\left(\ln\nu+\gamma_E\right)+\left[\frac{\lambda_2^2}{2}\left(\ln\nu+\gamma_E\right)^2
-\lambda_4\left(\ln\nu+\gamma_E\right)\right]\frac{\nu^2{\bar{\Lambda}}^4}{Q^4}\nonumber \\
&+&\left[ \frac{\lambda_2^3}{6}\left(\ln\nu+\gamma_E\right)^3
-\lambda_2^2\lambda_4\left(\ln\nu+\gamma_E\right)^2\right]\frac{\nu^3{\bar{\Lambda}}^6}{Q^6}+\cdots,
\end{eqnarray}
where we kept the sub-leading $\gamma_E$ constants for convenience (see below).
Next we use the fact that
\[{\Gamma(\epsilon)}\,\nu^{-\epsilon}= \int_0^{\infty}dt \frac{1}{t}\,t^\epsilon \, e^{-\nu t}\]
to write the following convolution integral in the ``+'' prescription, with a generic test function $h(t)$ whose
inverse Laplace transform is $H(\nu)$,
\begin{eqnarray}
\label{laplace_identity}
\left[{\Gamma(\epsilon)}\,\nu^{-\epsilon}-\frac{1}{\epsilon}\right]
\,H(\nu)&=&\int_0^{\infty}dt \, e^{-\nu t}\int_0^t d\tilde{t}
\,\left[\frac{1}{\tilde{t}}\,
e^{\epsilon\ln \tilde{t}}\right]_{+}\,h(t-\tilde{t})\\ \nonumber &=&
\int_0^{\infty}dt \, e^{-\nu t}\int_0^t d\tilde{t}
\,\frac{1}{\tilde{t}}\,
e^{\epsilon\ln \tilde{t}}\,\left[h(t-\tilde{t})-h(t)\right]
\end{eqnarray}
Now one can extract the inverse Laplace transform
of the terms in (\ref{col_exponent}), in the distribution sense, by expanding both sides of eq.~(\ref{laplace_identity})
in powers of $\epsilon$. In particular, one has \[\int_0^{\infty}dt\, \left[\frac1t\right]_+ \,e^{-\nu t} =-(\ln \nu+\gamma_E)
\,\,\,\,\,\,\,\,\,\,\,\,\,\,\,\,\,\,\,\,\,\,
\int_0^{\infty}dt\,  \left[\frac{\ln t}t\right]_+\,e^{-\nu t}=\frac12 (\ln \nu+\gamma_E)^2+\frac{\pi^2}{12}.\] Thus,
\begin{eqnarray}
\label{R_res_col_leading}
f_{\rm col}(Q^2,t)\!\!&=&\!\!\int_{\cal C} \frac{d\nu}{2\pi i\, }
e^{\nu t} \exp\left\{\left< e^{-\nu t}\right>_{\NP}^{\rm col}\right\} \\ \nonumber
&=&\!\! \delta(t)- \lambda_2\left[\frac{{\bar{\Lambda}}^2}{Q^2t}\right]_
+\!\!\!\!\!\star\delta^{(1)}(t)+\left[\lambda_2^2\frac{{\bar{\Lambda}}^4}{Q^4}\left(\frac{\ln t}{t}-\frac{\pi^2}{12}\delta(t)\right)
+\lambda_4\frac{{\bar{\Lambda}}^4}{Q^4}\frac{1}{t}\right]_+\!\!\!\!\!\star\delta^{(2)}(t)
+\cdots
\\ \nonumber
&=&\!\!
\left[1+\lambda_2\frac{{\bar{\Lambda}}^2}{Q^2t^2}-\left(2\lambda_2^2\,\ln\frac{1}{t}
+{\cal O}\left(1\right)\right)\,\frac{{\bar{\Lambda}}^4}{Q^4t^3}+{\cal O}\left(\frac{{\bar{\Lambda}}^6}{Q^6t^4}\right)\right]\,\delta(t),
\end{eqnarray}
where $\delta^{(n)}(t)$ is the $n$-th derivative of the
$\delta$-function\footnote{In the last line, the appropriate
integration prescription at $t\longrightarrow 0$ is understood.}.
Clearly, the leading correction $\frac{\lambda_2}{Q^2t^2}$ is as
important around $t\simeq {\bar{\Lambda}}/Q$ as the large-angle
corrections contained in the shape-function (\ref{SF}). The
sub-leading terms in (\ref{R_res_col_leading}) are less important
and can be ignored in a first approximation.

In conclusion, the most important power correction at
$t\gg{\bar{\Lambda}}/Q$ is the one associated with the simple renormalon
pole at $z=\frac12$. Its exponentiation amounts to a shift of the
resummed distribution by $\lambda_1{\bar{\Lambda}}/Q$. Closer to the
two-jet region sub-leading power corrections of the form
$\lambda_n({\bar{\Lambda}}/tQ)^n$ become important. Strictly based on
the large $\beta_0$ renormalon calculation, only odd values
associated with renormalon poles at half integers, are relevant.
However, since the large $\beta_0$ limit is not expected to
predict the singularity strength of the exact Borel transform, one
should expect $\lambda_n({\bar{\Lambda}}/tQ)^n$ with even $n$ to survive.
Assuming that the exact Borel function differs from the calculated
one only by a modification of the singularity strength, with the
$\nu$ dependence left unchanged, we predict that the even terms
will be suppressed by $1/\ln(Q^2/{\bar{\Lambda}}^2)$ compared to the odd
ones (see \ref{lambda_def}). Allowing both odd and even terms in the sum in
(\ref{SF_sum}) one recovers the shape-function description as far
as the large-angle gluon emission is concerned. Nevertheless, at
the same time a collinear correction proportional to
$\lambda_2({\bar{\Lambda}}/tQ)^2$ appears in~(\ref{R_res_col_leading}).
The latter has a sign apposite to the relevant term in the
large-angle shape-function, with a slightly different functional
form, thus reducing significantly the overall effect of the
non-perturbative parameter~$\lambda_2$.

Finally, we comment that an alternative way to derive the form of the power corrections from the 
large $\beta_0$ calculation is to use the characteristic function~(\ref{char_function_nu}).
Expanding the latter at small $\epsilon$ one obtains
\begin{eqnarray*}
\dot{\cal { F}}_{\nu}\left(\epsilon\right) &=&2\,\sqrt{\epsilon }\,\nu
 + \left( - {\frac {1}{2}} \,\nu
\,{\rm ln}\left({\frac {1}{\epsilon }} \right)
- 3\,\nu  - { \frac {1}{2}} \,\nu^{2}\right)\,
\epsilon
+ \left(\nu  + { \frac {1}{2}} \,\nu ^{2} + { \frac {1}{9}} \,\nu ^{3}\right)\,\epsilon ^{3/
2} \\
 &+&   \left({ \frac {1}{4}} \,\nu ^{2}\,{\rm ln}\left({ \frac {1}{\epsilon
}} \right) - { \frac {1}{12}} \,\nu ^{3} -
{ \frac {1}{48}} \,\nu ^{4}
\right)\,\epsilon ^{2} + \left( - { \frac {1}{6}} \,\nu
^{3} +{ \frac {1}{72}} \,\nu ^{4}+ { \frac {1}{300}} \,\nu ^{5}
\right)\,\epsilon ^{5/2} \\
 &+&   \left( { \frac {5}{36}} \,\nu ^{3}+ { \frac {1}{48}} \,\nu ^{4}- { \frac {1}{480}} \,\nu ^{5}
 - { \frac {1}{2160}} \,\nu ^{6} \right)
\,\epsilon ^{3} \\\nonumber &+& \left({- {
\frac {1}{360}} \,\nu ^{5}+ \frac {1}{3600}} \,\nu ^{6}
 + { \frac {1}{17640}} \,\nu ^{7} \right)\,\epsilon ^{7/2}+\cdots.
\end{eqnarray*}
Here only non-analytic terms are associated with ambiguity of the
perturbative sum~(\ref{char_function_rep_nu}), and therefore with
power corrections~\cite{DMW,beneke}. One can immediately identify the terms $(\nu\sqrt{\epsilon})^{n}$ where
$n=1,3,5,\ldots$ as the half-integer renormalon poles associated with large-angle soft
gluons. These are the most important power corrections. As for integer powers
$\epsilon^p$, the only non-analytic terms are at $p=1$ and $p=2$, due to the logarithm.
These ambiguities, however, are negligible for any $t\gg{\bar{\Lambda}}^2/Q^2$ since each power of $\epsilon$ is accompanied by $\nu$ rather than
by $\nu^2$.
Thus the picture is consistent with what we already learned based on the Borel analysis in the case that the
singularities at integer $z$ are just simple poles and are therefore cancelled by $\sin(\pi z)$ factor from
the analytic continuation.

%%%%%%%%%%%%%%%%%%%%%%%%%%%%%%%%%%%%%%%%%%%%%%%%%%%%%%%%%%%%%%%%%%%%%%%%%%%%%%%%

\section{Relation to fixed logarithmic accuracy and fixed-order calculations}

In order to prove consistency of DGE with the exact exponent to NLL accuracy we will now compare our
calculation to that of~\cite{CTTW}. This will lead to identification of coupling $\bar{A}$ as the 
``gluon bremsstrahlung'' coupling~\cite{CMW}. Next we briefly recall the way to match the
resummed result to the fixed-order calculation. 

\subsection{NLL resummation from an evolution equation}

The calculation of the differential cross-section by DGE requires to identify the 
coupling~$\bar{A}$. The analogy with the Abelian theory~\cite{BLM,average_thrust,conf} 
suggests that there is a unique all-order definition of $\bar{A}$ (analogous to the Gell-Mann--Low
effective charge) which does not depend on the observable considered, such that (\ref{dispersive})
is the leading term in the skeleton expansion. Since a systematic skeleton expansion 
has not yet been constructed in QCD other considerations must be used. 
The analogy with the Abelian theory allows one to fix the large $\beta_0$ coupling, 
as done in (\ref{A_bar}). To go beyond this limit we require consistency of the 
exponent to NLL accuracy. We will see that this amounts to identifying the coupling $\bar{A}$
as the ``gluon bremsstrahlung'' coupling~\cite{CMW}. 

The standard way of calculating the logarithmically enhanced
cross-section~\cite{CTTW} is based on an evolution equation for the jet
mass distribution $J(Q^2,k^2)$, where the kernel is the splitting function
of a gluon off a quark, $P(a,x)$. At leading order the splitting function
is 
\beq P(a,x)\,=\,C_F\, a \left[\frac{x}{1-x}+ \frac12 (1-x)\right]_+
=\,C_F\, a \left[\frac{1}{1-x}-\frac12 (1+x)\right]_+ \label{lo_splitting}
\eeq 
where $a(k^2)$ is the coupling ($\alpha_s(k^2)/\pi$) and $x$ is the
longitudinal momentum fraction in the branching. 
Fixing the coupling to be the ``gluon bremsstrahlung'' coupling~\cite{CMW},
\beq
a\longrightarrow \tilde{a}\equiv a_{\MSbar}+\left[\frac53\beta_0
+\left(\frac13-\frac{\pi^2}{12}\right)C_A\right] a_{\MSbar}^2+\cdots
\label{brem}
\eeq
this splitting function is correct to next-to-leading order, as far as the
singular \hbox{$1/(1-x)$} terms are concerned. 
At NLL accuracy, the evolution equation is~\cite{CTTW},
\beq
\frac{dJ_\nu(Q^2)}{d\ln Q^2} = 
\int_0^1dx P\left(\tilde{a}((1-x)Q^2),x\right)
\,\left[e^{-\nu(1-x)}-1\right]\,J_\nu(Q^2)
\eeq
where $J_\nu(Q^2)$ is the Laplace transform of $J(Q^2,k^2)$ with respect to $k^2/Q^2$.
The solution of this differential equation is~\cite{CTTW}
\beq
\ln J_\nu(Q^2)  =\int_0^1du \left[e^{-\nu u}-1\right] \frac{1}{u}\int_0^{1-u}
dx\,P\left(\tilde{a}((1-x)uQ^2),x\right).
\label{lnJ_nu}
\eeq
To this accuracy $1-\mbox{thrust}$, $t$, is the sum of invariant masses of the 
two hemispheres, so
\beq
R(Q^2,t)=\int_0^{\infty}dk^2 d\bar{k}^2J(Q^2,k^2)J(Q^2,{\bar{k}}^2)
\theta\left(tQ^2-k^2-{\bar{k}}^2\right),
\label{R_CTTW_1}
\eeq
and therefore,
\beq
\left.R(Q^2,t)\right\vert_{\res}=\int_{\cal C} \frac{d\nu}{2\pi i\, \nu}  e^{\nu t}\exp\left\{ 2 \ln J_\nu(Q^2)     \right\}.
\label{R_CTTW}
\eeq
From here, a straightforward calculation gives the leading and 
next-to-leading logs in the thrust distribution.

Let us now compare this calculation with ours. Specifically, $2\ln
J_{\nu}(Q^2)$ in (\ref{R_CTTW}) should be compared with $\left<
e^{-\nu t}\right>_{\SDG}$ in~(\ref{ds_res_1}). The two quantities
are defined in (\ref{lnJ_nu}) and (\ref{exp_nut_def}),
respectively. The first observation is that the integration
variable $u$ in (\ref{lnJ_nu}) can be identified with $t$, making
the two similar provided one identifies 
\beq
\left.\frac{1}{\sigma}\frac{d\sigma}{dt}(Q^2,t)\right\vert_{\SDG}
\longleftrightarrow \frac{2}{t}\int_0^{1-t} dx
\,P\left(\tilde{a}((1-x)tQ^2),x\right), \label{comparison_Sp} \eeq or,
explicitly (see~\ref{ours}), \beq
\int_{t^2}^{t}\frac{d\epsilon}{\epsilon}\,\bar{A}_{\eff} (\epsilon
Q^2)
\left[\frac{2}{t}-\frac{\epsilon}{t^2}-\frac{\epsilon^2}{t^3}\right]
\longleftrightarrow \frac{2}{t}\int_0^{1-t} dx \,{\beta_0}
\tilde{a}((1-x)tQ^2) \left[\frac{1}{1-x}-\frac12 (1+x)\right].
\label{comparison} 
\eeq 
If one identifies $\bar{A}_{\eff}(\epsilon Q^2)$ with ${\beta_0} \tilde{a}((1-x)tQ^2)$, 
and changes variables $\epsilon=(1-x)t$, the r.h.s. becomes
\[
\int_{t^2}^{t} \frac{d\epsilon}{\epsilon}\,\bar{A}_{\eff}
(\epsilon Q^2)\, \left[\frac2t-\frac{2\epsilon}{t^2}+\frac{\epsilon^2}{t^3}\right].
\]
One thus finds that the first term in the square brackets on both
sides is the same. Recall that this term is responsible for the
leading logs. The other terms, contributing to sub-leading logs,
are not the same. However, the NLL are: to NLL accuracy it is
enough to consider the conformal part~\cite{CTTW} by fixing the
scale of the coupling. The latter gives, in both cases,
$-\frac{1}{t}\frac32 \bar{A}(Q^2)$ from the collinear limit $\epsilon=t$.

We thus learned that DGE is consistent with the exact exponent to NLL accuracy 
provided one identifies the running coupling with the 
``gluon bremsstrahlung'' coupling~\cite{CMW} of eq.~(\ref{brem}).
In evaluating the Borel sum (\ref{Borel_res_nu}) we therefore use 
\beq
\bar{A}(k^2)=\frac{{A}(k^2)}{1-\left[\frac53
+\left(\frac13-\frac{\pi^2}{12}\right)C_A/\beta_0\right] {A}(k^2)}.
\label{A_bar_brem}
\eeq
instead of~(\ref{A_bar}). This simply amounts to fixing the definition of 
$\bar{\Lambda}$ (see~(\ref{A_Borel})).

The qualitative features of the SDG characteristic-function based
calculation appear also in the splitting-function based
calculation: there appear $\epsilon/t^2$ and $\epsilon^2/t^3$
terms which contribute to the log-enhanced cross-section due to
the collinear limit $\epsilon\simeq t$. As we saw in the previous
sections these terms lead to factorial enhancement of sub-leading
logs (already prior to exponentiation). However, the actual values
of sub-leading logs predicted by the splitting-function beyond the
NLL are different. Performing the integral on the r.h.s. 
of~(\ref{comparison}) explicitly we arrive at the following expansion
for $\frac{1}{\sigma}\frac{d\sigma}{dt}(Q^2,t)$ prior to
exponentiation:
\begin{eqnarray} \label{naive_sf}
&&\frac{1}{\sigma}\frac{d\sigma}{dt}(Q^2,t)=
\frac{C_F^{\,}}{t\,\beta_0}\,\left\{ \lefteqn{(2.\,L -
1.5)\,{{A(Q^2)}} + (3\,L^{2}+1.8333\,L - 4.2500 )\,
{{A(Q^2)}}^{2}} \right.\\ \nonumber
 & &+ \,\,   (4.6667\,L^{3} + 8.5000\,L^{2} - 9.5242\,L -
8.8151)\,{{A(Q^2)}}^{3} \\ \nonumber
 & &+\,\,    (7.5000\,L^{4} + 21.833\,L^{3} - 17.359\,L^{2} -
50.085\,L - 9.957)\,{{A(Q^2)}}^{4} \\ \nonumber
 & &+\,\,    (12.400\,L^{5} + 48.500\,L^{4} - 31.339\,L^{3} -
194.72\,L^{2} - 95.096\,L + 10.97)\,{{A(Q^2)}}^{5}+\cdots
\end{eqnarray}
This expansion can be compared directly with (\ref{our_res}). The LL
and NLL are of course the same. Sub-leading logs are similar in
many cases; note in particular the NNLL. Nevertheless, at lower
logs there are big variations, as one can deduced directly
from~(\ref{comparison}).

The fact that the two calculations differ beyond NLL is of no
surprise. Most importantly, the kinematic approximations made in
the splitting-function based calculation (e.g. the matrix element,
the definition of the observable), modify the logs beyond NLL
accuracy.

It should be emphasised that some approximations have been made
also in the characteristic-function based calculation. Although
this calculation is {\em exact} at the level of three partons with
an off-shell gluon, there are approximations beyond this level:
the massive-gluon characteristic-function does not take into
account the possible branching of the gluon into opposite
hemispheres, the so-called non-inclusive contribution. When the
gluon is roughly collinear with one of the quarks (the only limit
contributing to sub-leading logs in
$\left.\frac{1}{\sigma}\frac{d\sigma}{dt}(Q^2,t)\right\vert_{\SDG}$)
this type of branching is probably not significant. Indeed, the
comparison between
$\left.\frac{1}{\sigma}\frac{d\sigma}{dt}(Q^2,t)\right\vert_{\SDG}$
and the leading term in $\beta_0$ in the (numerically evaluated)
exact NLO coefficient in sec.~2.2 confirms this assertion. Another
important aspect is the validity of the exponentiation formula
(\ref{R_res}) beyond NLL accuracy. As discussed in sec.~3,
correlations in the emission of multiple soft and collinear gluons
may influence the sub-leading logs. This effect was neglected.

\subsection{Matching to the fixed-order calculation}

At large values of $t=1-\mbox{thrust}$, the resummed cross-section
(be it DGE or NLL) must be complemented by the exact fixed
order result. To do this we will use the $\log$-R matching scheme.
The basic idea of matching is to replace the low orders
in the resummed expression by the exact low order terms.
Schematically one writes~\hbox{ 
$R_{\rm matched}=R^{\NLO}+R_{\res}-R_{\res}^{\NLO}$} where $R_{\res}$ is the resummed 
cross-section and $R_{\res}^{\NLO}$ is its expansion to NLO.
There is no general argument for matching the integrated cross-section itself (R-matching) 
as compared to matching some function of $R$. Exponentiation may hold beyond the level of the 
logarithmically enhanced cross-section, so it is natural to match the logarithm of $R$ 
(log-R matching). In practice the difference between different matching procedures is small 
(see e.g.~\cite{CTTW}) and for simplicity we only consider $\log$-R matching.

Using the known NLO calculation~\cite{thrust-nlo,EVENT,EVENT2}, 
$$R(Q^2,t)\equiv 1+R_1(t)a+R_2(t)a^2+\cdots$$ where $R_1(t)$ is known 
analytically and $R_2(t)$ is parametrised based on a numerical 
computation by EVENT2~\cite{EVENT2}, we write
\beq
R(Q^2,t)=
\exp\left\{R_1a(Q^2)+\left(R_2-\frac12R_1^2\right)a(Q^2)^2
           +\left.\ln R(Q^2,t)\right\vert_{\res}
           -\left.\ln R(Q^2,t)\right\vert_{\res}^{\NLO}\right\}. 
\label{log_R_maching}
\eeq 
Here $\left. \ln R(Q^2,t)\right\vert_{\res}$ is our resummed
result for the logarithmically enhanced integrated cross-section,
given by eqs.~(\ref{R_res_PT}) and (\ref{R_res_sum_I}) and
$\left.\ln R(Q^2,t)\right\vert_{\res}^{\NLO}$ is its expansion, up
to NLO in $\rm {\overline{MS}}$,
\begin{eqnarray}
\label{NLO_exp_of_DGE}
\left.\ln R\left(Q^2,t\right)\right\vert_{\res}^{\NLO}\!\!&=&
\left[ - C_F\,L^{2} + {\frac {3}{2}} \,C_F\,L\right]\,a(Q^2)+
\biggl[ - C_F\,\beta_0\,L^{{3}^{\,}}  \\ \nonumber
&+& \left( - {\frac {11}{
12}} \,C_F\,\beta_0 - {\frac {1}{3}} \,\pi
^{2}\,C_F^{2} + \left({\frac {1}{12}} \,\pi ^{2}
 - {\frac {1}{3}} \right)\,C_A\,C_F\right)\,L^{2} \\ \nonumber
&+&  \left({\frac {15}{4}} \,C_F\,\beta_0
 + \left( - 4\,\zeta_3 + {\frac {1}{2}} \,\pi ^{2}\right)\,
C_F^{2} + \left({\frac {1}{2}}  -
{\frac {1}{8}} \,\pi ^{2}\right)\,C_A\,C_F\right)\,L \\ \nonumber
&+&  \left.
 \left(3\,\zeta_3 - {\frac {1}{180}} \,\pi^{4} - {\frac {3}{16}} \,\pi ^{2}\right)\,C_F^{2}\right]
a(Q^2)^{2}.
\end{eqnarray}
The matching~(\ref{log_R_maching}) is done in the $\rm {\overline
{MS}}$ scheme.

%%%%%%%%%%%%%%%%%%%%%%%%%%%%%%%%%%%%%%%%%%%%%%%%%%%%%%%%%%%%%%%%%%%%%%%%%%%%%%%%

\section {The phenomenological implication}

We saw that a careful treatment of running coupling effects (or
renormalons) in the logarithmically enhanced cross-section
requires resummation of sub-leading logs. Our investigation in
section~3.2 leaves no doubt: sub-leading logs contribute
significantly at all physically interesting values of the thrust.
On the other hand, it is well known that successful fits to the
distribution were obtained
in~\cite{DW_dist,Korchemsky_Tafat,Shape_function4} with a much
simpler treatment of the perturbative calculation, based on the
standard NLL resummation. Non-perturbative
corrections whose magnitude is not under control theoretically are
anyway important, and must be included in any fit. One may argue
that a more complete perturbative treatment is not necessary, and
that it amounts simply to a redefinition of the non-perturbative
correction. The answer is, of course, that as soon as quantitative
predictions are expected one should start with a reliable
perturbative calculation. Physically, the transition between the
perturbative and non-perturbative regimes is smooth, and one must
make sure that an arbitrary separation of the two does not have
too strong an impact on the results. Legitimate separations are
cut-off regularization of momentum integrals and principal-value
Borel summation~\cite{Contopanagos:1994yq}. 
It was shown in \cite{average_thrust} that these
two procedures can be fully consistent with each other provided
that renormalon resummation is performed. This means in particular
that the extracted value of $\alpha_s$ is the same using the two
prescriptions. Such consistency cannot be achieved if a truncated
perturbative expansion, e.g. NLO, is used instead. Our results in
sec.~3.2 imply that NLL truncation of the exponent is just as
dangerous. By performing renormalon resummation, such a truncation
is avoided and a more reliable comparison with data can be
achieved.

Here we shall perform global fits to the thrust distribution at all energies,
based on our perturbative calculation of section~3. As discussed in section 3.3
we use Borel summation. An equivalent calculation can be done
using cut-off regularization. The translation of our results to this language
is straightforward and may be useful once comparison can be made to a
non-perturbative calculation of the shape-function with a hard cut-off as an
ultraviolet regulator.

The perturbative calculation must be supplemented by
non-perturbative corrections in order to be compared with the
data. Our guiding principle is that the functional form of the
non-perturbative corrections is dominated by the ambiguities
identified in the calculation of the perturbative sum. We assume
that the SDG based calculation is sufficient to
identify these ambiguities. Of course, there are inherent
limitations to this calculation: splitting of a gluon into partons
that end up in opposite hemispheres is not taken into
account, nor are correlations between the emitted gluons.
These effects are not expected to be important in the two-jet region, so it is
reasonable to neglect them.

\subsection{Perturbative calculation}

\begin{figure}[t]
\begin{center}
\mbox{\kern-0.5cm\epsfig{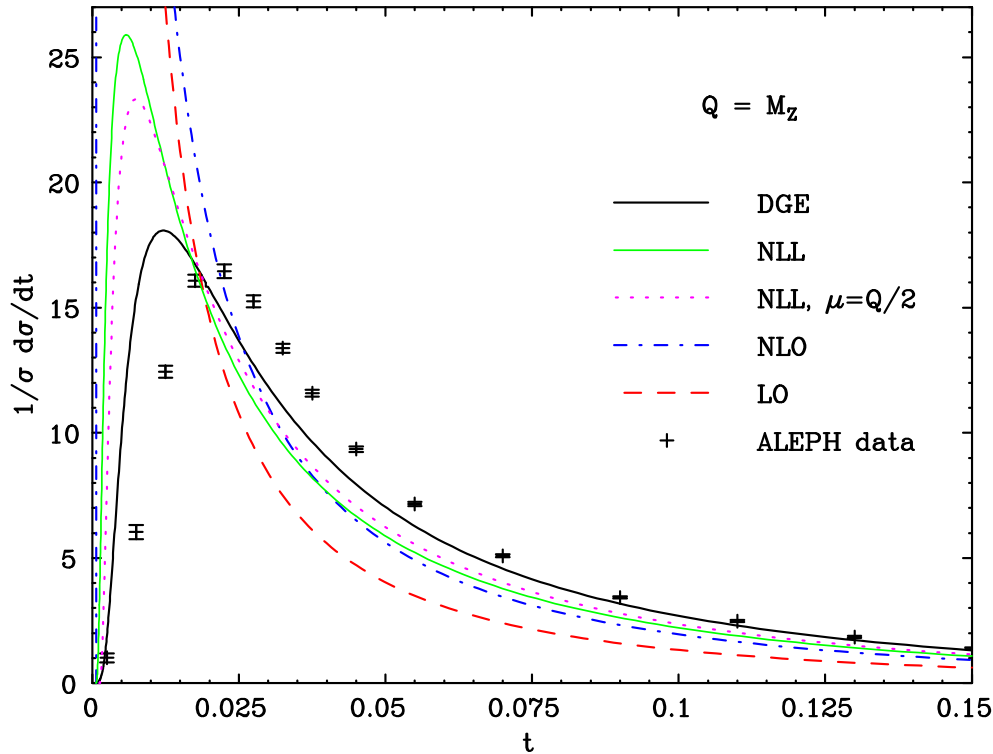}
}
\end{center}
\caption{DGE, the standard NLL and the LO and NLO
results as a function of~$t$ in the two-jet region, at
$Q={\rm M_Z}$. Fixed-order and NLL results are in the
$\overline{\rm MS}$ scheme. As an example of scale dependence
the NLL result is shown at two different renormalization 
points $\mu_R=Q$ and $\mu_R=Q/2$.
For the LO and NLO results $\mu_R=Q$.
We assume $\alpha_s^{\MSbar}({\rm M_Z})=0.110$. $\aleph$
data~\cite{ALEPH91} is shown for orientation.}
\label{naive_vs_res}
\end{figure}

\setcounter{footnote}{0}
Before attempting any fits to the data involving non-perturbative
corrections, it is useful to examine the impact of the resummation
simply by looking at the perturbatively calculated cross-section
in fig.~\ref{naive_vs_res}. Needless to say, the fixed-order
calculations, LO and NLO, do not describe the physical cross
section in the two-jet region. The large difference between the
NLO and the LO signals large higher-order corrections. So does the
renormalization scale dependence of the result (not shown in the
plot). The qualitatively different behaviour of the NLL result is
clearly seen: thanks to exponentiation it vanishes at $t=0$.
Nevertheless, as fig.~\ref{naive_vs_res} shows, the renormalization
scale dependence is still appreciable\footnote{It has been shown~\cite{average_thrust,conf} that renormalon resummation 
in single scale observables can be imitated by an appropriate choice of the renormalization 
scale, similarly to the original formulation of the BLM approach~\cite{BLM}.
One might wonder to what extent DGE can be imitated by simply tuning the
scale of the coupling in the standard NLL formula. Physically, the naive scale-setting approach 
is bound to fail for multi-scale observables such as differential cross-sections, since the typical
scale of emitted gluons is not anymore a fixed fraction of the center-of-mass energy, but rather some
function of all the external parameters. In our case, it is a function of both $Q$ and $t$. 
Indeed, choosing lower scales for the coupling in the NLL formula 
one cannot reproduce the DGE result.}. We stress that the factor
$2$ between the renormalization points is
completely arbitrary. This difference should {\em not} be considered as
a reliable estimate of the error due to running coupling effects
that have been neglected, since the physical scale, (e.g. the
transverse momentum) in the two-jet region is significantly
smaller than $Q$. As anticipated in sec.~3.2, the impact of the
additional resummation of sub-leading logs by DGE is quite
significant. Having performed this resummation, the renormalization
scale dependence (at the level of the logarithmically enhanced
cross-section) is avoided.

In fig.~\ref{naive_vs_res}, as in all forthcoming calculations we
use the so-called log-R matching scheme~\cite{CTTW}
, defined in Eq.~(\ref{log_R_maching}), to complement
the calculation of the logarithmically enhanced cross-section.
Renormalization scale dependence appears only
beyond the level of the logarithmically enhanced cross-section and
it is therefore suppressed by $t$.
Numerically, the residual scale dependence (see table~\ref{residual_scale_dep})
is very small as long as the log-enhanced cross-section dominates. The difference is
much too small to be observed in the resolution of fig.~\ref{naive_vs_res}.
At large $t$ ($t\gsim 0.2$) the terms which are suppressed by $t$ start making an impact and eventually
the scale dependence becomes comparable to its values in the NLL and the NLO
results.
\begin{table}[H]
\caption{Residual renormalization scale dependence: the relative difference between the
differential cross-section at $Q={\rm M_Z}$ as calculated from~(\ref{log_R_maching}) when the
matching with the NLO result is performed at $\mu_R=Q$ and at $Q/2$. \label{residual_scale_dep}}
\begin{center}
\begin{tabular}{ccccccccc}
\hline
$t$&0.02&    0.05&    0.10&    0.15&    0.20&    0.25&    0.30&    0.33\\
$\%$& 0.20&    0.15&    0.01&    0.29&    0.58&   1.1&    2.8&    7.3\\
\hline
\end{tabular}
\end{center}
\end{table}

\begin{figure}[t]
\begin{center}
\mbox{\kern-0.5cm\epsfig{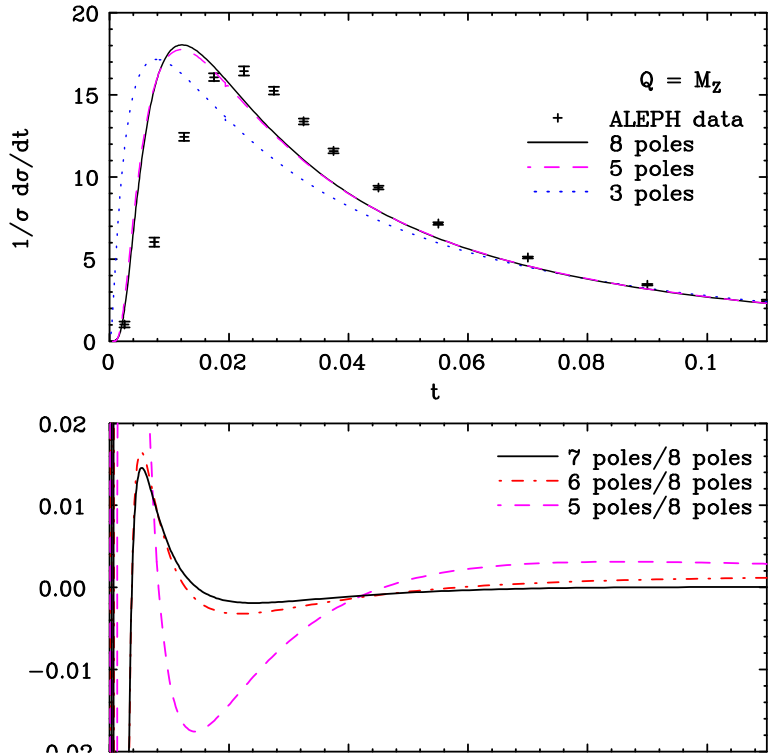}
}
\end{center}
\caption{The convergence of increasing approximants to the
perturbative cross-section, according to the technique of
section~3.3, eq.~(\ref{R_res_sum_I}). The upper plot shows the
differential cross-section for $p_{\rm max}=3,5$ and $8$ and the
lower plot shows the relative error of the differential
cross-section with respect to the $p_{\rm max}=8$ calculation, for
$p_{\rm max}=5,6$ and $7$. We fix $\alpha_s^{\MSbar}({\rm
M_Z})=0.110$ and use two-loop running coupling; log-R matching is
applied.
 }
\label{convergence}
\end{figure}

Next, we confront the question of reliability of the approximation
technique we used in section~3.3 in evaluating the Borel sum. In
the calculation presented in fig.~\ref{naive_vs_res} we evaluated
the Borel sum in (\ref{R_res_sum_I}) using $p_{\rm max}=8$. This
means that the function $V(z)$ in eq.~(\ref{V_def}) was
approximated by $8$ poles. Note that $p_{\rm max}$ is also the
largest power of $\ln \nu$ respected by this approximation
(see~(\ref{r_p_def})). In addition, the renormalon
integrals $\tilde{I}_p(\ln Q^2/{\bar{\Lambda}}^2)$ were evaluated using
the two-loop running coupling in the ``gluon bremsstrahlung''
scheme (see section~4). How sensitive are the results to these
approximations? To answer this question we repeat the
calculation varying the choice of $p_{\rm max}$ and the
running coupling. Considering first the choice of~$p_{\rm max}$,
we show in fig.~\ref{convergence} approximants of increasing
order.
The convergence is quite remarkable. In the upper plot the $p_{\rm
max}=5$ curve (dashed) can hardly be distinguished from the
$p_{\rm max}=8$ curve. Indeed, as the lower plot shows, the
difference between them is very small and it exceeds $1\%$ only
around the distribution peak. For our fits, we choose $p_{\rm
max}=8$ which guarantees a truncation 
error\footnote{This truncation error must not be confused
with the ambiguity of the perturbative sum.} of less than $1\%$
down to $t=0.01$ at ${\rm M_Z}$ (or $0.01\, {\rm M_Z}/Q$ at $Q$).

\begin{figure}[t]
\begin{center}
\mbox{\kern-0.5cm\epsfig{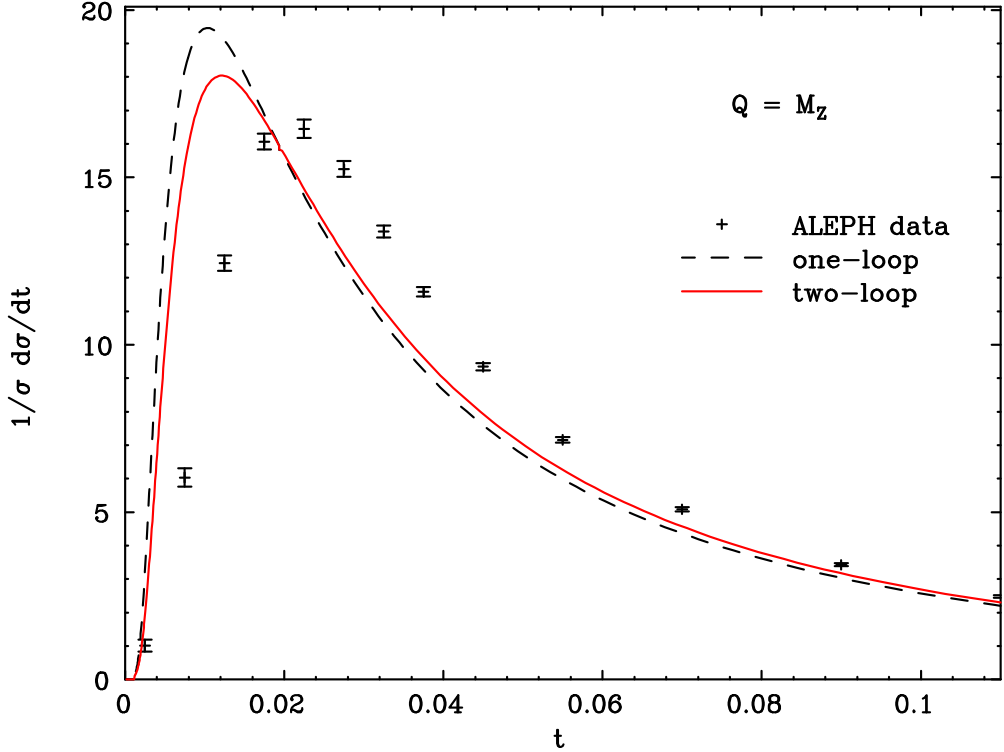}
}
\end{center}
\caption{The resummed cross-section at ${\rm M_Z}$ based on
one-loop (dashed) and two-loop (full line) running coupling. We
fix $\alpha_s^{\MSbar}({\rm M_Z})=0.110$ and use $p_{\max}=8$;
log-R matching is applied. $\aleph$ data~\cite{ALEPH91} is shown for orientation.
 }
\label{12_loop}
\end{figure}

The stability of the calculation with respect to the
running coupling can be checked by repeating the calculation with
a different renormalization-group equation. In principle, one may
also consider changing the definition of $\Lambda$, thus replacing
the ``gluon bremsstrahlung'' coupling by some other coupling that
matches the same Abelian limit. This was done in a similar context
in the case of the average thrust~\cite{average_thrust}. It was
found there that such a replacement introduces a non-negligible
variation in the calculated observable, but the final impact on
the extracted value of $\alpha_s$ was rather small (of order
$1\%$). In the context of the distribution, the choice of the
``gluon bremsstrahlung'' coupling is almost imposed on us by
consistency with the NLL exponentiation kernel. Therefore we shall
not make such a modification. The simplest variation in the
renormalization-group equation is to replace it by the one-loop.
In this case the renormalon integrals~$\tilde{I}_p(\ln
Q^2/{\bar{\Lambda}}^2)$ are evaluated by eq.~(\ref{Ip_1loop}) instead of
eq.~(\ref{Ip_2loop}). Such a comparison is presented in
fig.~\ref{12_loop}.
The difference between the two curves is quite significant, in
particular in the peak region. Note, however, that contrary to the
two-loop case, the one-loop resummed result does not include all
the next-to-leading logs (the terms proportional to $\beta_1$ are
omitted). The difference between two- and three-loop\footnote{In
the spirit of \cite{CMW} and our discussion in sec. 4, the
appropriate value for $\beta_2$ should probably be set based on
the NNLO coefficient of the singular term $1/(1-x)$ in the
splitting function.} running coupling is expected to be much
smaller. However, it may well be non-negligible close to the peak,
so it is certainly worth consideration in the future. In the following
we will use the two-loop running coupling.

\subsection{Non-perturbative corrections}

It was realized
before~\cite{Shape_function1,Shape_function2,Shape_function3,Shape_function4}
that non-perturbative power corrections of the form $\lambda_n
(\Lambda/tQ)^n$ to any power $n$ become relevant around the
distribution peak. As shown in section~3.5, such corrections are
indeed expected based on the form of the ambiguity of the
perturbative result.  If all the powers are as important, one
would need an infinite number of parameters -- or a completely
arbitrary shape-function -- to bridge the gap between the
perturbative result and the data. In practice one assumes that
there is a region in $t$ where a power corrections expansion such
as~(\ref{SF_sum}) is dominated by its first few terms, or in other
words, that the shape-function can be described by its first few
moments. It is therefore essential that $\lambda_n$ do not grow
too fast with $n$. Having made this assumption, one can put a cut
$t_L$ on the fitted range in $t$ for any given energy, and fit the
data for $t>t_L$ using a reasonably flexible parametrisation of
the shape-function. This cut should scale like the distribution
peak, i.e. as $1/Q$, \beq t_L(Q)=t_L(M_Z)\,\frac{M_Z}{Q}. \eeq In
any case, the dependence of the results on the precise location of
this cut and on the specific functional form of the shape-function
must not be too large.

The simplest possibility is to place the cut at large enough
values of $t$ such that the only relevant non-perturbative
parameter will be $\lambda_1$, which controls the shift of the
distribution~(\ref{dist_shift}). This approach was suggested
in~\cite{DW_dist}. The advantage is that the perturbative
calculation is better under control and the ``model dependence'' in
the parametrisation of non-perturbative corrections is minimized.
The price, however, is that one does not use a large
part of the data, and in particular, one excludes the region where
the distribution changes sharply and is therefore expected to be
most constraining and most informative.

Here we use various different minimal $t$ cuts and perform fits
where a flexible shape-function is convoluted with the
perturbative distribution as well as fits with just a shift. For the
shape-function-based fits the minimal cut can be quite low, thus
using almost all available data. At the low end we choose
$t_L({\rm M_Z})=0.01$. This is equivalent to $Qt\gsim 4{\bar{\Lambda}}$
and it guarantees convergence of the Borel function approximation
within $1\%$, as shown in fig.~\ref{convergence}. Less ambitious cuts
$t_L({\rm M_Z})=0.02$ and $0.03$ are useful to check the
stability, which is indicative of whether the power-correction
expansion is convergent enough. For
the shift-based fit higher cuts must be used. Also here stability of
the results is important to guarantee that sub-leading power
corrections are not too relevant in the fitted range.

For simplicity we shall assume that the parameters of the shift or the shape-function are $Q$ independent,
even though we have shown that logarithmic dependence is expected in general. In principle, a fit that
allows logarithmic dependence of $\lambda_n$ according to eq.~(\ref{lambda_def}) is more appropriate.
But, since $\kappa_p$ are not known, the price will be having more parameters which can only be constrained
by very precise data at {\em several different energies}.
The available data is not constraining enough for such a fit.

\subsubsection{Fits with a shifted distribution}

Let us consider first a fit based on shifting the perturbative distribution.
We use experimental data from all energies in the range $14\, {\rm GeV} \leq Q \leq 189\, {\rm GeV}$
(summarised in table~\ref{chi_sq_exp} below). When available, the systematic errors have been added
in quadrature to the statistical errors.
Table \ref{shift_based_fit} summarizes the fit results for  $\alpha_s$ and $\lambda_1$ as a
function of the low cut.
\begin{table}[htb]
\caption{ Fits of $\alpha_s$ and
$\lambda_1$ to data in the range $14\, {\rm GeV} \leq Q \leq 189\, {\rm
GeV}$ with a lower limit $t_{L}({\rm M_Z})\, {\rm M_Z}/Q$ and
upper limit $t_{H}=0.32$. The last two columns show the obtained value of $\chi^2$ per degree
of freedom and the total number of data points used in the fit, respectively.
\label{shift_based_fit}}
\begin{center}
\begin{tabular}{ccccc}
\hline
$t_{L}({\rm M_Z})$ & $\alpha_s^{\MSbar}({\rm M_Z})$ & $\lambda_1$ &
 $\chi^2/{\rm dof} $ & points \\
     0.02       &  0.1088 $\pm$ 0.0002    &  2.71  $\pm$  0.11  & 0.97  & 231 \\
     0.03       &  0.1081 $\pm$ 0.0003    &  3.11  $\pm$  0.16  & 0.80  & 213 \\
     0.04       &  0.1085 $\pm$ 0.0004    &  2.81  $\pm$  0.23  & 0.68  & 193 \\
     0.05       &  0.1092 $\pm$ 0.0005    &  2.32  $\pm$  0.30  & 0.66  & 177 \\
     0.06       &  0.1100 $\pm$ 0.0006    &  1.76  $\pm$  0.40  & 0.64  & 165 \\
     0.07       &  0.1106 $\pm$ 0.0008    &  1.39  $\pm$  0.48  & 0.65  & 152 \\
     0.08       &  0.1108 $\pm$ 0.0009    &  1.27  $\pm$  0.53  & 0.62  & 140 \\
     0.09       &  0.1114 $\pm$ 0.0012    &  0.84  $\pm$  0.69  & 0.62  & 127 \\
     0.10       &  0.1119 $\pm$ 0.0014    &  0.57  $\pm$  0.81  & 0.64  & 119 \\
     0.11       &  0.1126 $\pm$ 0.0019    &  0.18  $\pm$  1.09  & 0.64  & 109 \\
     0.12       &  0.1130 $\pm$ 0.0029    & -0.08  $\pm$  1.26  & 0.61  & 101 \\
     0.13       &  0.1156 $\pm$ 0.0037    & -1.26  $\pm$  1.67  & 0.60  & 94 \\
     0.14       &  0.1123 $\pm$ 0.0045    &  0.32  $\pm$  2.48  & 0.60  & 91 \\
\hline
\end{tabular}
\end{center}
\end{table}
\begin{figure}[t]
\begin{center}
\mbox{\kern-0.5cm\epsfig{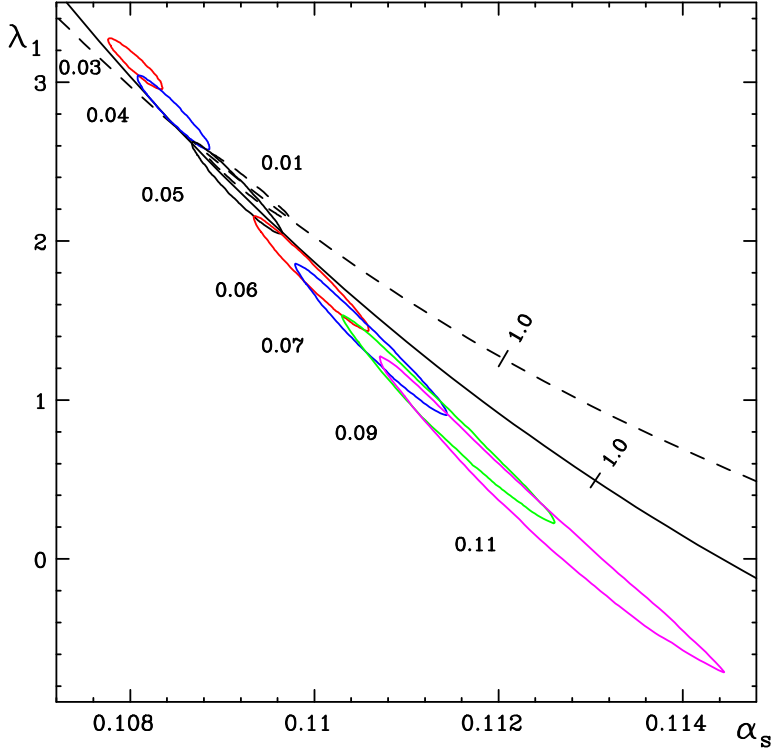}
}
\end{center}
\caption{The correlation between $\alpha_s$ and $\lambda_1$, as determined by a
shift-based fit with various cuts (full one-sigma contours) and by a
shape-function-based fit with a cut $t_L({\rm M_Z})=0.01$ (dashed one-sigma
contour). The additional lines indicate the values of $\lambda_1$ in fits with
a fixed $\alpha_s$. The full line corresponds to the shift-based fit with
$t_L({\rm M_Z})=0.05$ with and the dashed to the shape-function-based fit with
$t_L({\rm M_Z})=0.01$. When $\alpha_s$ is free (the centre of the corresponding
contour) \hbox{$\chi^2/{\rm dof}=0.66$} and $0.84$, respectively. The points
where $\chi^2/{\rm dof}=1$  are~indicated. \label{correlation} }
\end{figure}

First of all we note that the fits are very good: the $\chi^2/{\rm dof}$ 
values are unusually low. Next, we see that the results depend on the low cut.
Particularly stable results, as far as the extracted value of $\alpha_s$ is
concerned, are obtained for $t_{L}({\rm M_Z})\gsim~0.06$. Above this value
consecutive cuts yield $\alpha_s$ values that are within errors of each other.
This stability means that sub-leading power corrections in this region are
small enough to be neglected. Thus we regard the result 
$\alpha_s^{\MSbar}({\rm M_Z})\simeq 0.110$ as reliable. 
Too high cuts $t_{L}({\rm M_Z})\geq~0.10$ are
disfavoured because the number of fitted points decreases significantly (note
that the propagated experimental error increases accordingly). Cuts like
$t_{L}({\rm M_Z})= 0.05$  have the advantage of increased sensitivity to
$\lambda_1$. Lower cuts are probably too sensitive to sub-leading power
corrections. This is reflected in the stronger dependence on the cut and in the
deterioration of $\chi^2/{\rm dof}$. In the following we shall use 
$t_{L}({\rm M_Z})= 0.05$ for illustrations. 
It has the additional advantage of consistency
with the shape-function-based fit (see below).

The table and fig.~\ref{correlation} show that
there is a strong correlation between $\alpha_s$ and $\lambda_1$.
A similar correlation was found in \cite{DW_dist}. This means that
quantitative discussion of the values of non-perturbative
parameters, like $\lambda_n$ is relevant only for a fixed
$\alpha_s$.

In all the fits discussed so far we used an upper cut $t_{H}=0.32$ which is $Q$
independent. We checked that the location of this cut does not influence the
fit. For example, keeping the lower cut fixed $t_{L}({\rm M_Z})=0.05$ and
changing the upper cut between $t_H=0.29$ and $t_H=0.35$ (corresponding to a
total of $154$ to $184$ fitted points, respectively) the central value of
$\alpha_s$ changes by just $0.2\%$ (and that of $\lambda_1$ by $4\%$).

\begin{figure}[t]
\begin{center}
\mbox{\kern-0.5cm\epsfig{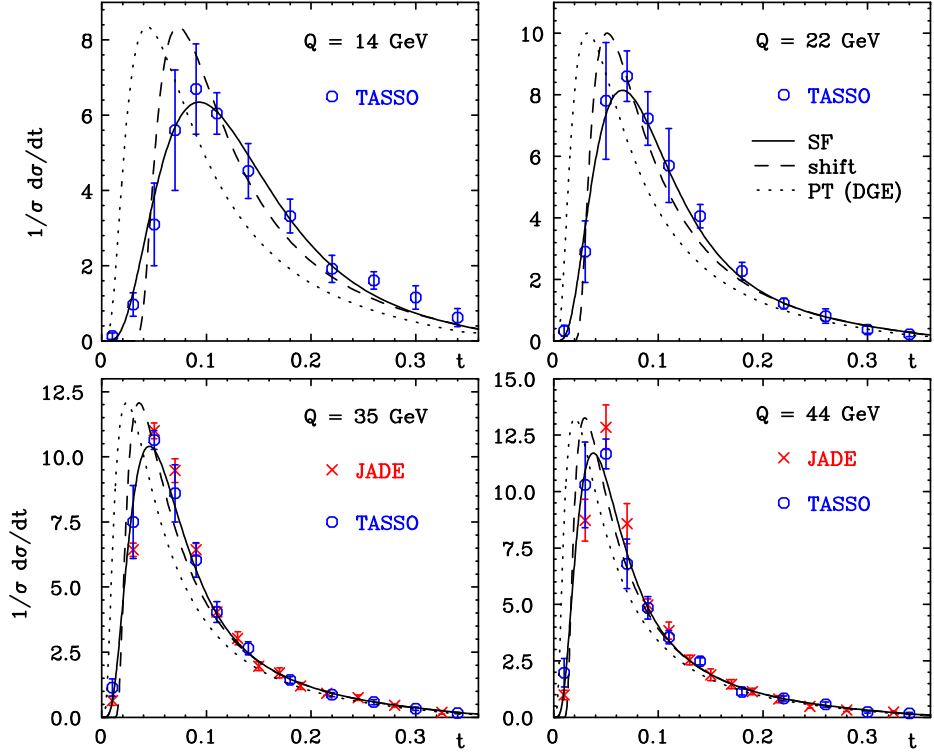}
}
\end{center}
\caption{Comparison with experimental data at $Q=14,\, 22,\,
35$ and $44$ GeV with the perturbative calculation (dotted) and
the corresponding fits using a shift (dashed) or a shape-function
(full line). The perturbative calculation (DGE) is performed using
Borel summation with two-loop runing coupling.
The fits are based on the range $t_L({\rm M_Z})\,\frac{\rm
M_Z}{Q}\,<t\,<\, t_H=0.32$, where $t_L({\rm M_Z})=0.05$ in the
case of a shift and $t_L({\rm M_Z})=0.01$ in the case of the
shape-function. The coupling is $\alpha_s^{\MSbar}({\rm
M_Z})=0.1093$ ($0.1092$) which is the best fit value in the
shape-function (shift) based fit.
 }
\label{fit1}
\end{figure}

\begin{figure}[t]
\begin{center}
\mbox{\kern-0.5cm\epsfig{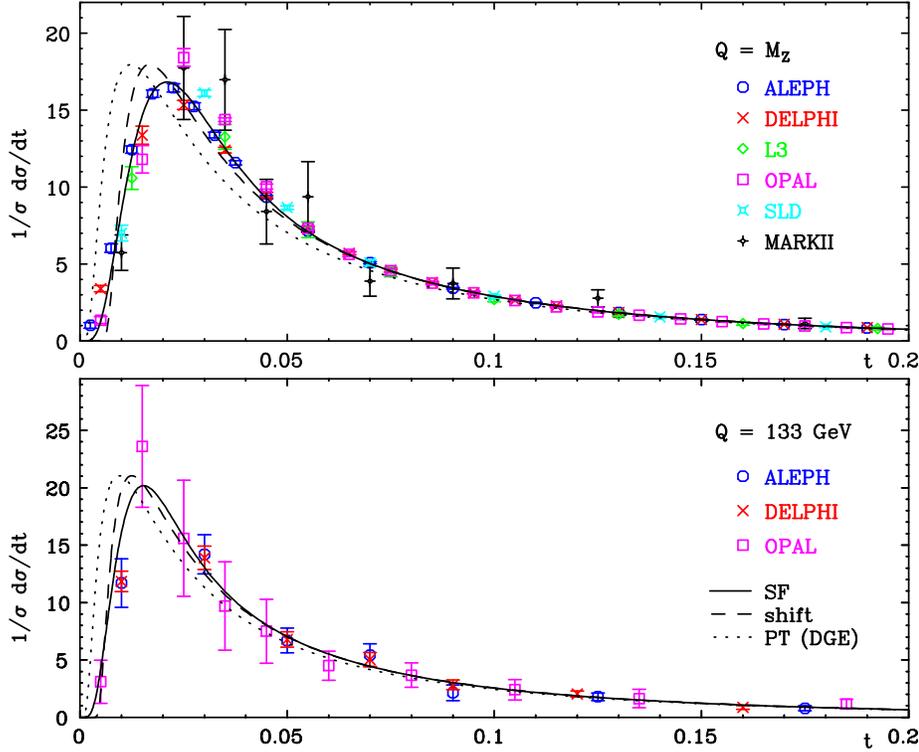}
}
\end{center}
\caption{Comparison with experimental data at $Q=91$ and
$133$ GeV with the perturbative calculation (dotted) and the
corresponding fits using a shift (dashed) or a shape-function
(full line).  See fig.~\ref{fit1} for further details.
 }
\label{fit2}
\end{figure}

\begin{figure}[t]
\begin{center}
\mbox{\kern-0.5cm\epsfig{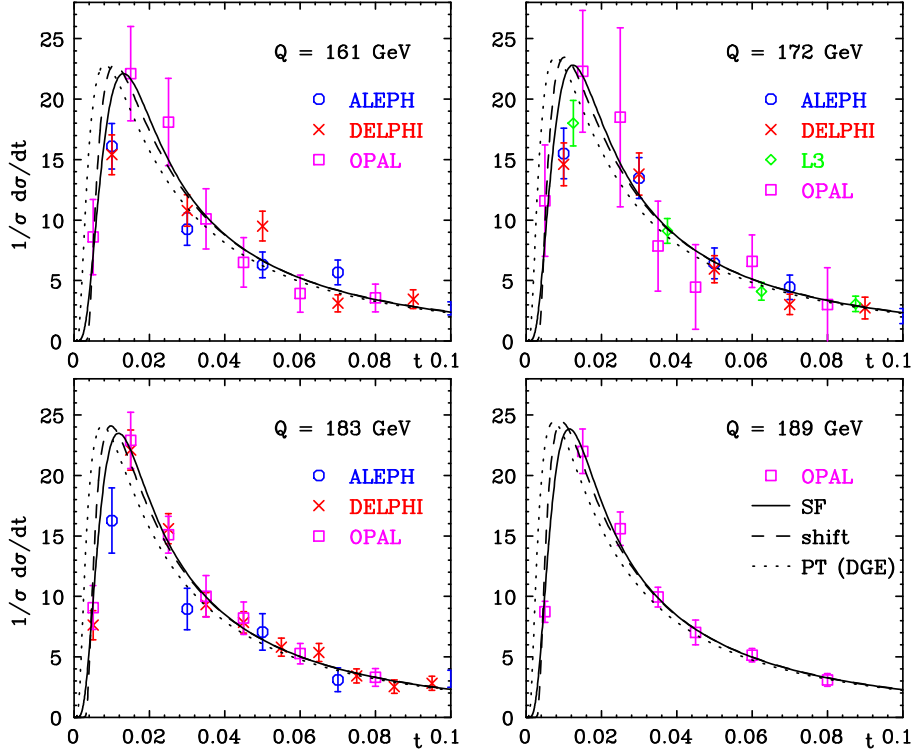}
}
\end{center}
\caption{Comparison with experimental data at $Q=161,\, 172,\, 183$ and 
$189$ GeV with the perturbative calculation (dotted) and the corresponding 
fits using a shift (dashed) or a shape-function (full line). 
See fig.~\ref{fit1} for further details.
 }
\label{fit3}
\end{figure}

Figures~\ref{fit1} through~\ref{fit3} show the fitted distribution and the
original perturbative distribution together with the experimental data. For
large $Q$ we show only the small $t$ region, which is more interesting. The
agreement between the shifted distribution and the data is good. Nevertheless,
it is clear that around the peak additional corrections are required (in the
case of ${\rm M_Z}$, for example). Indeed we saw that such corrections are
expected based on the renormalon ambiguity.

\subsubsection{Fits with a shape-function}

Consider now a fit where the perturbative distribution is
convoluted with a shape-function. According to sec.~3.5, this
means, in fact, two convolutions: one with the large-angle
shape-function $f_{\rm la}(\epsilon)$, defined in eqs.~(\ref{SF})
and (\ref{SF_sum}), and another which corrects for the collinear
contribution where we only include the leading term
in~(\ref{R_res_col_leading}),
\[
f_{\rm col}(Q^2,t)
=\delta(t)-\lambda_2\left[\frac{{\bar{\Lambda}}^2}{Q^2t}\right]_{+}\!\!\!\!\star\delta^{(1)}(t).
\]
The main difference with previous works where a shape-function was
used~\cite{Shape_function4,Korchemsky_Tafat} is in the starting
point: the perturbative distribution there is the standard NLL
result (with some cut-off regularization) while we start with the
DGE result avoiding the truncation of sub-leading logs. The latter provides
an analytic continuation of the perturbative treatment beyond its
original range of applicability.

In spite of the significant difference in the perturbative
distribution as compared
to~\cite{Shape_function4,Korchemsky_Tafat}, we consider their
ansatz for the shape-function appropriate also in our case,
barring the fact that it is positive definite. To
allow in principle the suppression of even moments as implied by
our analysis in sec. 3.5, we consider a function that is flexible
enough: \beq \label{col_cor} f_{\rm la}(\epsilon)=n_0
\epsilon^q\,(1+k_1\epsilon+k_2\epsilon^2)\,e^{-b_1\epsilon-b_2\epsilon^2}
\eeq where $n_0$ is fixed by the normalization requirement $\int
f(\epsilon) d\epsilon =1$ and $q, k_1,k_2,b_1$ and $b_2$ are free
parameters to be fixed by the fit to the data. We assume $q>0$ so
that $f_{\rm la}(0)=0$ and $b_2>0$ (or $b_2=0$ and $b_1>0$)
so that the shape-function decreases at large $\epsilon$.

Note that with a shape-function of this form, the central moments $\lambda_n$
do not have a definite sign. In particular both positive and negative values
for $\lambda_1$ are possible. When a cut-off regularization is used it is
intuitively clear that the shift is to the right: soft gluons widen the jets.
Indeed $\lambda_1$ was found to be positive in both the shift-based fit in
\cite{DW_dist} and in the shape-function
\cite{Shape_function4,Korchemsky_Tafat} based fit. The shift parameter
$\lambda_1$ was shown to be roughly consistent with the leading power
correction to average event-shapes \cite{DW_dist,Mil}, which is always
positive. The intuitive argument is not anymore valid when principal value
regularization is used instead. According to \cite{average_thrust}, in this
regularization $\lambda_1$ is still positive. We stress, however, that there is
no general reason to exclude a priori the possibility of a negative
$\lambda_1$.

\begin{figure}[t]
\begin{center}
\mbox{\kern-0.5cm\epsfig{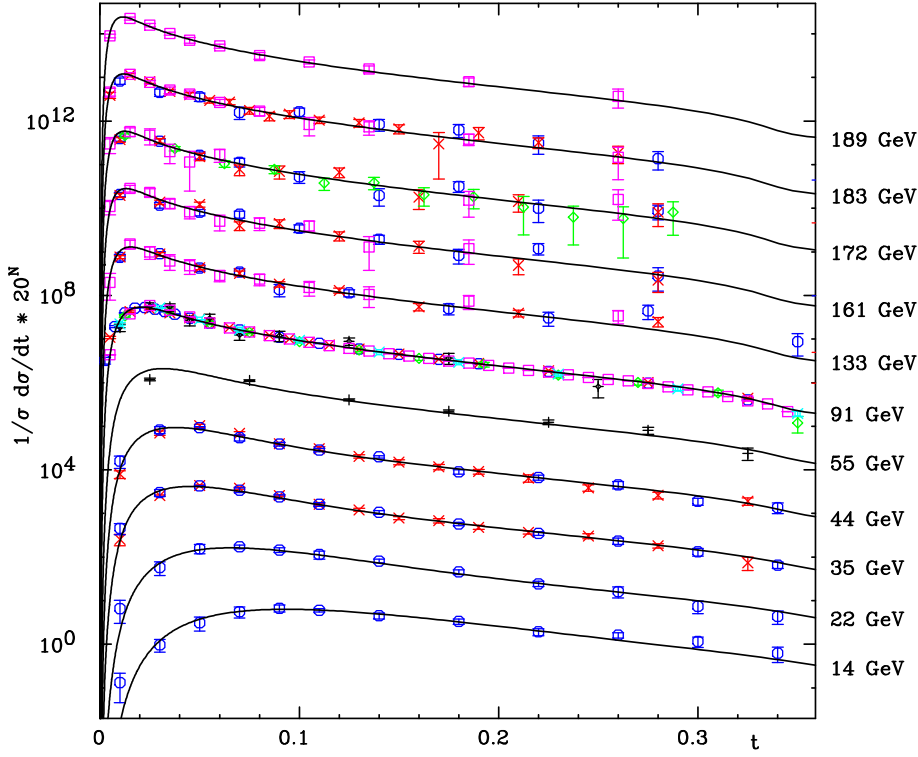}
}
\end{center}
\caption{Comparison of DGE convoluted with the best fit shape-function 
($b_2 = 0$) with experimental data for $Q=14$ up to $189$ GeV. Both
the DGE result and the data have been multiplied with a factor $20^N$,
with $N=0..10$ for $Q=14..189$ GeV.
Here $\alpha_s^{\MSbar}({\rm M_Z}) = 0.1093$. Data points with 
large errors ($\geq 100\%$) are not shown.}
\label{all_t_log}
\end{figure}

Performing a fit with~$f_{\rm la}(\epsilon)$ and the collinear 
correction~(\ref{col_cor}) with a lower cut $t_L({\rm M_Z})=0.01$ and an upper
cut $t_H=0.32$, we obtained a very good fit with $\chi^2/{\rm dof} =0.84$ (for
a total of 246 fitted data points). The fit results are shown in
figures~\ref{fit1} through~\ref{fit3}. The entire range of $t$ is shown for all
energies in fig.~\ref{all_t_log}.
The parameters in this case are
\begin{eqnarray}
\label{fit_f1}
\begin{tabular}{ll}
$\alpha_s^{\MSbar}({\rm M_Z}) = 0.1093 \pm 0.0004$&  \\
$q = 0.580  \pm 0.115 $&\\
$b_1 = 0.408 \pm  0.045$  &$k_1 = -0.099 \pm 0.034$\\
$b_2 = 0$ &$ k_2 = 2.5\cdot10^{-6} \pm 0.003$
\end{tabular}
\end{eqnarray}
corresponding to
\begin{eqnarray}
\label{mom_f1}
\lambda_1 &=& 2.34\pm 0.21\\ \nonumber
\lambda_2 &=& 0.37\pm 1.38 \,.
\end{eqnarray}
It turns out that the fit has a strong preference for a linear term in the 
exponent, so that if both $b_1$ and $b_2$ are free, then $b_2$ is set by the
fit to zero. In order to try another functional form, thus testing the
sensitivity of the fit results to the properties of the shape-function, we also
fitted the data in the same range with the function where the linear term in
the exponent is missing: $b_1\equiv 0$. In this case we got a best fit with
$\chi^2/{\rm dof} =0.88$, i.e. slightly worse than the previous one, with the
following parameters,
\begin{eqnarray}
\label{fit_f2}
\begin{tabular}{ll}
$\alpha_s^{\MSbar}({\rm M_Z}) = 0.1082 \pm 0.0003$&\\
$q =   1.53   \pm     1.23$ &\\
$b_1\equiv 0$& $k_1 = -0.422 \pm 0.052 $\\
$b_2 = 0.099 \pm 0.058 $& $k_2 = 0.052 \pm 0.010$
\end{tabular}\end{eqnarray}
corresponding to
\begin{eqnarray}
\label{mom_f2}
\lambda_1 &=& 2.92\pm 0.14\\ \nonumber
\lambda_2 &=& -3.52\pm 0.57 \, .
\end{eqnarray}

We see that most of the parameters of the shape-function are fairly well
constrained by the fit. Unfortunately, this does not imply that the central
moments of the shape-function are well determined. As
explained above, due to the correlation between $\alpha_s$ and the
non-perturbative parameters, it makes sense to fix $\alpha_s$ in the fit when
comparing the results for the two possible exponents. Doing so with
$\alpha_s=0.1093$ we still obtain good fits in both cases. For a linear term in
the exponent we get $\lambda_1 =  2.35 \pm 0.05$ ($\chi^2/{\rm dof} = 0.84$)
and for a square in the exponent we get $\lambda_1 = 2.45\pm 0.04$
($\chi^2/{\rm dof}=0.92$). Now the values of $\lambda_1$ are much closer. Note
that also the propagated errors become smaller, since $\alpha_s$ and
$\lambda_1$ are so strongly correlated.

Figure~\ref{correlation} shows how $\lambda_1$ varies when the fits with the
shift and with the shape-function (with a linear term in the exponent) are
performed with a fixed value of $\alpha_s$. As intuitively expected, forcing a
larger value of $\alpha_s$, $\lambda_1$ decreases. 
For $\alpha_s^{\MSbar}({\rm M_Z})\simeq 0.109$ 
similar values are obtained for $\lambda_1$  from the shift
and from the shape-function (note that this depends on the chosen lower cut),
however, for higher values of the coupling the values of $\lambda_1$ start
differing. In this case the shape-function is not so well approximated by a
$\delta$-function (a shift) and its higher central moments $\lambda_n$ play a
role.

To allow a quantitative discussion concerning the higher central moments let us
fix also $\lambda_1$ and compare the fits with the two exponents for the same
$\alpha_s$ and $\lambda_1$. We choose them to be the values obtained in our
best bit $(\ref{fit_f1})$: $\alpha_s^{\MSbar}({\rm M_Z})=0.1093$ and
$\lambda_1=2.34$.

Quantitative information on higher moments is theoretically important for
several reasons. First, it would be interesting to see the effect of the
collinear contribution. At the same time it would be interesting to check our
prediction that $\lambda_2$ (like other even central moments of the
shape-function) is suppressed. Even ignoring the strong dependence on
$\alpha_s$, these two ingredients are hard to address by fitting the data: if
$\lambda_2$ is small then the collinear correction is not important, and
conversely -- if the collinear correction is included a major part of the
effect of $\lambda_2$ is cancelled, so $\lambda_2$ would be very hard to
access. To deal with this we make three different fits with the same fixed
values of $\alpha_s$ and $\lambda_1$ mentioned above: with the collinear
correction, without it, and finally a fit where $\lambda_2$ is fixed to zero.
The results are summarised in table~\ref{lambda_2}.
\begin{table}[htb]
\caption{Determination of $\lambda_2$ and $\lambda_3$ for fixed 
$\alpha_s^{\MSbar}({\rm M_Z})=0.1093$ and $\lambda_1=2.34$ \label{lambda_2}}
\begin{center}
\begin{tabular}{c|ccc|ccc}
\hline
         &       \multicolumn{3}{c|}{$b_2=0$}                     & \multicolumn{3}{c}{$b_1=0$}  \\
\hline
                   &  $\lambda_2$   &  $\lambda_3$  & $\chi^2/{\rm dof} $ & $\lambda_2$      &  $\lambda_3$ & $\chi^2/{\rm dof} $  \\
with collinear     & $0.42\pm 0.54$ &  $-52\pm 12$  &  0.84               & $-0.75 \pm 0.43$ &  $-20\pm 7 $ &      0.95\\
no collinear       & $0.48\pm 0.52$ &  $-53\pm 12$  &  0.84               & $-0.69 \pm 0.42$ &  $-21\pm 7 $ &      0.95\\
$\lambda_2\equiv 0$&       $0$      &  $-44\pm 5$   &  0.84               &      $0$         &  $-31\pm 3 $ &      0.96 \\
\hline
\end{tabular}
\end{center}
\end{table}

The first conclusion is that $\lambda_2$ is not well determined by the fits:
even for fixed values of $\alpha_s$ and $\lambda_1$ the two functional forms
yield somewhat different results. Nevertheless, a vanishing $\lambda_2$ is
certainly not inconsistent with the data: the fit does not deteriorate much by
fixing it to zero. One also sees that the collinear correction is not
important: including it or not leaves $\lambda_2$ almost identically the same.
Had the value of $\lambda_2$ been significantly different from zero, such a
situation would have been impossible, but for a small (and not very well
constrained) $\lambda_2$ it is quite reasonable.

Table~\ref{lambda_2} also gives the $\lambda_3$ values. It is clear that
$\lambda_3$ is not constrained enough unless  $\alpha_s$, $\lambda_1$ and 
$\lambda_2$ are all fixed. Still, one thing that can be said in general
(varying the other parameters, including $\alpha_s$) is that negative values
for $\lambda_3$ are strongly preferred. Thus, a positive-definite ansatz for
the shape-function (in the principal-value regularisation!) is not adequate.
Note that the preferred signs of $\lambda_1$ (positive) and $\lambda_3$
(negative) are compatible with $\rho_{2p}$ being all positive (see
eq.~(\ref{rho_PV}) and (\ref{lambda_def})).

The shape-function-based fit can be considered an alternative way
to extract the value of $\alpha_s$. We saw already, that in spite
of the fact that the determination of the non-perturbative
parameters is far from being complete, $\alpha_s$ does not change
much. In particular, for the two functions used,
$\alpha_s^{\MSbar}({\rm M_Z})$ varies only from~$0.1082$
to~$0.1093$. A further stability check is to what extent it is
sensitive to the location of the low cut $t_L(Q)$. The answer is
provided by table~\ref{t_L_sens_SF}.
\begin{table}[htb]
\caption{
Fits of $\alpha_s$ and the shape-function (with a linear term in the exponent, $b_2=0$) as a function of the lower cut
$t_{L}({\rm Q})$. The upper cut is $t_{H}=0.32$.\label{t_L_sens_SF}}
\begin{center}
\begin{tabular}{cccccc}
\hline
 $t_{L}({\rm M_Z})$ & $\alpha_s^{\MSbar}({\rm M_Z})$ & $\lambda_1$  & $\lambda_2$ & $\chi^2/{\rm dof} $ & points \\
     0.010       &  0.1093 $\pm$ 0.0004    &  2.34  $\pm$  0.21  &  0.37  $\pm$  1.38  & 0.84 & 246 \\
     0.020       &  0.1101 $\pm$ 0.0005    &  2.06  $\pm$  0.22  &  2.61  $\pm$  1.56  & 0.73 & 231 \\
     0.030       &  0.1109 $\pm$ 0.0007    &  1.84  $\pm$  0.25  &  3.41  $\pm$  1.71  & 0.70 & 213 \\
     0.040       &  0.1108 $\pm$ 0.0011    &  1.88  $\pm$  0.31  &  2.80  $\pm$  1.84  & 0.63 & 193 \\
     0.050       &  0.1102 $\pm$ 0.0012    &  2.04  $\pm$  0.45  &  4.23  $\pm$  2.56  & 0.63 & 177 \\
\hline
\end{tabular}
\end{center}
\end{table}
The sensitivity to $t_L$ in the determination of $\alpha_s$ is rather small. In
particular it is smaller than in the case of a shift-based fit. There is still
some tendency that $\alpha_s$ becomes larger for higher (thus less ambitious)
cuts. While higher cuts are safer from the point of view suppressing
sub-leading power corrections, they are less constraining as far as the
multi-parameter shape-function is concerned. Note also that the propagated
experimental error on $\alpha_s$ and $\lambda_1$ increases for higher cuts.
``Ambitious cuts'' where the peak region itself is included in the fit , e.g.
$t_{L}({\rm M_Z})=0.010$, are therefore favoured. We conclude that according to
the shape-function-based fit $\alpha_s^{\MSbar}({\rm M_Z})$ ranges between
$0.108$ and $0.111$. These variations should be considered part of the
theoretical uncertainty.

Finally, we summarise in table~\ref{chi_sq_exp} the contribution to $\chi^2$ in the fit
by each experiment separately. The middle columns correspond to a fit of the
shifted distribution for $\alpha_s$ and $\lambda_1$ with a lower cut
$t_{L}({\rm M_Z})=0.05$ and an upper cut $t_{H}=0.32$. The columns on the right
correspond to a fit for $\alpha_s$ and the parameters of the shape-function
with a lower cut $t_{L}({\rm M_Z})=0.01$ and upper limit $t_{H}=0.32$. The OPAL
and SLD data sets at $91.2$ GeV are not included in the fit since they are not
consistent with the other data sets at this energy for small $t$. We note that
there are no particular trends in the $\chi^2/{\rm point}$ as a function of
energy.
\begin{table}[H]
\caption{Contribution to $\chi^2$ in the two fitting procedures. \label{chi_sq_exp}}
\begin{center}
\begin{tabular}{|lcc|cc|cc|}
\hline
 Experiment    & Reference  & Q [GeV]&  $\chi^2$\,(shift)&points(shift) &  $\chi^2\,({\rm SF})$& points\,(SF)\\
\hline
  TASSO & \cite{TASSO}                &  14.0 &    0.00  &    0 &    4.58  &  7    \\
  TASSO & \cite{TASSO}                &  22.0 &    1.43  &    2 &    3.17  &  8    \\
   JADE & \cite{JADE}                 &  35.0 &    3.80  &    6 &   29.31  & 11    \\
  TASSO & \cite{TASSO}                &  35.0 &    1.54  &    4 &    4.10  &  9    \\
   JADE & \cite{JADE}                 &  44.0 &    5.57  &    7 &   14.47  & 11    \\
  TASSO & \cite{TASSO}                &  44.0 &    9.11  &    5 &   12.86  &  9    \\
    AMY & \cite{AMY}                  &  55.0 &    2.49  &    4 &   10.75  &  5    \\
  ALEPH & \cite{ALEPH91}              &  91.2 &    4.03  &   10 &    8.12  & 17    \\
 DELPHI & \cite{DELPHI91}             &  91.2 &    9.68  &   13 &   17.18  & 17    \\
     L3 & \cite{L391}                 &  91.2 &    5.74  &    7 &    6.13  &  9    \\
 MARKII & \cite{MARKII91}             &  91.2 &    7.29  &    6 &    9.54  &  9    \\
   OPAL & \cite{OPAL91}               &  91.2 &  (27.07) &  (27)& (286.09) &(31)   \\
    SLD & \cite{SLD91}                &  91.2 &   (9.10) &   (6)&  (96.14) & (8)   \\
  ALEPH & \cite{ALEPH133}             & 133.0 &    6.08  &    7 &    6.28  &  8    \\
 DELPHI & \cite{DELPHI133_161_172_183}& 133.0 &    5.80  &    7 &    6.02  &  8    \\
   OPAL & \cite{OPAL133}              & 133.0 &    2.66  &    7 &    3.44  & 10    \\
  ALEPH & \cite{ALEPH161_172}         & 161.0 &    5.62  &    7 &   12.34  &  8    \\
 DELPHI & \cite{DELPHI133_161_172_183}& 161.0 &    9.96  &    7 &   12.19  &  8    \\
   OPAL & \cite{OPAL161}              & 161.0 &    2.64  &    8 &    3.41  & 10    \\
  ALEPH & \cite{ALEPH161_172}         & 172.0 &    5.94  &    7 &    6.22  &  8    \\
 DELPHI & \cite{DELPHI133_161_172_183}& 172.0 &    5.08  &    7 &    5.58  &  8    \\
     L3 & \cite{L3172}                & 172.0 &    3.81  &   10 &    4.14  & 11    \\
   OPAL & \cite{OPAL172_183_189}      & 172.0 &    3.50  &    8 &    3.77  & 10    \\
  ALEPH & \cite{ALEPH183}             & 183.0 &    4.53  &    7 &    8.35  &  8    \\
 DELPHI & \cite{DELPHI133_161_172_183}& 183.0 &    7.54  &   15 &    8.40  & 17    \\
   OPAL & \cite{OPAL172_183_189}      & 183.0 &    0.97  &    8 &    1.45  & 10    \\
   OPAL & \cite{OPAL172_183_189}      & 189.0 &    0.96  &    8 &    1.68  & 10    \\
\hline
  Sum  &  &      &  115.8  &  177  &  203.5  &  246\\
\hline
\end{tabular}
\end{center}
\end{table}

\subsection{Moments of the thrust distribution}

Given the assumption that the first few moments of the thrust 
$\left<t^n\right>$ are dominated by the two-jet configuration, 
and therefore by the logarithmically enhanced cross-section, one can
relate~\cite{Shape_function3,Shape_function4} the parameters $\lambda_n$ to the
power corrections of the moments:
\begin{eqnarray}
\label{shift_effect}
\left<t\right>_{{\rm two-jet}}&\simeq &\left<t\right>_{\PT}+\lambda_1{\bar{\Lambda}}/Q \\ \nonumber
\left<t^2\right>_{{\rm two-jet}}&\simeq&\left<t^2\right>_{\PT}
+2\lambda_1\left<t\right>_{\PT}{\bar{\Lambda}}/Q+\lambda_2{\bar{\Lambda}}^2/Q^2.
\end{eqnarray}
These predictions have been compared in \cite{higher_moments} with the power
correction from a {\em single} gluon emission, based on the characteristic
function~(\ref{F}),
\begin{eqnarray}
\label{SDG_effect}
\left<t\right>_{\SDG}&\simeq&\left<t\right>_{\PT}+\lambda{\bar{\Lambda}}/Q
 +{\cal O}({\bar{\Lambda}}^3/Q^3)\\ \nonumber
\left<t^2\right>_{\SDG}&\simeq&\left<t^2\right>_{\PT}+{\cal O}({{\bar{\Lambda}}^3/Q^3}).
\end{eqnarray}
The assumption that the two-jet region is dominant probably holds for the
average thrust. The immediate conclusion is that the shift of the distribution
$\lambda_1$ must coincide with the leading power-correction for the average,
$\lambda$. The latter was determined from a fit in~\cite{average_thrust} using
the principal value Borel sum regularization. 
For $\alpha_s^{\MSbar}({\rm M_Z})=0.110$, 
which is the best fit value in the case of the average thrust,
$\lambda\bar{\Lambda}=0.62\pm0.12$ GeV. This number can be compared with our
current fits to the distribution with a fixed 
$\alpha_s^{\MSbar}({\rm M_Z})=0.110$. We get 
$\lambda_1\bar{\Lambda}= 0.44 \pm 0.02$ GeV in a
shift-based fit with a cut $t_L({\rm M_Z})=0.05$ 
(corresponding to $\chi^2/{\rm dof}=0.67$ for 177 points) and 
$\lambda_1\bar{\Lambda}= 0.48 \pm 0.01$ GeV with
a shape-function-based fit with a cut $t_L({\rm M_Z})=0.01$ (corresponding to
$\chi^2/{\rm dof}=0.85$ for 246 points). The power correction extracted from
the distribution is somewhat lower than the corresponding power correction
extracted from $\left<t\right>$, however the discrepancy is not large.
Note that there is a difference between the resummation procedures used in the two cases. 
In the case of the distribution DGE is used, where the exponentiation plays a major
role but terms that are suppressed by $t$ are neglected. On the other hand in
case of the average a {\em single} dressed gluon renormalon sum is used, where
terms that are suppressed by $t$ (and higher powers) are included but multiple
gluon emission is neglected (apart from the NLO term).

An even more difficult case is that of the second moment $\left<t^2\right>$.
As explained in \cite{higher_moments}, the absence of the term
$2\lambda_1\left<t\right>_{\PT}{\bar{\Lambda}}/Q$ in (\ref{SDG_effect}) is an artifact
of using the single gluon approximation. On the other hand the absence of the
term $\lambda_2{\bar{\Lambda}}^2/Q^2$ is physically meaningful (a SDG analysis is the appropriate tool
to identify such a power-correction, if it exists). Our current finding that
the effect of $\lambda_2$ is cancelled at the level of the distribution
is consistent with that of eq.~(\ref{SDG_effect}). It is interesting to note, however, that
eq.~(\ref{SDG_effect}) was derived in~\cite{higher_moments} based on the entire
characteristic function, while our current conclusion concerning $\lambda_2$ is based on
the logarithmically enhanced cross-section alone.
If two-jet dominance applies to $\left<t^2\right>$, fixing $\lambda_1$ and
$\lambda_2$ by the distribution would allow the determination of the power-corrections to
$\left<t^2\right>$ based on~(\ref{shift_effect}). Unfortunately, this is
too optimistic: as one can verify, e.g.
by looking at the data for the distribution scaled by $t^2$ (see \cite{Moriond}), at ${\rm M_Z}$
this observable gets a significant contribution from the region $t\gsim 0.2$. As we saw at various
stages of the current investigation (see e.g. fig.~\ref{log_mom_fig} and
table~\ref{residual_scale_dep})
the logarithmically enhanced cross-section ceases to dominate at $t\gsim 0.2$.
It follows
that the second moment of the thrust depends on contributions we neglected here. Perturbative and
non-perturbative physics of three jets\footnote{Recently, non-perturbative effects in three-jet
observables were addressed~\cite{3jet}.}, including terms that are
sub-leading in $t$ at NNLO, must be taken into account for a reliable analysis of this quantity.

\subsection{Theoretical uncertainty}

As always, there are unproven assumptions and various limitations
to our approach. Here we address several issues shortly.
A quantitative analysis is restricted to those approximations we
have made where a particularly significant impact on the extracted
value of $\alpha_s$ is expected.

First in line is the uncertainty in the perturbative calculation.
Here there are two major issues: the first is the uncertainty due
to missing NNLO calculations.
The significance of NNLO corrections was emphasized 
recently in~\cite{Burby:2000bu}.
In our analysis it is reflected for instance in the large
renormalization scale dependence in the region $t\sim 1/3$.
Once such a calculation is available, matching with the DGE would allow a
significant reduction in the uncertainty, in particular,
concerning the extraction of $\alpha_s$. The second issue concerns
exponentiation: the DGE performed here is only approximate beyond
NLL. An exact NNLL calculation would allow one to test certain aspects of this
approach and complement the resummation.

Concentrating on the two-jet region we neglected effects that are
explicitly suppressed by $t$, but renormalon factorial growth is not
restricted to the logarithmically enhanced terms. Large numerical
coefficients appear also in the ``remainder function''.
Fig.~\ref{log_mom_fig} shows that at $t\simeq 0.2$, such
sub-leading terms become important. Eventually, DGE may be
generalized to included them. Currently it is not known whether
these terms exponentiate or not (and therefore they are treated
differently in different ``matching schemes''). One can give an
estimate of the possible impact of such terms by making a
modification of the log, as suggested in~\cite{CTTW}, e.g.  $\ln
1/t\longrightarrow \ln (1/t -1)$. Such a modification ensures that
the logarithmically enhanced cross-section will vanish at $t=1/2$,
like the physical cross-section, rather than at $t=1$. Some
results for a shift-based fit are summarized in
table~\ref{mod_log}.
\begin{table}[htb]
\caption{
Fits of $\alpha_s$ and $\lambda_1$ to data with a modified log.
The upper limit is $t_{H}=0.32$.
 The results are presented as a function of the lower limit. \label{mod_log}}
\begin{center}
\begin{tabular}{ccccc}
\hline
 $t_L({\rm M_Z})$ & $\alpha_s^{\MSbar}({\rm M_Z})$ & $\lambda_1$ & $\chi^2/{\rm dof} $ &  points  \\
     0.02       &  0.1100 $\pm$ 0.0003    &  2.24  $\pm$  0.10  & 1.16  & 231 \\
     0.03       &  0.1095 $\pm$ 0.0003    &  2.42  $\pm$  0.15  & 1.08  & 213 \\
     0.04       &  0.1106 $\pm$ 0.0005    &  1.80  $\pm$  0.25  & 0.94  & 193 \\
     0.05       &  0.1119 $\pm$ 0.0006    &  1.10  $\pm$  0.26  & 0.88  & 177 \\
     0.06       &  0.1134 $\pm$ 0.0007    &  0.42  $\pm$  0.29  & 0.83  & 165 \\
     0.07       &  0.1146 $\pm$ 0.0010    & -0.11  $\pm$  0.36  & 0.85  & 152 \\
     0.08       &  0.1151 $\pm$ 0.0012    & -0.31  $\pm$  0.41  & 0.84  & 140 \\
     0.09       &  0.1154 $\pm$ 0.0014    & -0.43  $\pm$  0.49  & 0.87  & 127 \\
     0.10       &  0.1160 $\pm$ 0.0017    & -0.64  $\pm$  0.58  & 0.88  & 119 \\
     0.11       &  0.1154 $\pm$ 0.0022    & -0.39  $\pm$  0.84  & 0.89  & 109 \\
\hline
\end{tabular}
\end{center}
\end{table}

The general quality of these fits is worse than in the case of the
non-modified log, table~\ref{shift_based_fit}. In addition there
is a stronger dependence of $\alpha_s$ on both cuts.
These two facts suggest that such a
modification is not quite the appropriate way to take into account
the fact that the physical cross-section vanishes at $t=1/2$.
Still, as a rough measure of the impact of the terms we neglected
these results are useful. For the favoured cuts
the difference in $\alpha_s$ is quite
appreciable. It changes from $\sim 0.110$ to $\sim 0.114$.
A similar study for the shape-function-based fit (with $t_L({\rm M_Z})$)
shows a somewhat
smaller change in $\alpha_s$: the central value changes from $\sim
0.109$ to $\sim 0.112$. Altogether we see that NNLO (and higher
order) terms which are not logarithmically enhanced have a
relatively significant impact on $\alpha_s$. We consider this as
the largest source of uncertainty.

One aspect in which the DGE may be improved is the running coupling formula.
Our entire analysis was performed with a two-loop running coupling.
The comparison with the one-loop case in fig.~\ref{12_loop} shows a relatively large variation.
We expect the difference between three-loop and two-loop to be much smaller, but
it may be non-negligible.

Next, there are uncertainties in the way hadronization corrections
are treated. Here we assumed that the dominant non-perturbative
corrections appear in a way that matches the ambiguity in the
perturbative calculation. The success this general approach has had in
event-shapes and elsewhere is encouraging. However, here we take
it one step further: we assume that the $\nu$ dependence of each
renormalon ambiguity (not only the first!) is reflected in the
dependence of the corresponding non-perturbative corrections. Of
course, any assumption we took in the perturbative calculation may
have an impact on the prediction. In particular,
\begin{itemize}
\item {} Having neglected effects that are explicitly suppressed by $t$ in the perturbative calculation,
we cannot expect the power correction analysis to be reliable beyond ${\cal O}(1/t)$ accuracy.
\item { } Since exponentiation is based on two-jet kinematics, non-perturbative effects that are
associated with three jets (where the recoil of the quark must be
taken into account) are not included.
\item { } The normalization of the thrust variable by the sum of energies ($Q$) in~(\ref{T_def2})
(rather than by the sum of the absolute values of the
three-momenta) was shown to affect the coefficient of the leading
power-correction in the case of the average
thrust~\cite{BB_DY,beneke}. There is a similar impact on the
$1/Qt$ correction in the case of the distribution, and probably
also on higher power corrections.
\item {} Having neglected the non-inclusive decay of the gluon in the perturbative calculation
in sec.~2, the effect is missed also on the non-perturbative
level. See discussion below eq.~(\ref{our_res}).
\item { } In our framework it is very natural to treat non-perturbative corrections as if they all
exponentiate, since the perturbative calculation indicating their necessity is a calculation of the exponent.
This is most clearly put in eq.~(\ref{R_res_NP}): the ambiguity appears in the
exponent and so should be the power corrections. It is probably true that in the peak region, the leading power
corrections indeed exponentiate, but in general there are power corrections that do not.
\end{itemize}

There are also hadronization effects that cannot be associated with ambiguities in the
perturbative calculation. For example, the latter does not contain information
on the spectrum of the hadrons produced. In a recent work Salam
and Wicke~\cite{Salam} showed that hadron mass effects introduce
important corrections, for average values of event-shape
variables.

Another fact we ignored here is that in reality also heavy quarks are produced. Since the data
contains all events whereas in the calculation the quarks are treated as massless, 
some systematic error is expected.

\section {Conclusions}

Many infrared safe observables in QCD are sensitive to soft and
collinear gluon radiation. This sensitivity appears at the
perturbative level in the form of large Sudakov logs. In order to
deal with this situation one must perform resummation of the
logarithmically enhanced cross-section. The resummation has two
aspects: exponentiation and integrals over the running coupling.
The Dressed Gluon Exponentiation (DGE) differs from previous resummation methods in the way the
second aspect is addressed. The approach makes a direct link
between the resummation of Sudakov logs and that of infrared
renormalons. We showed that a careful treatment of running coupling
integrals makes it possible to take into account a large class of logs, which
is fully consistent with both the exact exponent up to NLL
accuracy and renormalization group invariance. The latter is
realized only when all the powers of the log are resummed.
Moreover, sub-leading logs are factorially enhanced compared to
the leading ones. As a consequence the standard approximation
based on keeping the leading and the next-to-leading logs alone is
not always justified.

DGE is a method to calculate a well defined class of log-enhanced
terms based on exponentiation of the Single Dressed Gluon (SDG)
renormalon-sum. The basic idea is that the exponent can be
represented in terms of an integral over an observable-dependent
characteristic function times the running coupling, similarly to
the standard dispersive approach in renormalon
calculations~\cite{Beneke:1995qe,Ball:1995ni,DMW,beneke}. 
This representation is natural
since the exponentiation kernel is primarily associated with a
single gluon emission.  The method is quite general and can readily be
applied to other physical quantities. The first stage is to
calculate analytically the large $N_f$ or massive gluon
characteristic function, and then to identify the phase-space
limits in the integration over the gluon virtuality as well as the
specific terms in the characteristic function that contribute to
logarithmically enhanced terms in the perturbative coefficients.
This allows to obtain the logarithmically enhanced part of the SDG
cross-section to all orders. At the next stage the physical cross
section is obtained by exponentiating the SDG result. In the case
considered here this exponentiation is rather straightforward,
given the assumption that gluons are emitted independently from
the quarks and contribute additively to $1\,-$ thrust.
In making this assumption we ignored correlations in multi-gluon
emission that occur at higher orders. As a result, the
exponentiation is only approximate as far as sub-leading logs
(NNLL and beyond) are concerned. DGE cannot replace the
construction of the exact exponent as a solution of the evolution
equation up to a given logarithmic accuracy. It is rather
complementary to it, similarly to the way resummation in general
is complementary to fixed-order calculations.

One of the main results of this work is the observation that sub-leading Sudakov logs are factorially
enhanced compared to the leading logs. This observation was made first at the level of the SDG
cross-section, before exponentiation. At this level factorial growth appears only due to
the collinear limit of phase space.
The Borel function (in the large $\beta_0$ limit) has simple poles at $z=1$ and $2$.
However, the Borel integral
is still well defined thanks to the analytic continuation. A much more dramatic factorial growth appears at
the level of the exponentiated result in the Laplace conjugate variable~$\nu$. Here both the collinear
and large-angle (soft) limits of phase space contribute to
infrared renormalons. As a result of the large-angle contributions, the first infrared renormalon
appears at $z=1/2$, thus enhancing the divergence of the perturbative expansion as well as the
relative significance of sub-leading logs compared to the leading logs.
In the standard exponentiation
formula~\cite{CTTW} the most prominent enhancement of sub-leading logs, associated with the $z=1/2$
renormalon, appear due to the leading term in the splitting function, the same term that generates the leading
logs.

The traditional formulation of resummation is based on writing the
exponent as an expansion in the coupling, with coefficients $g_n(\xi)$ that
are functions of the combination $\xi\equiv L\beta_0
\alpha_s/\pi$, as in~(\ref{Exponentiation}) (or in
(\ref{log_expansion})). This way of organising the expansion is
convenient when working with a fixed logarithmic accuracy. In
this formulation, the enhancement of sub-leading logs translates
into the properties of $g_n(\xi)$: these functions become more
singular at $\xi=1/2$ and $\xi=1$ as $n$ increases, and in
addition they increase factorially. As was shown in
fig.~\ref{Sub_logs}, the convergence of the expansion~(\ref{log_expansion})
depends strongly on the coupling (and thus on $Q^2$). In the case
of the thrust sub-leading logs are quite important at all relevant
energies.  When using DGE and keeping all the logs the standard way of
organizing the expansion loses its attractiveness: as a
consequence of the factorial divergence of the functions $g_n(\xi)$ at large $n$
the series must be truncated. Truncation at the minimal term would
minimise the error, but this is not practical since the
enhanced singularity of increasing order functions implies that
the truncation order must itself be a function of $L$.
An elegant alternative was presented in section.~3.3: the exponent can be written in
terms of a finite set of analytic functions $I_p(\ln
Q^2/{\bar{\Lambda}}^2)$ with coefficients $\bar{r}_p$ that are polynomials
in $L$. Each function $I_p(\ln Q^2/{\bar{\Lambda}}^2)$ corresponds to a
single renormalon integral over a simple or a double pole. In this
way running coupling effects are resummed first and the logs only later.
As was shown in fig.~\ref{convergence}, the convergence of this resummation
method is quite remarkable.

The ambiguities of the perturbative result obtained by DGE can be used as a
template for the non-perturbative power corrections. In particular, we
saw that it is possible to identify the $t$ dependence associated
with each renormalon singularity in the exponent.
The leading power-correction for $t\gg \bar{\Lambda}/Q$ scales as
$1/Q$. It corresponds to a simple renormalon pole at $z=1/2$,
with an ambiguity which is proportional to $ 1/t$.
This results, upon exponentiation, in a shift of the
distribution, as predicted by~\cite{Shape_function1,Shape_function2,DW_dist}.
The physical origin of this correction is
the sensitivity of the thrust to large-angle soft gluon emission.
The same sensitivity gives rise to other renormalon singularities that are further
away from the origin. The corresponding ambiguities scale as~$\sim\bar{\Lambda}^n/Q^n$ and are
therefore less important at large $t$.
However, at small $t$ they behave as $1/t^n$, implying
non-perturbative corrections of the form $\lambda_n{\bar{\Lambda}}^n/(Qt)^n$.
As emphasised by Korchemsky and Sterman, these corrections are all important in the
distribution peak region. Since they exponentiate, it is natural to describe them by a non-perturbative
shape-function. The parameters $\lambda_n$ then correspond to the central moments of
this function.

Our calculation shows that the ambiguities~$\sim\bar{\Lambda}^n/Q^n$ with $n=2,4,\ldots$
vanish in the large~$\beta_0$ limit. This strongly suggests
that the parameters $\lambda_n$ for even $n$ are suppressed according to eq.~(\ref{lambda_def}).
In addition, the appearance of a double renormalon pole at $z=1$ leads to a collinear power
correction proportional to $\lambda_2\, {\bar{\Lambda}}^2/(Qt)^2$, which has the opposite sign to the
$\lambda_2$ term in the large-angle contribution. This leads to an almost complete cancellation
of the overall effect of $\lambda_2$.

The immediate phenomenological implication of these findings
is that by shifting the distribution
one should be able to fit the data in a wider range (closer to the peak) than a priori expected:
there is no new effect beyond that of the shift so long
as~$\frac{\lambda_3}{\lambda_1}\,\frac{{\bar{\Lambda}}^2}{Q^2t^2}\ll 1$, instead
of~$\frac{\lambda_2}{\lambda_1}\,\frac{{\bar{\Lambda}}}{Qt}\ll 1$.
The phenomenological analysis indeed shows very successful shift-based fits, even for rather low
$t$ values (this was first observed in~\cite{DW_dist}). Clearly, in the peak region
itself, a shift is not enough, and a convolution with a shape-function must replace it.

Finally, our main conclusions from the phenomenological analysis
can be summarised as follows: we obtained good fits based on the
DGE result both with a shifted distribution and with a shape
function. The two methods are fairly consistent with each other.
This demonstrates that the data in the peak region can be used to
extract $\alpha_s$ provided the appropriate resummation is
performed and the power corrections are included. In the fits, the central
value of $\alpha_s^{\MSbar}({\rm M_Z})$ ranges between $0.108$ and
$0.115$. The most important factor which limits the determination of
$\alpha_s$ is the absence of a NNLO calculation. In particular,
terms that are explicitly suppressed by $t$ at NNLO are important,
as can be deduced from the fact that the modification of the log:
$\ln 1/t \longrightarrow \ln (1/t-1)$ can change the extracted
$\alpha_s^{\MSbar}({\rm M_Z})$ by~$4\%$. 
There are also non-negligible differences~($3\%$) between different 
fitting procedures, such as dependence on the low $t$ 
cut, the use of a shift or a shape-function, and the 
parametric form of the latter. Other factors
contributing to the uncertainty are discussed in sec.~5.4.

The correlation between the non-perturbative parameters
$\lambda_n$ and $\alpha_s$ is strong. Consequently a quantitative
discussion is restricted to the case where $\alpha_s$ is fixed. 
Then the shift
parameter, or the first moment of the shape function, $\lambda_1$
is reasonably well determined (see fig.~\ref{correlation}). 
Even when a modified log is used,
the value of $\lambda_1$ as a function of $\alpha_s$ remains
roughly the same. The sub-leading power-corrections are harder to
constrain by the data. To determine the higher central moments of
the shape-function it is useful to fix also $\lambda_1$. 
As an example, fixing $\alpha_s$ and $\lambda_1$ to 
their best fit values we find that $\lambda_2$ is small and that
the fit is not affected much by fixing it to zero (see table~\ref{lambda_2}). 
Thus our theoretical prediction that the effect of $\lambda_2$ is small can
be consistent with the data. A clear preference of the fits for a
shape-function which is not positive-definite was observed.

\vspace{50pt}

\noindent
{\large {\bf Acknowledgments}}

\vspace{30pt}

\noindent
We would like to thank Stefano Catani, Gregory Korchemsky and Gavin Salam for very interesting and
useful discussions and Otmar Biebel and Stefan Kluth for useful correspondence.
\newpage

\begin{appendix}
\section{Numerical calculation of the NNLL part of the NLO coefficient}

To calculate the NNLL part of the NLO coefficient we have used the EVENT2~\cite{EVENT2}
program. Writing the thrust-distribution on the form
\begin{eqnarray}\label{eq:dsig_nlo}
\left.\frac{1}{\sigma_0}\frac{d\sigma}{dL}(Q^2,L)\right|_{\mbox{\tiny NLO}}
& = & {\cal A}(L)a(Q^2)+{\cal B}(L)a(Q^2)^2,
\end{eqnarray}
where $L=\ln(1/t)$, $a=\alpha_s^{\MSbar}/\pi$, and ${\cal A}(L)$ is
the analytically known leading order coefficient, the program
calculates the next-to-leading order coefficient ${\cal B}(L)$ in
the $\overline{\mbox{MS}}$ scheme using Monte Carlo integration.
Note that the normalisation in Eq.~(\ref{eq:dsig_nlo}) is to the
Born cross-section $\sigma_0$ and not to the total cross-section 
$\sigma=\left(1+\frac 34 C_F \,a(Q^2)+\ldots\right)\sigma_0$.
The program is written such that the different colour factor parts
can be calculated separately (this was done before using the EVENT 
program~\cite{EVENT} in \cite{Magnoli:1990mt,Kluth_Biebel}) 
in the following way,
\begin{eqnarray*}
{\cal B}(L)& = & C_F^2{\cal B}_{C_F}(L)+C_FC_A{\cal B}_{C_A}(L)+C_FT_FN_F{\cal B}_{T_F}(L).
\end{eqnarray*}

Since we are interested in the large $L$ (small $t$) region we
decreased the invariant mass cut-off on any pair of partons,
$m_{ij}^2>\mbox{CUTOFF}Q^2$, in the program to $10^{-14}$ from the
default $10^{-8}$ in order not to restrict the phase-space for
small $t$. To increase the sampling of the small $t$ region we
also changed the parameters $\mbox{NPOW1}$ and $\mbox{NPOW2}$ to 4
instead of the default values $2$. The calculations presented below
are based on $10^9$ events (samplings of the integral). We used a
linear binning in $L$ and the bin size was chosen to be $\Delta
L=0.5$.

\begin{figure}[t]
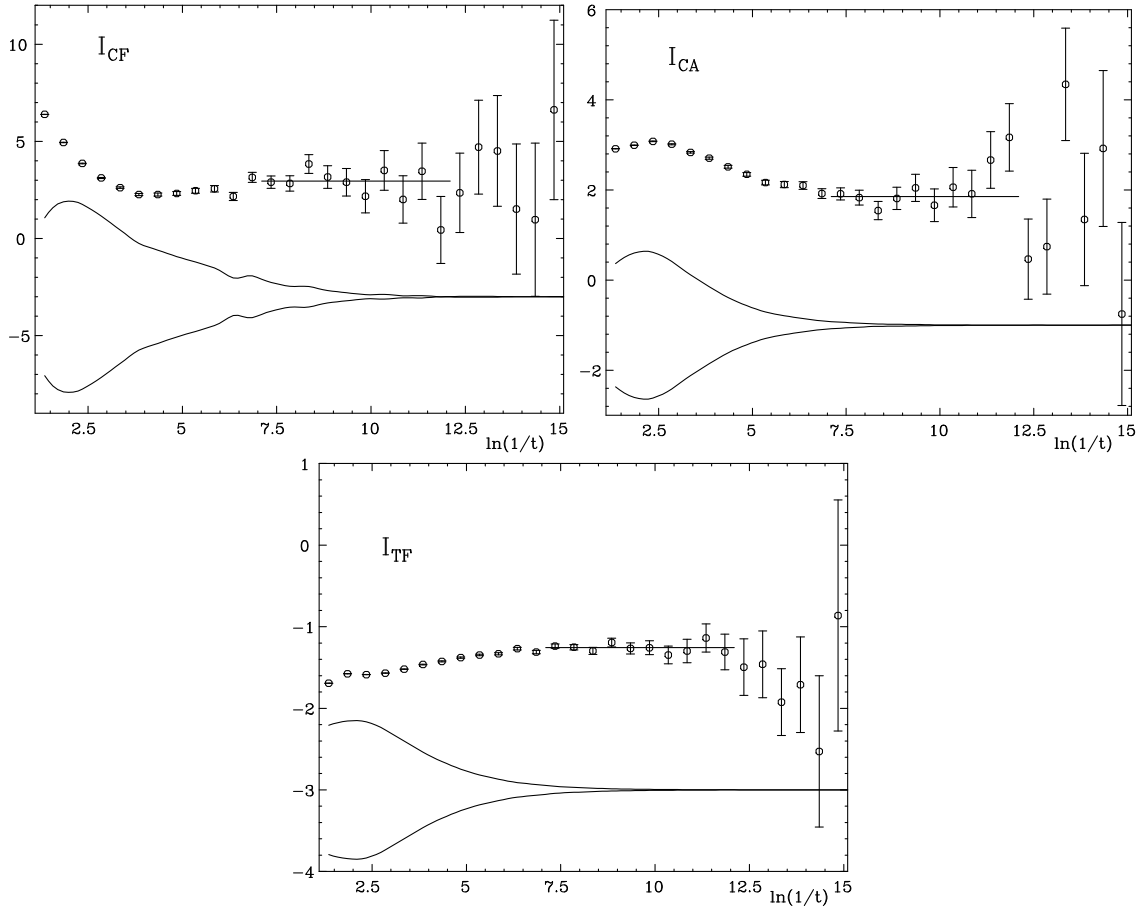

\begin{center}
\epsfig{file=cfpart.ps,width=6cm,angle=90}
\epsfig{file=capart.ps,width=6cm,angle=90}
\epsfig{file=tfpart.ps,width=6cm,angle=90}
\end{center}
\caption[0]{Results of the numerical calculation of the NLO
coefficients ${\cal B}_{i}$ with the NLL terms subtracted for the
different colour factors. The band shows the estimated systematic
error from sub-leading corrections when extracting the NNLL
coefficients. The results of the fits are shown as straight lines
in the regions fitted.} \label{fig:integrals}
\end{figure}

For each bin in $L$ the Monte Carlo integration gives the integrated
cross-section.  From these integrals we then subtract the known NLL
parts (except the constant contributions which will be dealt with separately) 
and normalise to the bin size giving the following integrals for each bin,
\begin{eqnarray*}
I_{i} & = &
\frac{1}{\Delta L}\int_{L_{\min}}^{L_{\max}}
\left({\cal B}_{i}(L)-{\cal B}_{i}^{\mbox{\tiny NLL}}(L)\right) dL  \,  ,
\end{eqnarray*}
where $i=C_F,\, C_A,\, T_F$, and
\begin{eqnarray*}
{\cal B}_{C_F}^{\mbox{\tiny NLL}}(L)& = &
2L^3-\frac{9}{2}L^2+\left(\frac{13}{4}-\pi^2\right)L  \, ,\\
{\cal B}_{C_A}^{\mbox{\tiny NLL}}(L)& = &
-\frac{11}{4}L^2+\left(\frac{\pi^2}{6}-\frac{169}{72}\right)L  \, , \\
{\cal B}_{T_F}^{\mbox{\tiny NLL}}(L)& =&
L^2+\frac{11}{18}L \, .
\end{eqnarray*}
The results for the different integrals are shown in
fig.~\ref{fig:integrals}.
If it were not for sub-leading corrections, the integrals $I_i$ would
coincide with the NNLL coefficients ${\cal B}_{i,1}$.
In principle one could take into account sub-leading terms by fitting
a more general form to the integrals $I_i$. We choose instead
to estimate a systematic error from sub-leading corrections for each bin
in $L$. The estimate is based on assuming that the most important
form of sub-leading corrections has the same power of logs as the leading
log multiplied by one factor of $t$. The size of the coefficient
multiplying this  term is then assumed to be the same as
${\cal B}_{i,1}$ estimated from the integral $I_i$. The systematic error
for each bin is thus estimated by, e.g.
\begin{eqnarray*}
\Delta I_{C_F}& = &
\frac{I_{C_F}}{\Delta L}\int_{L_{\min}}^{L_{\max}} L^3\exp(-L) dL \,  .
\end{eqnarray*}
These errors are illustrated as bands in fig.~\ref{fig:integrals}.
When making the fit these systematic errors are added in
quadrature to the statistical errors from the Monte Carlo
integration.

From fig.~\ref{fig:integrals} it is clear that for $L \lsim 7$ the
sub-leading corrections start to become important. It is also
clear that for $L\gsim 12$ the statistical errors start to
increase significantly\footnote{For even larger $L$ the
phase-space limitation imposed by the invariant mass cut-off becomes
clearly noticeable.}. Thus we choose the fit range $7.1<L<12.1$.
Fitting a constant in this range
 gives the following results,
\begin{eqnarray*}
{\cal B}_{C_F,1} & = & 2.95 \pm 0.27 \pm 0.11 \, , \\
{\cal B}_{C_A,1} & = & 1.855 \pm 0.081 \pm 0.093  \, ,\\
{\cal B}_{T_F,1} & = & -1.255 \pm 0.022 \pm 0.010 \, .
\end{eqnarray*}
where the first error is from the fit and the second from the maximal
variation when varying the upper and lower limits for the fit by
$\pm 1$.

From these results we can now get the NNLL term ${\cal G}_{21}$ in the
order $a^2$ contribution to $\ln R$,
$$\ln R = \left[{\cal G}_{12}L^2+{\cal G}_{11}L\right]a+\left[{\cal G}_{23}L^3+
{\cal G}_{22}L^2+{\cal G}_{21}L\right]a^2+\cdots.$$
For the $C_A$ and $T_F$ parts
these are just the same as the ${\cal B}_{i,1}$ but for the $C_F$ part
one has to subtract the contribution coming from the expansion
of the exponent and the different normalisation used in
Eq.~(\ref{eq:dsig_nlo}), $(\pi^2-3)/4$.
Adding the errors in quadrature and writing the result as
\begin{eqnarray*}
{\cal G}_{21} & =&
{\cal G}_{\beta_0,1}C_F\beta_0+{\cal G}_{C_F,1}C_F^2+{\cal G}_{C_A,1}C_AC_F
\end{eqnarray*}
we get
\begin{eqnarray*}
{\cal G}_{C_F,1} & = & 1.23 \pm 0.29 \, ,  \\
{\cal G}_{C_A,1} & = & -1.60 \pm 0.14 \, ,  \\
{\cal G}_{\beta_0,1} & = & 3.766 \pm 0.070 \, .
\end{eqnarray*}

These results can be compared with Eq.~(\ref{NLO_exp_of_DGE}) where
\begin{eqnarray*}
{\cal G}_{C_F,1} & \simeq & 0.13 \, ,  \\
{\cal G}_{C_A,1} & \simeq & -0.73 \, ,  \\
{\cal G}_{\beta_0,1} & = & 3.75 \, ,
\end{eqnarray*}
in order to examine to what accuracy the DGE generates the
sub-leading logs. The coefficient which is directly calculated in
our approach is ${\cal G}_{\beta_0,1}$. In this case the agreement
is very good. As expected, the other coefficients which are
generated by the exponentiation do not agree. It is encouraging,
however, that their values are smaller and that they have the
correct signs. Thus, including the corresponding terms at NNLO and
beyond is certainly an improvement.

For easy comparison with earlier results~\cite{CTTW,OPAL91,Akers:1995rx} 
we also write the result in the expansion
$$\ln R =
\left[G_{12}L^2+G_{11}L\right]a/2+
\left[G_{23}L^3+G_{22}L^2+G_{21}L\right]a^2/4+\cdots,$$
with $G_{21}= G_{T_F,1}+G_{C_F,1}+G_{C_A,1}$.
Our results (for $N_F=5$) are then given by
\begin{eqnarray*}
G_{C_F,1} & = & 9 \pm 2  \, , \\
G_{C_A,1} & = & 30 \pm 2  \, , \\
G_{T_F,1} & = & -16.7 \pm 0.3 \, .
\end{eqnarray*}
The result for the sum $G_{21}=22\pm 3$ agrees within errors with
earlier results, but the results for the colour factors $C_A$ and
$T_F$ are significantly different \cite{Akers:1995rx}. The reason
for these discrepancies can probably be found in the different
ranges in $L$ used for the fit, namely $7.1 < L < 12.1$ and $3.1 < L < 6.7$ respectively.

\end{appendix}

\end{document}